\documentclass[useAMS,usenatbib,psfig]{mn2e}

\usepackage{epsfig,amssymb,appendix}

\newcommand{\HI}{H\,{\sc i}}

\newcommand{\kms}{~km\,s$^{-1}$}

\newcommand{\MHI}{$M_{\rm HI}$}

\newcommand{\Msun}{$M_{\odot}$}

\title[HI in groups]{Southern GEMS groups II:  HI distribution, mass functions and HI deficient galaxies\thanks{The
       observations were obtained with the Australia Telescope which
       is funded by the Commonwealth of Australia for operations as a
       National Facility managed by CSIRO.}}
       \author[Kilborn et al. ]{Virginia A. Kilborn$^{1,2}$\thanks{E-mail:
       vkilborn@swin.edu.au}, 
       Duncan A. Forbes$^{1}$,
       David G. Barnes$^{1,3}$, 
       B\"arbel S. Koribalski$^{2}$,\newauthor       
       Sarah Brough$^{1}$,
       and Katie Kern$^{1,2}$\\
       $^1$Centre for Astrophysics \& Supercomputing, Swinburne University of Technology,  Mail H39, PO Box 218, Hawthorn, VIC 3122, Australia\\
       $^2$Australia Telescope National Facility, CSIRO, 
       P.O. Box 76, Epping, NSW 1710, Australia\\
       $^3$School of Physics, University of Melbourne,  Parkville, VIC 3010, Australia\\}

\begin{document}

\date{}

\pagerange{\pageref{firstpage}--\pageref{lastpage}} \pubyear{2005}

\maketitle

\label{firstpage}

\begin{abstract}

We investigate the neutral hydrogen (\HI) content of sixteen groups
for which we have multi-wavelength data including X-ray
observations. Wide-field imaging of the groups was obtained with the
20-cm multibeam system on the 64-m Parkes telescope. We have detected
ten previously uncatalogued \HI\ sources, one of which has no visible
optical counterpart. We examine the \HI\ properties of the groups,
compared to their X-ray characteristics, finding that those groups
with a higher X-ray temperature and luminosity contain less \HI\ per
galaxy.  The \HI\ content of a group depends on its morphological
make-up, with those groups dominated by early-type galaxies containing
the least total \HI. We determined the expected \HI\ for the spiral
galaxies in the groups, and found that a number of the galaxies were
\HI\ deficient. The \HI\ deficient spirals were found both in groups
with and without a hot intra-group medium. The \HI\ deficient galaxies
were not necessarily found at the centre of the groups, however, we
did find that two thirds of \HI\ deficient galaxies were found within
about 1 Mpc from the group centre, indicating that the group
environment is affecting the gas-loss from these galaxies. We
determined the \HI\ mass function for a composite sample of 15 groups,
and found that it is significantly flatter than the field \HI\ mass
function. We also find a lack of high \HI-mass galaxies in groups. One
possible cause of this effect is the tidal stripping of \HI\ gas from
spiral galaxies as they are pre-processed in groups.

\end{abstract}


\section{Introduction}

Environment has an important effect on the evolution of galaxies. The
observation of \citet{butcher78} that high-redshift clusters contain a
much higher percentage of blue spiral galaxies than present-day
clusters indicates possible evolution and transformation of spirals in
the densest regions of the Universe. Further evidence is seen in the
star formation rates of galaxies in clusters which are much lower than
those galaxies of similar morphology in the field
(e.g. \citealt{lewis2002,gomez2003}). One outstanding question is
whether spiral galaxies are somehow transformed into the present-day
lenticular, or S0, galaxies. The mechanism for such a transformation
is still not clear: Stripping, interactions, and the cessation of star
formation leading to the fading of a spiral disk have all been
proposed \citep{wilman09}.

It is now accepted that the galaxy group environment plays an
important influence on the evolution of galaxies. Star formation is
suppressed with respect to the field, in galaxies which lie in
densities equivalent to the group environment
\citep{lewis2002,gomez2003}, and over half of all galaxies live in the
group environment \citep{eke2004}. Intra-group X-rays are observed in
dynamically evolved groups, and these groups also tend to have a
central, bright elliptical galaxy and high early-type fraction in the
group (e.g. \citealt{mulchaey1998}), similar as is seen in the much
denser, and rarer cluster environment.

One of the first signs of a spiral galaxy undergoing a transformation
is the cessation of star formation, and the loss of cool gas. The
latter is detectable via observations of the neutral hydrogen (HI) in
spiral galaxies. It is well-known that spiral galaxies in rich
clusters tend to contain less \HI\ than field spirals of similar
morphology and size - these galaxies are ``\HI\ deficient''
(e.g. \citealt{magri1988,cayatte1990,solanes2001,gavazzi05}). There
have been several claims of a similar \HI\ deficiency detected in
loose groups of galaxies
\citep{sengupta2007,sengupta2006,kilborn05,omar2005,chamaraux2004}. Galaxy
groups are the ideal environment for galaxy-galaxy interactions due to
the low relative velocities of the galaxies, and coupled with the
observation of \HI\ deficient galaxies in groups where there is no hot
intra-group gas, galaxy interactions could be a way of removing \HI\
from spiral galaxies. However, the detection of ram pressure
stripping of a galaxy in a loose group \citep{rasmussen2006} begs the
question of the frequency and significance of this process. Such a
question can only be answered by a consistent blind \HI\ survey of
loose groups, in combination with multi-wavelength data, including
X-ray data.

As the \HI\ distribution of galaxies changes with environment, so does
the \HI\ mass distribution (hereafter \HI\ mass function). The \HI\
Parkes All Sky Survey (HIPASS) shows a trend for a steepening low-mass
slope in denser environments \citep{zwaan2005}. Conversely,
\cite{springob2005} find the low-mass slope flattens with denser
environment. Both studies have biases however - the HIPASS sample used
only \HI\ detected galaxies to measure the density of the environment,
and the \cite{springob2005} study was based on an optically selected
sample of galaxies. There have been a limited number of blind \HI\
studies of groups and clusters. \cite{freeland09} found an \HI\ mass
function with a low-mass slope consistent with being flat in a blind
survey of eight groups, and a flat low-mass slope has also been observed
in both the Virgo cluster \citep{davies2004,rosenberg2002} and the
Canis Venatici region \citep{kovac2005}. Conversly, a steep low-mass
slope was found for the Centaurus A group \citep{banks1999}.

To address some of the issues and inconsistencies detailed above, we
are conducting a study on the effect of the group environment on
galaxies, the Group Evolution Multi-Wavelength Study (GEMS; \citealt{o4,f6}). In
particular, we have conducted a wide-field \HI\ imaging study of sixteen
nearby GEMS groups, using the Parkes radiotelescope. Our data gives us
optically unbiased \HI\ observations of a $\sim 5.5^{\rm o} \times
5.5^{\rm o} $ region around the groups, and a velocity coverage of
$\sim 2000$ \kms. \cite{brough06b} studied the dynamics, group
membership, and optical properties of the sixteen groups. In this paper we detail the \HI\
properties of the sixteen groups, and  we probe the
\HI\ mass function in groups and surrounding environment.

All velocities quoted are heliocentric, in the optical convention, and
we use $H_0$ = 70 \kms\ Mpc$^{-1}$.

\section{Sample Selection and Data Reduction}
 The full GEMS sample includes 60 galaxy groups, chosen from ten
 optical catalogues, the majority with recession velocities between
 1000--3000 \kms\ and with a ROSAT PSPC exposure time of more than
 10,000s within 20$^{\prime}$ of the group position. The full
 selection criteria are given in Osmond \& Ponman (2004) and Forbes et
 al. (2006). We chose 16 groups that were near or South of Declination
 0$^{\rm o}$, for wide-field \HI\ imaging at Parkes.  The groups
 selected for \HI\ observations have a variety of X-ray properties,
 from just the central galaxy being detected (6 groups), to diffuse
 group-scale X-ray emission (8 groups), to no X-ray detection at all
 (2 groups). Properties of the sixteen observed groups are given in
 Table~\ref{tab:groups_properties}, including the number of members,
 X-ray and \HI\ parameters, and early-type fraction. Observing
 parameters for the groups, including central observing position,
 central observing velocity, and number of galaxies detected in \HI\,
 are given in Table~\ref{tab:groups}. We also list the distance to
 each group from \citet{brough06b}, who find the most accurate
 distance available to each group either from surface brightness
 fluctuation measurements, globular cluster luminosity functions, or
 the corrected mean group velocity.

\begin{table*} 
\caption{Details of the sixteen groups}
\label{tab:groups_properties} 
\begin{tabular}{lrlllcccccc}
\hline
Group    & No. of gals &  log$_{10}L_X$($R_{500}$) &$T_X$& $HI_{tot}$ &X-ray Emission& Central gal. type&$f_{early}$ &  & \\
         &             & [erg s$^{-1}]$             &[keV] &[$10^8$\Msun]  &  & & \\
(1) & (2) & (3) & (4) & (5) &(6)& (7) & (8)  \\
\hline
NGC 524 & 16  &$41.33\pm 0.05$ &$0.65\pm 0.07   $&$1.6\pm 0.5    $ &Galaxy     &E & $0.56\pm 0.15$\\
NGC 720 &  6  &$41.43\pm 0.02$ &$0.52\pm 0.03   $&$\cdots        $ &Group      &E & $0.67\pm 0.19$\\
NGC 1052& 29  &$40.53\pm 0.15$ &$0.41\pm 0.15   $&$166.5\pm 3.1  $ &Galaxy     &E & $0.17\pm 0.34$\\
NGC 1332& 10  &$40.93\pm 0.02$ &$0.56\pm 0.03   $&$40.3\pm 1.7   $ &Galaxy     &E & $0.80\pm 0.09$\\
NGC 1407& 24  &$41.92\pm 0.02$ &$1.02\pm 0.04   $&$5.2\pm 0.7    $ &Group      &E & $0.91\pm 0.03$\\
NGC 1566&  4  &$40.85\pm 0.05$ &$0.70\pm 0.11   $&$41.8\pm 2.0   $ &Galaxy     &E & $0.75\pm 0.12$\\
NGC 1808&  6  &$< 40.59       $ &$\cdots      $&$132.2\pm 3.1  $ &Undetected &L & $0.00\pm 0.71$\\
NGC 3557& 14  &$42.11\pm 0.04$ &$0.24\pm 0.02   $&$117.5\pm 4.6  $ &Group      &E & $0.60\pm 0.18$\\
NGC 3783&  9  &$40.94\pm 0.11$ &$\cdots         $&$118.5\pm 6.9  $ &Group      &L & $0.00\pm 0.50$\\
NGC 3923& 30  &$41.07\pm 0.02$ &$0.52\pm 0.03   $&$182.5\pm 4.3  $ &Galaxy     &E & $0.33\pm 0.27$\\
NGC 4636& 17  &$41.71\pm 0.02$ &$0.84\pm 0.02   $&$12.6\pm 1.0   $ &Group      &E & $0.42\pm 0.17$\\
NGC 5044& 32  &$43.09\pm 0.01$ &$1.21\pm 0.02   $&$96.1\pm 3.3   $ &Group      &E & $0.75\pm 0.05$\\
NGC 7144$^{\dagger}$&$\cdots$& $\cdots$  &$\cdots$ &$40.71\pm 0.13$   &Galaxy     &$\cdots $&$\cdots        $  \\
NGC 7714&  4  &$< 40.48       $ &$0.46\pm 0.07$&$114.7\pm 10.4 $ &Undetected &L & $0.00\pm 1.00$\\
HCG 90  & 38  &$41.79\pm 0.05$ &$0.39\pm 0.04   $&$112.4\pm 4.7  $ &Group      &E & $0.36\pm 0.17$\\
IC 1459 & 10  &$41.46\pm 0.04$ &$\cdots         $&$221.5\pm 5.5  $ &Group      &E & $0.29\pm 0.27$\\\hline

\hline
\end{tabular}
\flushleft 
$^{\dagger}$\cite{brough06b} did not find any members of this group using the friends-of-friends algorithm.

The columns are (1) Group name; (2) Number of group members according to the \cite{brough06b} friends-of-friends algorithm; (3) X-ray luminosity within $R_{500}$ from \cite{o4} (erg s$^{-1}$); (4) X-ray temperature of the group from \cite{o4} (keV); (5) Total \HI\ mass of the group from this work ($10^8$\Msun ); (6) X-ray emission type of the group;
(7) Central galaxy type; (8) Early-type fraction of the group \citep{brough06b}.\\

\end{table*}

\begin{table*} 
\caption{Summary of the \HI\ observations and results.}
\label{tab:groups} 
\begin{tabular}{llclcccccc}
\hline
Group  & $\alpha,\delta$(J2000)&  $v_{sys}$ &Dist.& Obs time & rms & Detectability&  \#\HI\ &\#New & \#New $z$\\
         & [$^{\rm h\,m}$], [\degr\,\arcmin]
            &[\kms] & [Mpc] & [$h$] & [mJy beam$^{-1}$] &  [mJy beam$^{-1}$]\\
(1) & (2) & (3) & (4) & (5) &(6)& (7) & (8) & (9) & (10) \\
\hline
NGC 524  & 01 24,   +09 26  & 2632 & 22.3&22  & 6.6 &25&9   &0&0\\
NGC 720  & 01 52, $-$13 34  & 1717 & 25.7&20  & 6.9 &22&8   &1&0\\
NGC 1052 & 02 40, $-$08 08  & 1438 & 18.0&28  & 5.6 &18&15  &1&0\\
NGC 1332 & 03 26, $-$21 17  & 1449 & 20.9&20  & 7.7 &25&16  &0&0\\
NGC 1407 & 03 40, $-$18 37  & 1695 & 20.9&21  & 9.1 &25&8   &0&0\\
NGC 1566 & 04 16, $-$55 35  & 1246 & 21  &20  & 7.8 &20&13  &0&2\\
NGC 1808 & 05 08, $-$37 36  & 1141 & 17  &24  & 6.0 &18&7   &0&0\\
NGC 3557 & 11 09, $-$37 30  & 2979 & 42.5&27  & 6.0 &20&13  &0&3\\
NGC 3783 & 11 37, $-$37 53  & 2819 & 36  &13  & 15.7&28&12  &3$^{\dagger}$&1\\
NGC 3923 & 11 50, $-$28 46  & 1497 & 21.3&20  & 7.0 &25&13  &1&0\\
NGC 4636 & 12 42,   +02 41  & 1696 & 13.6&31  & 5.0 &15&22  &0&0\\
NGC 5044 & 13 15, $-$16 23  & 2379 & 29.0&24  & 6.4 &18&23  &3&3\\
NGC 7144 & 21 53, $-$48 28  & 1880 & 22.8&22  & 5.7 &18& 8  &0&6\\
NGC 7714 & 23 36,   +02 05  & 2908 & 39  &20  & 14.0&$\cdots$&7 &1&0\\
HCG 90   & 22 02, $-$32 06  & 2557 & 36  &21  & 6.9 &25& 12 &0&0\\
IC 1459  & 22 57, $-$36 36  & 1672 & 27.2&23  & 5.8 &18&18  &0&0\\\hline
Total    &                 &            &      &     &     &&204&10&15\\     

\hline
\end{tabular}

\flushleft
$^{\dagger}$One of these newly detected galaxies was found in follow-up ATCA data, not the original Parkes survey.

The columns are (1) Group name; (2) Central position of the \HI\ cube;
(3) Central velocity of the group (km s$^{-1}$); (4) Distance to the group in Mpc
from \cite{brough06b}; (5) Approximate integration time (hours); (6)
rms noise per channel in the final \HI\ cube, smoothed to a velocity
resolution of 13.2 \kms (mJy beam$^{-1}$); (7) Detectability of synthetic Gaussian
sources as described in the text (mJy beam$^{-1}$); (8) Number of \HI\
detections; (9) Number of previously uncatalogued (``new'')
detections; (10) Number of new redshifts of previously catalogued
galaxies.\\

\end{table*}

\subsection{Parkes observations and data reduction}

Wide-field \HI\ imaging of the sixteen groups was undertaken at the
Parkes radio-telescope from 2000-2003. A 5.5$^{\rm o} \times 5.5^{\rm
o}$ region around each group centre was surveyed, and $\sim 20h$ of
data was obtained for each group. A basket-weave observing pattern was
used for the observations (e.g. \citealt{barnes01}), where scans
were made at a rate of one degree per minute, along lines of
equatorial latitude and longitude, separated by 4 arcmin. We used 8MHz
filters, with 2048 channels, providing unsmoothed spectral resolution
of 3.9 kHz, or 0.83 \kms.

The data were reduced using the AIPS++ packages {\sc livedata} and
{\sc gridzilla}.  {\sc livedata} was used to correct for the bandpass, and
a tukey filter was applied to reduce ringing and noise in the data
(Barnes et al. 2001). Bandpass correction was applied on a per-beam,
per-scan basis, by iteratively clipping the data and fitting a 2nd
degree polynomial to the time series of each channel. We masked out
data within 20 arcmin of known \HI\ sources in the field (based on a
first, quick process and search step), preventing contamination of the
calculated bandpass. Each bandpass-corrected spectrum was frame-shifted to
the barycenter of the Earth-Sun system, and then baseline corrected by
subtracting a source-masked, iterative, clipped 2nd-degree polynomial
fit, this time in the spectral domain.

The {\sc gridzilla} task was used to image the data. We used the {\sc
wgtmed} statistic, which calculates mesh pixel values by taking the
weighted median of data within 6 arcmin of the centre of each
pixel. The weight values were directly proportional to the canonical,
Gaussian observing beam profile which has a FWHM of 14.4 arcmin. The pixel width in the final cubes is 4 arcmin. The
final cubes were smoothed to a spectral resolution of 13.2 \kms\
for the searching process, and the rms noise level for each cube is
given in Table~\ref{tab:groups}. The average rms noise in the cubes is
7.6$\pm2.9$ mJy beam$^{-1}$ per channel, but two cubes had
significantly greater rms noise of 15.7 mJy beam$^{-1}$ per channel
for the NGC 3783 cube (see Kilborn et al. 2006), and 14.0 mJy
beam$^{-1}$ per channel for the NGC 7714 cube, due to solar
interference in the data.

\subsection{Source finding and completeness}

The Parkes \HI\ cubes were searched visually for sources using the
{\sc karma} suite of visualisation packages. Each cube was
searched three times in R.A--Dec., R.A.--velocity and Dec.--velocity
projections, and two people searched each cube for sources. The
number of \HI\ sources detected in each cube is given in
Table~\ref{tab:groups}. Sources visible in each projection, and with a
peak flux of $\gtrsim 3\sigma$ were catalogued as sources.

To determine the completeness in each cube, we added synthetic
sources to each \HI\ cube, and noted the detection rate for
different peak flux and velocity widths of the sources. The synthetic
sources were Gaussian in profile, and had peak fluxes of 10, 15,
18, 20, 22, 25, 30 mJy. For each peak flux, three
synthetic sources were added to the cube, with velocity width of
50, 150 and 300 \kms. We define the detectability of the cube, as the
peak flux at above which all synthetic sources were recovered,
regardless of velocity width. Extra synthetic sources were inserted to
the NGC 3783 cube, as its rms noise was much higher than the other
cubes. In this case, the synthetic sources had the peak flux values of
25, 28, 30, 32 and 35 mJy. We were unable to determine a
detectability limit for the NGC 7714 cube. This cube suffers
from significant solar interference, and the rms noise varies across the
cube. We therefore exclude this cube from the \HI\ mass function
calculations. The detectability of each cube is listed in
Table~\ref{tab:groups}. In general the detectability is $\sim
3\sigma$ above the noise, but not in every case.

\begin{table*} 
\caption{New galaxies found in the GEMS HI survey. The columns are as follows: (1) Galaxy name; (2) R.A.(J2000), Dec.(J2000); (3) Heliocentric velocity in the optical convention (\kms); (4) \HI\ mass; (5) Status of ATCA follow-up observations. (6) Reference where the ATCA data are published and discussed in detail. All quantities are measured from the Parkes data, unless otherwise specified.}
\label{tab:newgals} 
\begin{tabular}{llclll}
\hline
GEMS name  & $\alpha,\delta$(J2000)& V$_sys$ & $M_{HI}$ & ATCA & Reference \\
         & [$^{\rm h\,m\,s}$], [\degr\,\arcmin\,\arcsec]
            &[\kms] & [$10^8$ \Msun] & status &  \\
(1) & (2) & (3) & (4) & (5) \\
\hline
GEMS\_N720\_6   &01:48:58,-12:48:19  & $1770 \pm 7 $ & $5.8  \pm 0.9 $   & N & $\cdots$\\
$[$MMB2004$]$ J0249-0806 & 02:49:13,-08:07:18 & $1419 \pm 4$ & $5.8 \pm 0.5 $ & Y &  \citet{mckay04}\\
GEMS\_N3783\_2  &11:31:32,-36:18:53  & $2733 \pm 11$ & $5.8  \pm 1.7$    & Y & \citet{kilborn06}\\
GEMS\_N3783\_8  &11:37:54,-37:56:04  & $2947 \pm 6 $ & $21.1 \pm 2.9$    & Y & \citet{kilborn06}\\
GEMS\_N3923\_11 &11:50:17,-30:04:33 & $1600 \pm 3$& $1.4 \pm 0.4$ & Y &  \citet{kern07}\\
GEMS\_N5044\_7  &13:10:59,-15:20:44  & $2828 \pm  8$ & $8.5  \pm 1.3$    & N & $\cdots$\\
$[$MMB2004$]$ J1320-1427 &13:20:14,-14:25:41 & $2748 \pm  2$ & $9.7  \pm 1.1$ & Y & \citet{mckay04}\\
GEMS\_N5044\_18 &13:11:36,-14:40:40  & $2488 \pm  4$ & $6.3  \pm 0.9$    & Y & \citet{kern07}\\
GEMS\_N7714\_7 & 23:45:19,+03:41:28 & $2755 \pm 7 $ & $86.0 \pm 4.8$ & N & $\cdots$\\ 
ATCA\_1134-37   &11:34:02,-37:14:15  & $3141 \pm  5$ & $2.4  \pm 0.6^1$  & Y & \citet{kilborn06}\\
\hline
\end{tabular}
\flushleft $^1$ Parameters from ATCA observations.\\

\end{table*}

\subsection{Parameterisation of the sources, and flux comparison with HICAT}

Once the sources were catalogued as above, an interactive fitting
program using {\sc miriad} routines was run to determine the central
position, heliocentric velocity, velocity widths (50\% and 20\%),
integrated flux, peak flux density, and rms noise of the
spectrum. Firstly {\sc moment} was used to make a flux integrated \HI\
distribution map of each source, and {\sc imfit} was then invoked to
determine the central position and extent of the source by fitting a
2-dimensional Gaussian profile to the \HI\ distribution. This central
position was then used in the task {\sc mbspect} to determine all the
other parameters for the source. The resolution of the cube for
this step was 6.6 \kms. The flux of the source was calculated using 3
boxes of 3, 5 and 7 pixels in width around the source, to determine if
any sources were extended rather than point sources. Those sources
with an increase of $>20\%$ in flux between the fit from 5 and 7
pixels were flagged as extended, and the box width was increased in the
fit of these sources until a stable flux was determined for the
source.  The \HI\ parameters for the sources in each group are listed
in Appendix~\ref{tab:app1}.

A number of the GEMS sources overlap with the HIPASS catalogue, HICAT
\citep{meyer04}. We compared all \HI\ parameters for the sources
common to the two surveys, and found excellent agreement in systemic
velocity, flux, position and velocity widths.
The majority of fluxes agree within
$< 30$\%, and those that have a larger error than this can generally
be explained by confusion, or poor baselines in the HICAT spectra. We
note, that the velocity resolution for fitting the GEMS sources is 6.6
\kms, compared to 26.4 \kms\ for HICAT.

\subsection{Optical matching to \HI\ sources and definition of the groups}

 The 6dFGS 2DR \citep{jones05} and NED  databases were used to obtain
 the previously catalogued optical galaxies, with optical redshifts in
 a $\sim$5.5$^{\rm o} \times 5.5^{\rm o}$ region around each group
 centre to match the \HI\ detections to optical galaxies. Full details
 of the optical detections in the groups are given in
 \citet{brough06b}.  We matched our \HI\ sources to optical galaxies
 by searching for galaxies nearby in position and velocity. The
 matched optical counterparts to the \HI\ detections are given in
 Appendix~\ref{tab:app2}. In the case where there were two or more optically
 catalogued galaxies near the position of the \HI\ detection, and at
 the same velocity, then all of these are listed.  In some cases there
 were previously catalogued galaxies with no known redshift - we
 tentatively include these with a redshift in our list, but they need
 to be confirmed with higher resolution \HI\ observations or optical
 confirmation of the redshift.  Table~\ref{tab:groups} lists the
 number of \HI\ detections, new detections, and new redshifts found in
 each of the cubes.

\begin{figure*}
\begin{tabular}{ccc}
\mbox{\psfig{file=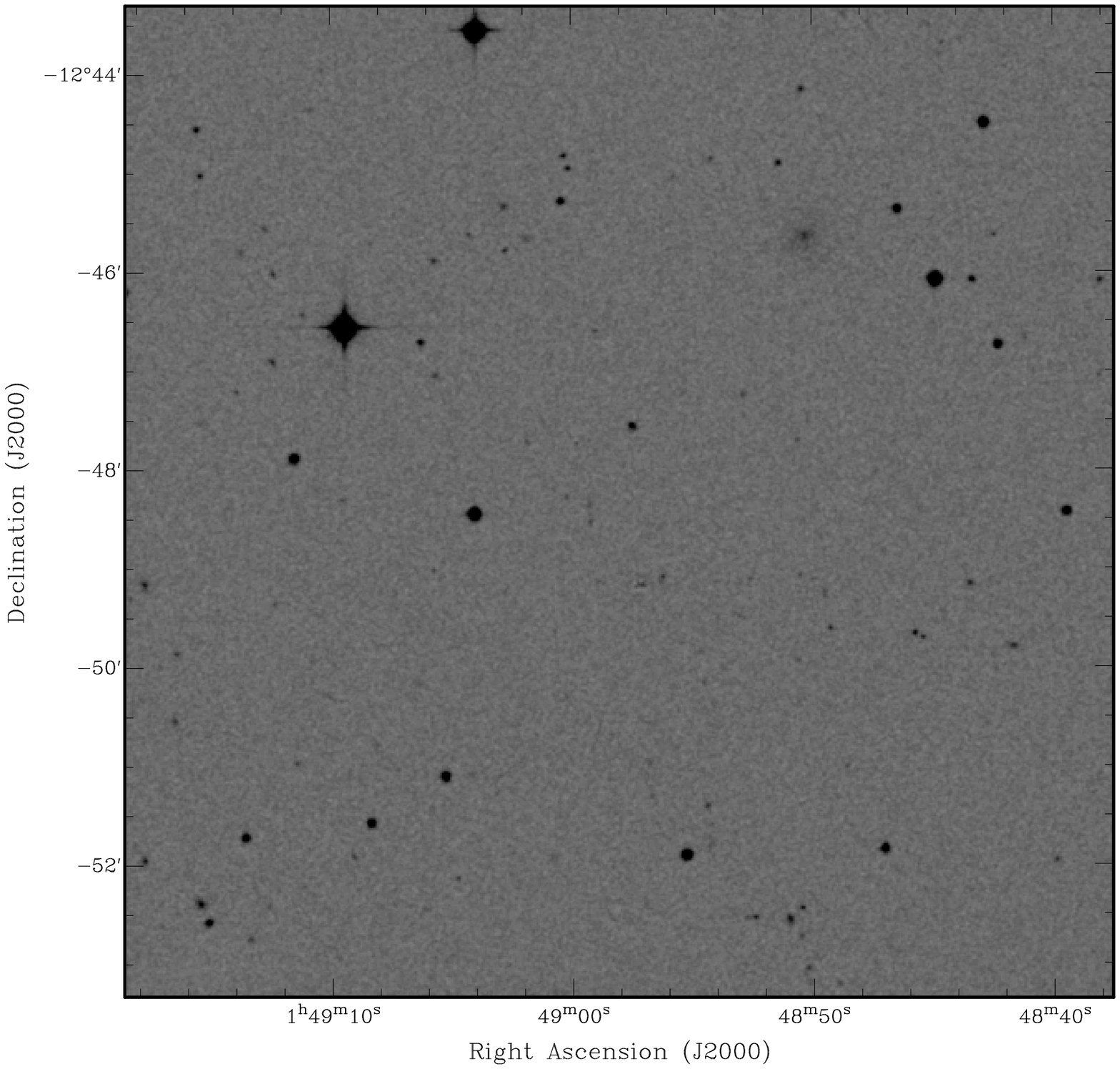,width=5.5cm,angle=-90}}&
\mbox{\psfig{file=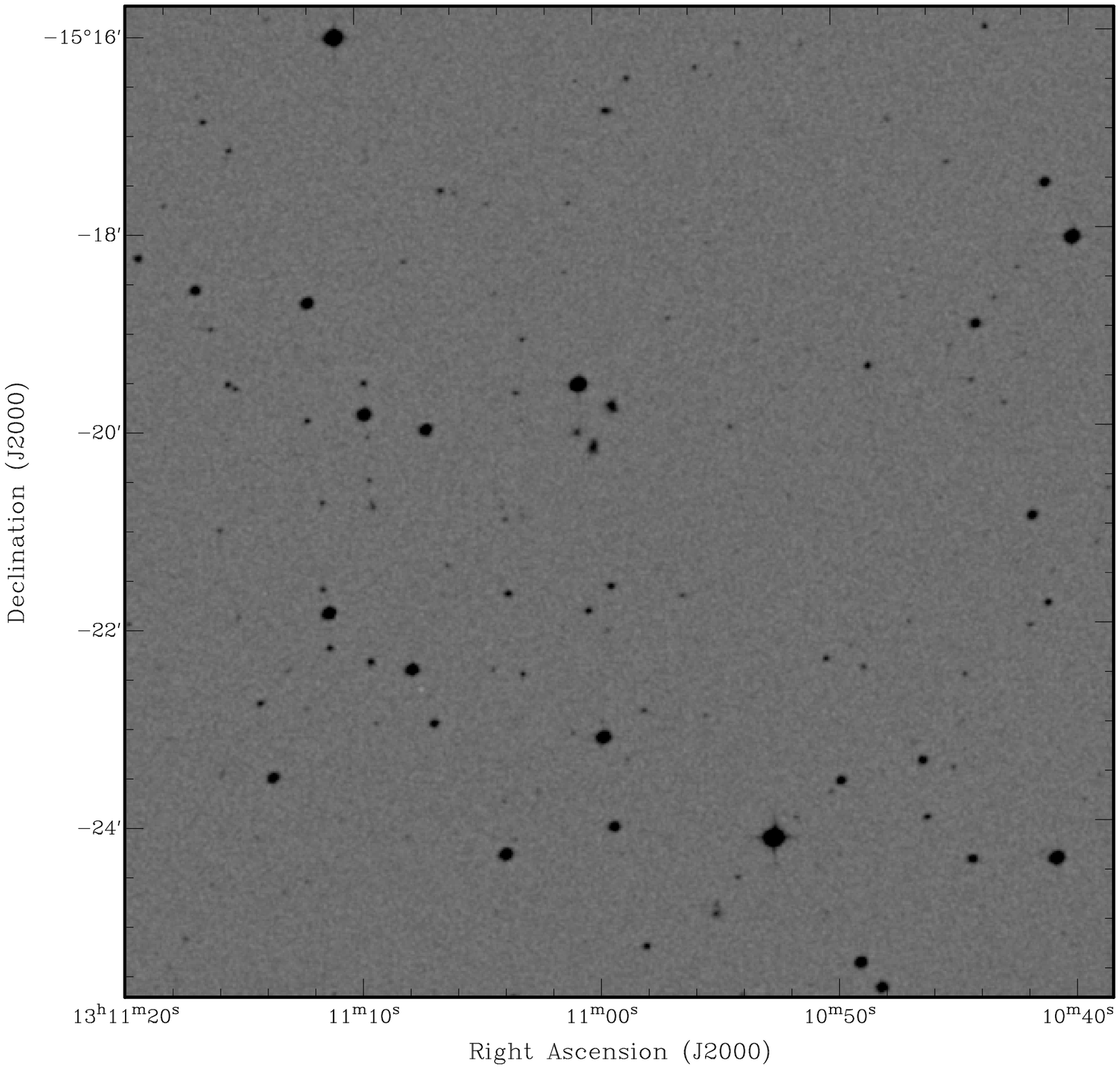,width=5.5cm,angle=-90}}&
\mbox{\psfig{file=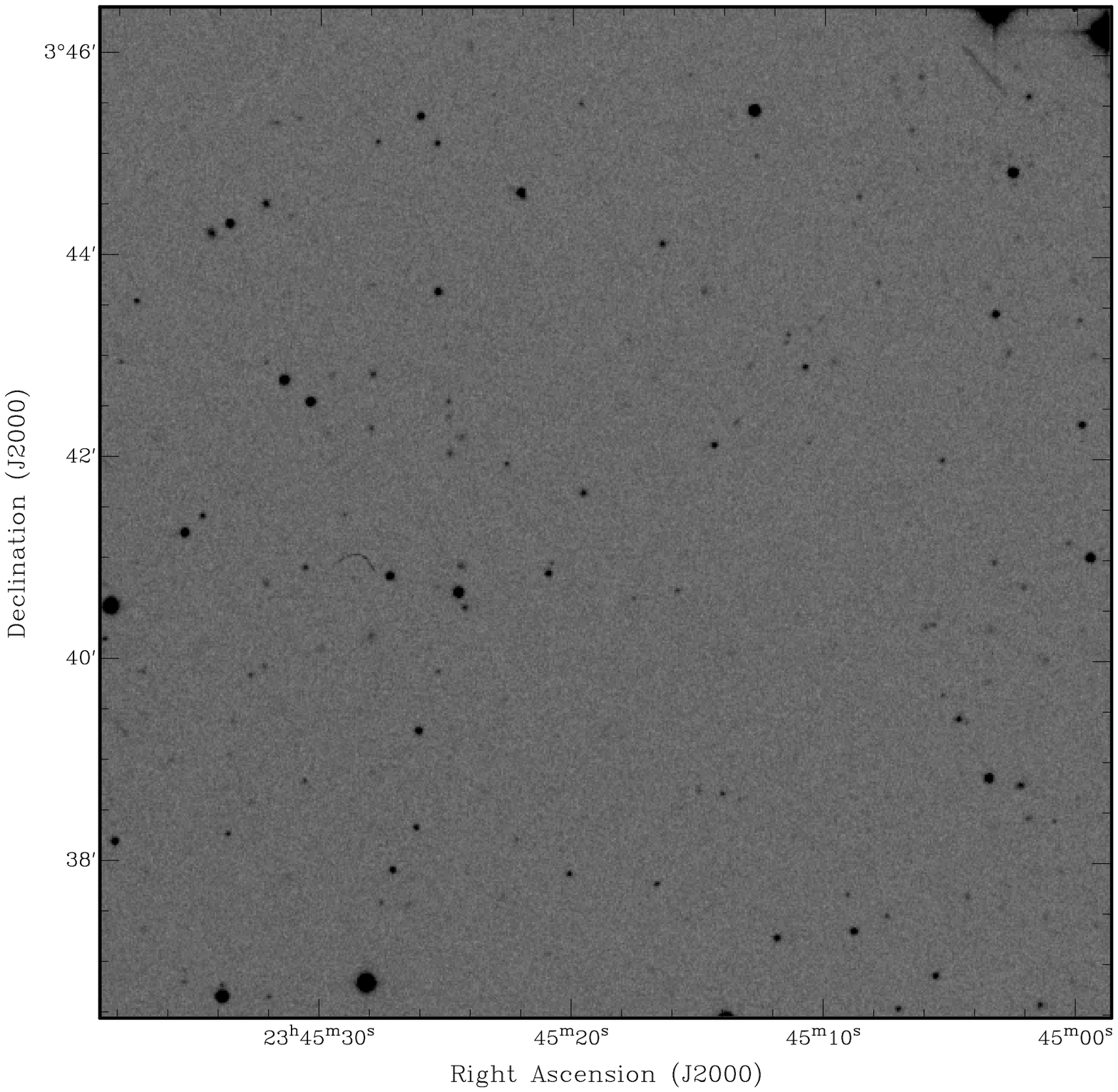,width=5.5cm,angle=-90}}\\
\end{tabular}
\caption{A 10$\times$10 arcminute optical DSSII (R-band) image centred on the Parkes position of the new galaxies in the GEMS HI survey without ATCA follow-up. From left to right they are: GEMS\_N720\_6, GEMS\_N5044\_7, GEMS\_N7714\_7.}
\label{fig:newgals}
\end{figure*}

\subsection{ATCA follow-up observations}

We obtained high resolution images of a number of new galaxies and
confused sources at the Australia Telescope Compact Array between
2003--2006. The results from the follow-up survey are presented in
\citet{mckay04,kilborn05,kilborn06} and \citet{kern07}. For this
paper, we use the Parkes \HI\ positions and fluxes. Those sources with
ATCA follow-up observations are noted in Appendix~\ref{tab:app2}.

\subsection{Previously uncatalogued galaxies}

We detected ten previously uncatalogued galaxies, and their GEMS names
and details are listed in Table~\ref{tab:newgals}, along with a
reference to detailed maps and descriptions where available. We have
made high resolution ATCA observations of seven of the new Parkes
detections. In addition to these new detections, we also identified a
new dwarf galaxy in follow-up ATCA observations of the spiral galaxy
ESO 378-G 011 in the NGC 3783 group. This galaxy was named
ATCA\_1134-37, and the data for this detection is detailed in Kilborn
et al. (2006). Figure~\ref{fig:newgals} shows a $10 \times 10$
arcminute square region around the position of the Parkes centre for
the three previously uncatalogued detections where we do not have ATCA
observations.  Two out of three of these new detections have a probable
optical counterpart visible in the field, whereas there is no obvious
optical counterpart in the GEMS\_N7714\_7 field. Further high
resolution \HI\ observations of this object are required to determine
its nature.

The average \HI\ mass of these previously uncatalogued galaxies is $
\sim 8 \times 10^8$ \Msun, so they do not contribute a large amount to
the total \HI\ mass of the groups.  In general, the previously
uncatalogued \HI\ sources correspond to faint dwarf galaxies. The
exception is GEMS\_N3783\_2, for which, despite deep optical imaging
and an accurate position from ATCA observations, we are unable to find
an optical counterpart for.  This detection is discussed in detail in
Kilborn et al. (2006).

We determined a redshift for the first time for sixteen previously
catalogued galaxies. We have obtained ATCA observations for 12 of
these galaxies confirming the optical counterpart. Included in these
12 detections are several instances where there was more than one possible
optical counterpart to the original Parkes source, which turned out to
 comprise two or more galaxies containing neutral
hydrogen. Galaxies with newly determined redshifts are listed in
Appendix~\ref{tab:app2}, along with a reference to the ATCA maps and
description of the galaxies where available.


\begin{figure*}
\begin{tabular}{cc}
\mbox{\psfig{file=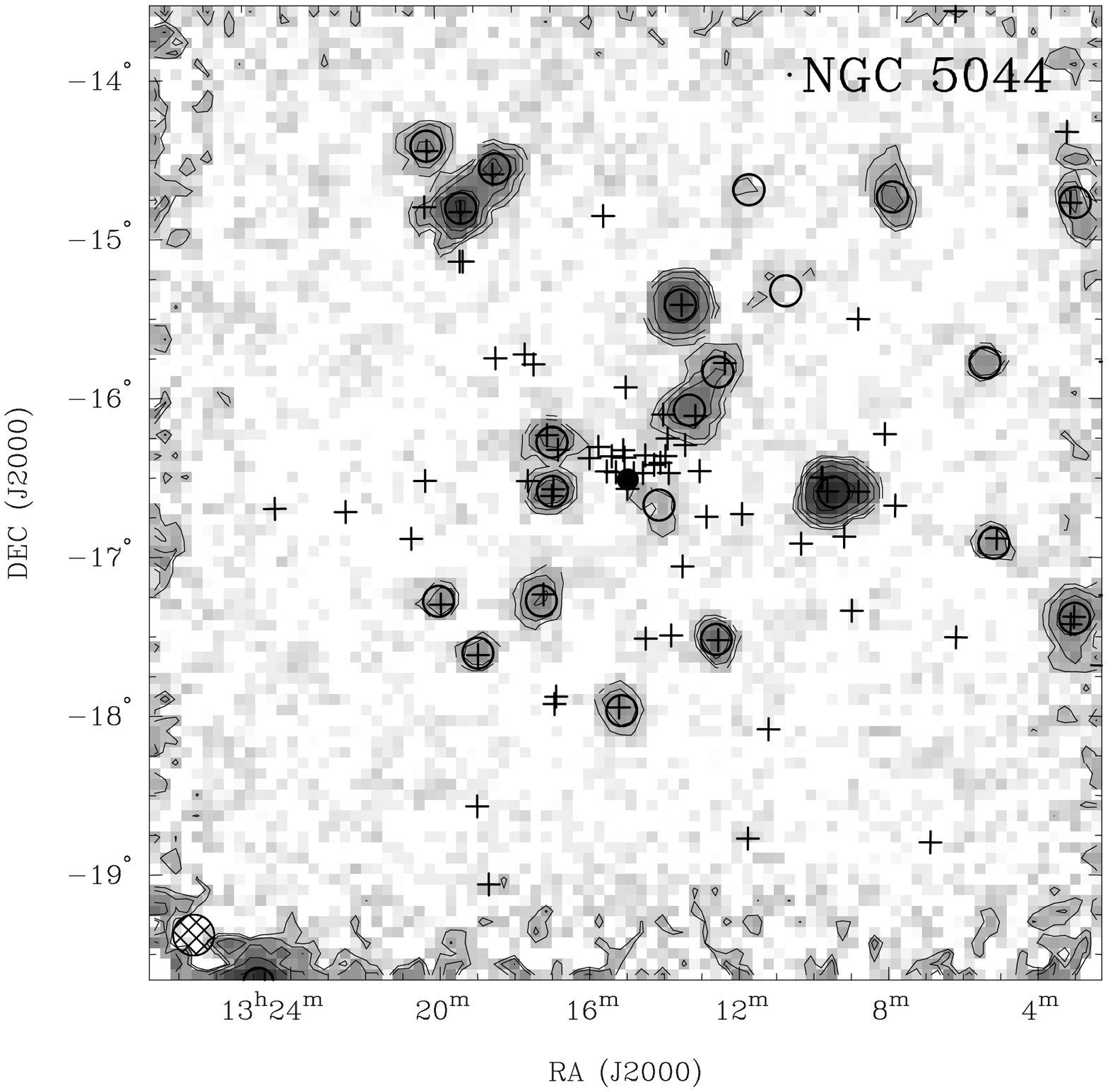,width=7.2cm, angle=-90}}& 
\mbox{\psfig{file=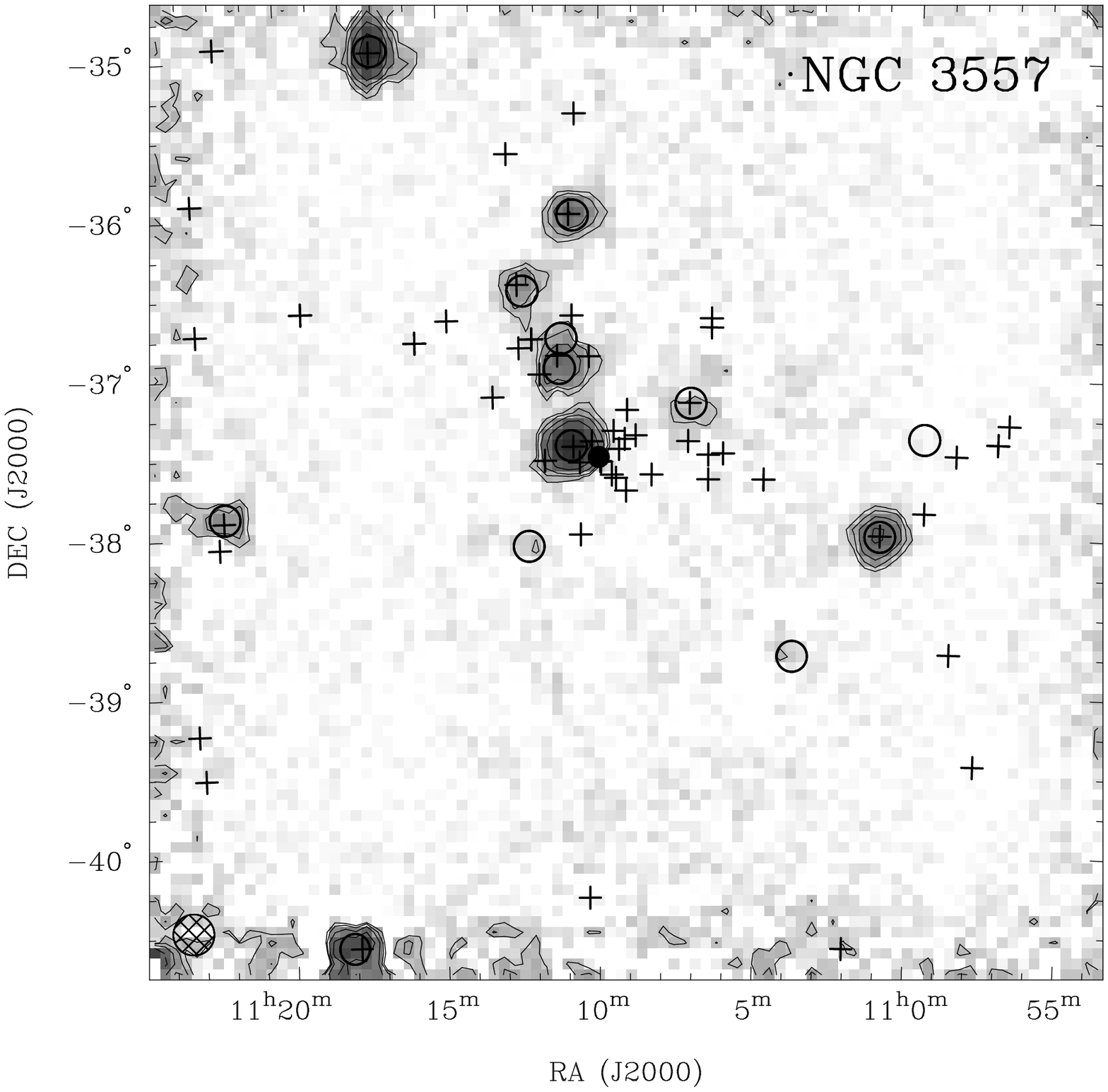,width=7.2cm, angle=-90}}\\
\mbox{\psfig{file=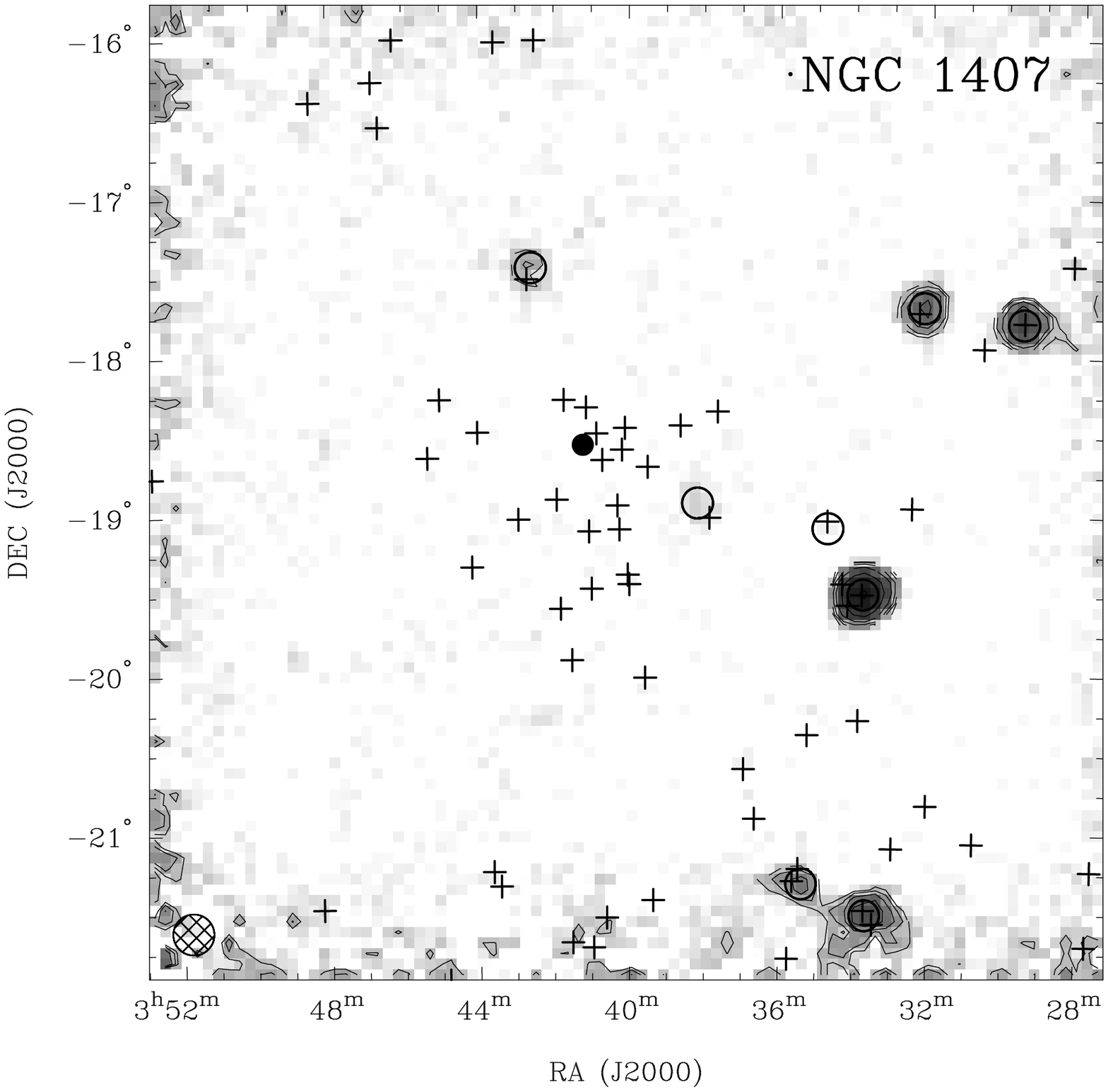,width=7.2cm, angle=-90}}&
\mbox{\psfig{file=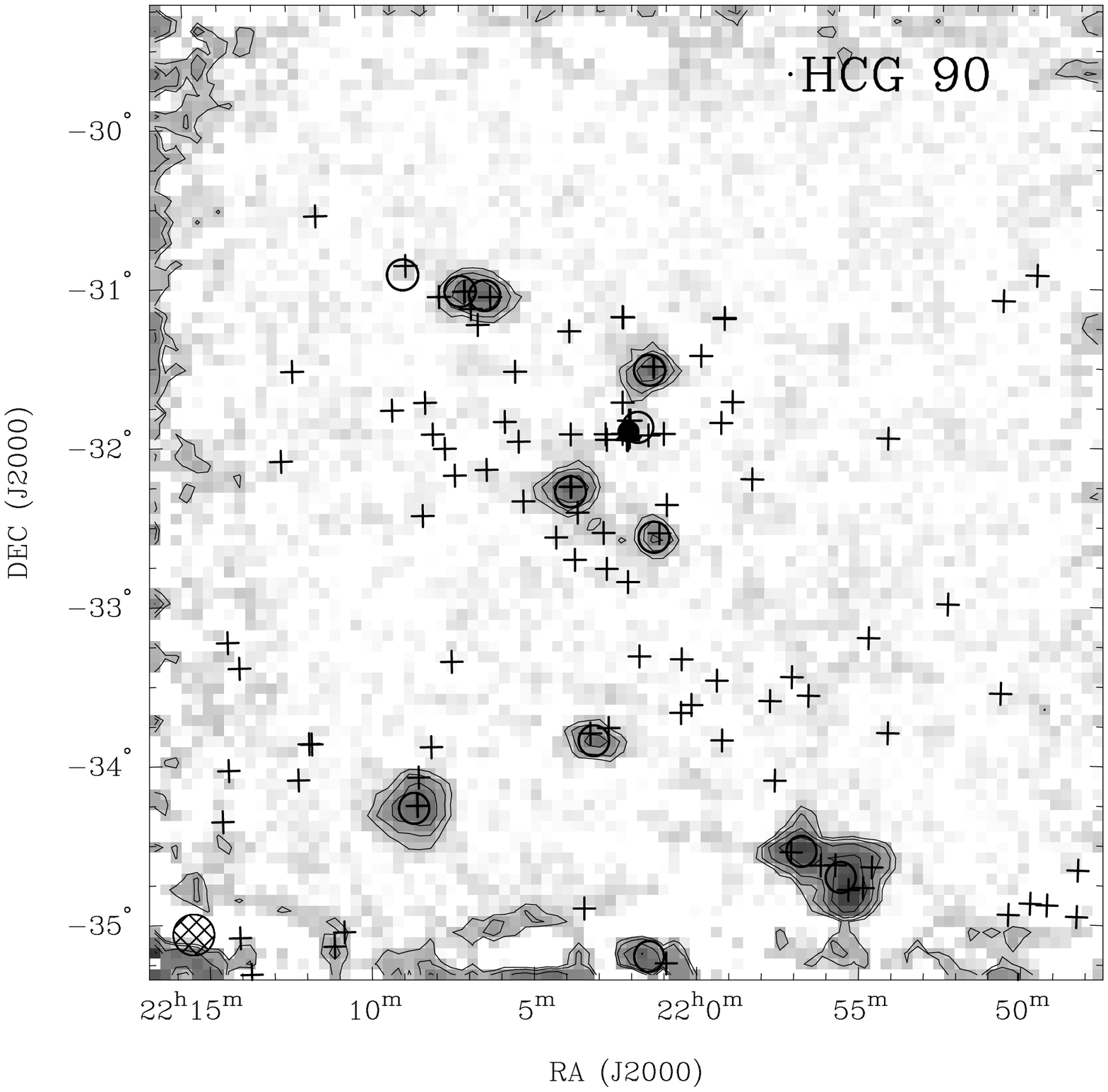,width=7.2cm, angle=-90}}\\
\mbox{\psfig{file=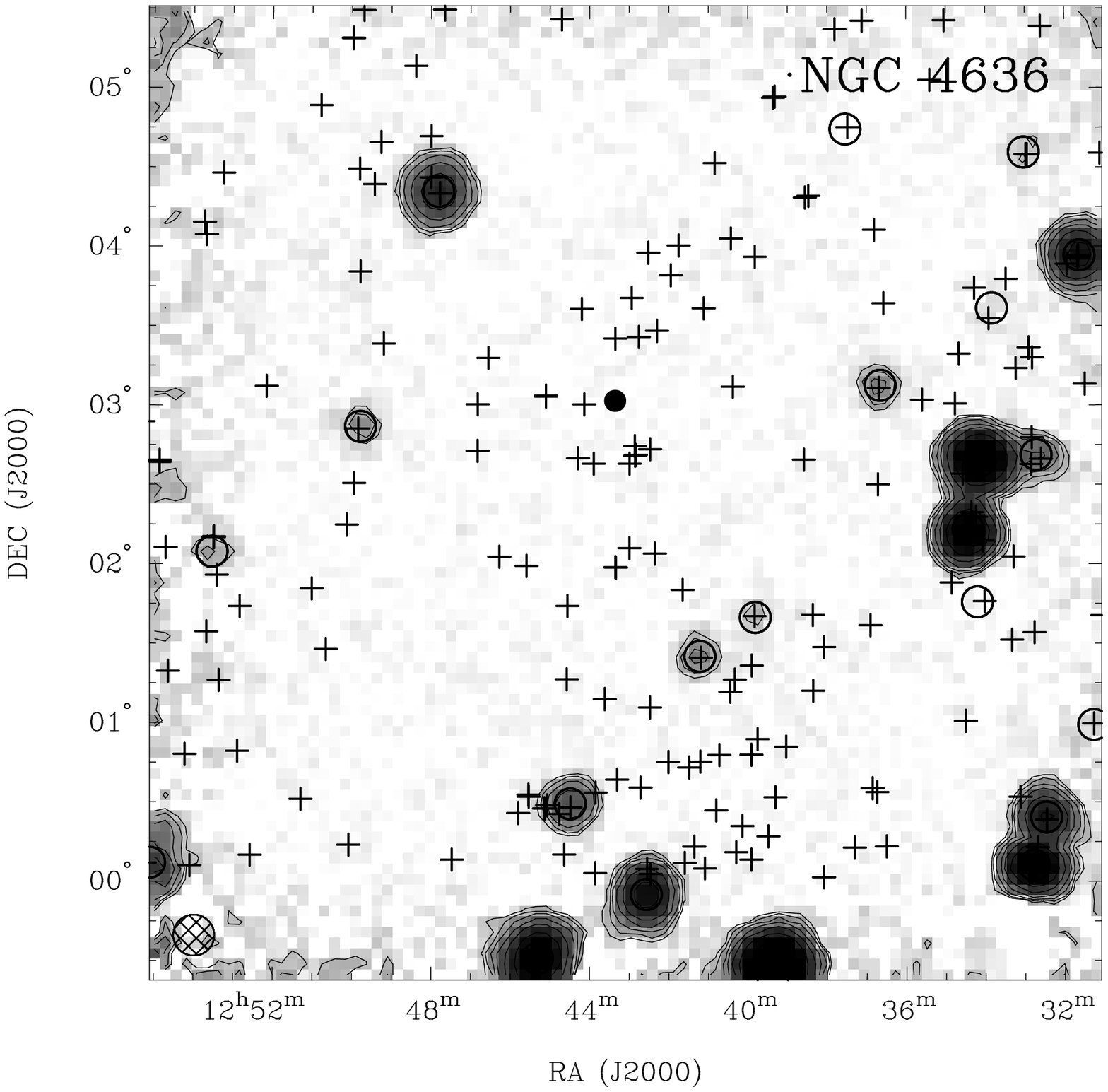,width=7.2cm, angle=-90}}&
\mbox{\psfig{file=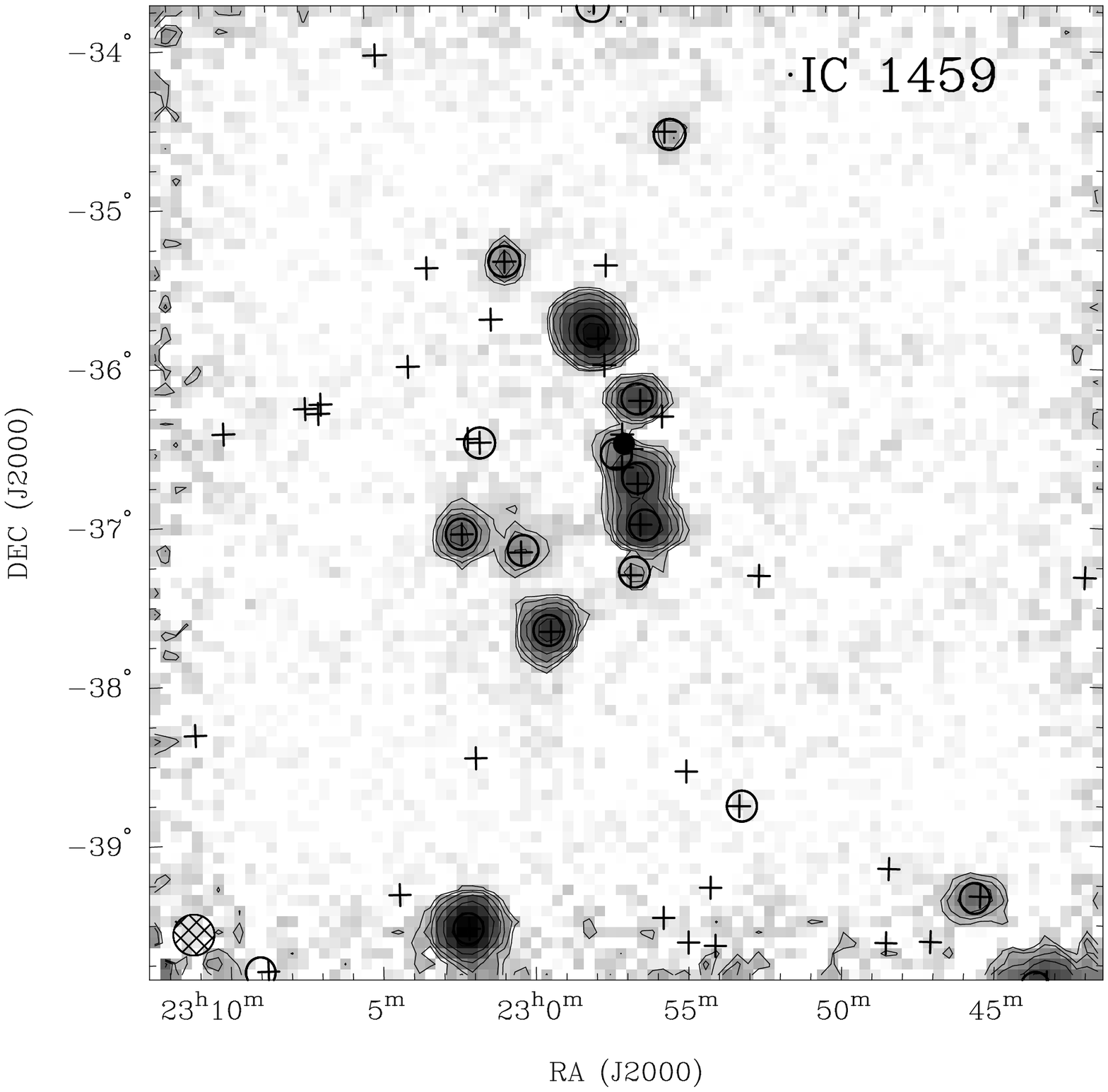,width=7.2cm, angle=-90}}\\
\end{tabular}
\caption{Groups with diffuse group-scale X-ray emission. The contours
and greyscale represent the \HI\ intensity, where the contour levels
are 1, 2, 4, 8, 16, 32 Jy beam$^{-1}$, apart from NGC 3783 where the
contour levels start at 2 Jy beam$^{-1}$. Open circles mark the
position of the sources we detected in \HI, and the crosses show the
position of previously catalogued galaxies in NED or the 6dFGS with
known velocity in the region of the \HI\ cube. The solid circle
marks the luminosity weighted centre for each group
\citep{brough06b}. The Parkes 15.5\arcmin\ beam is shown in the bottom
left hand corner of each image. The images are plotted in order of
descending X-ray luminosity.}
\label{fig:xray1}
\end{figure*}

\begin{figure*}
\begin{tabular}{cc}

\mbox{\psfig{file=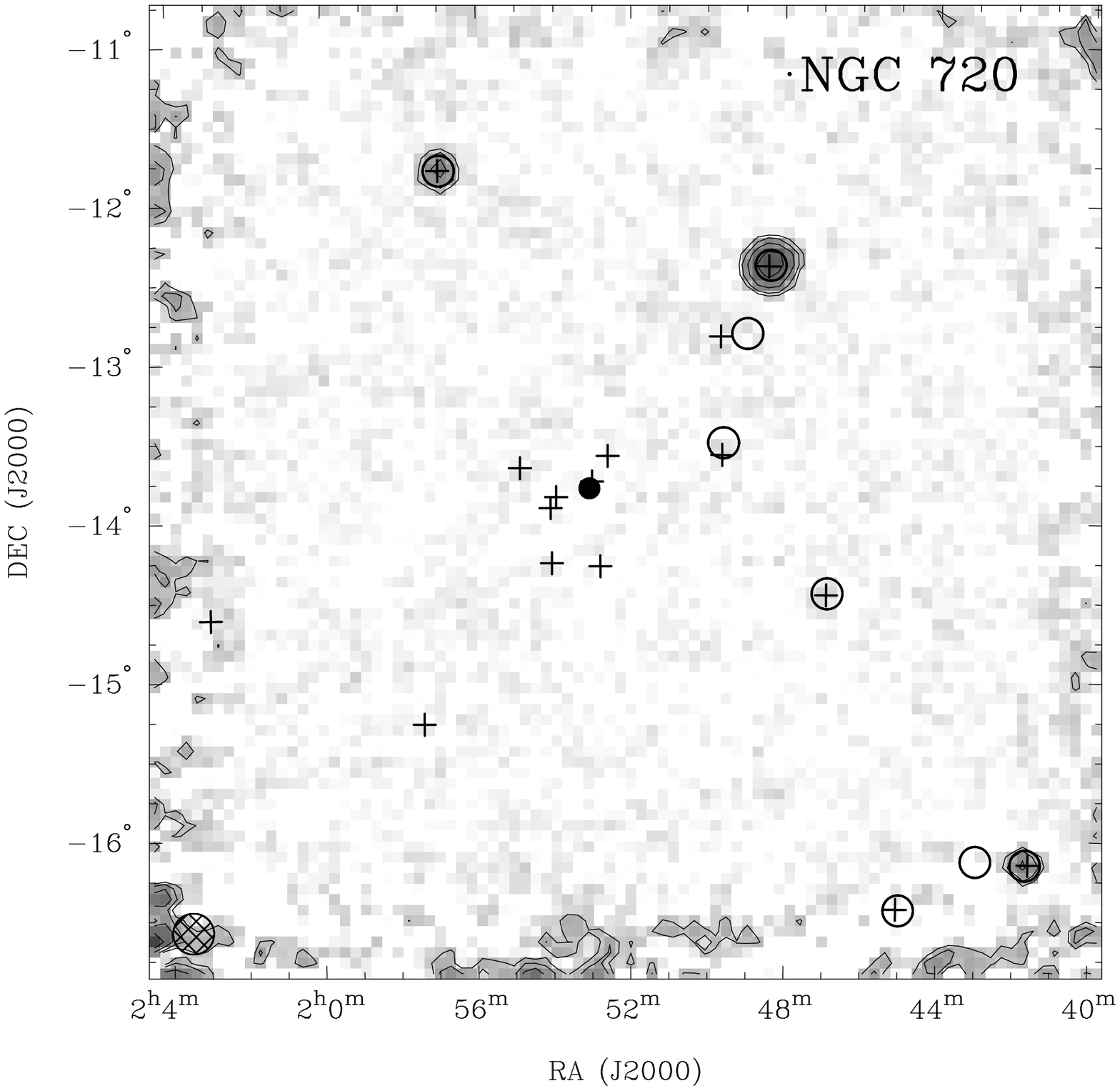,width=7.2cm, angle=-90}}&
\mbox{\psfig{file=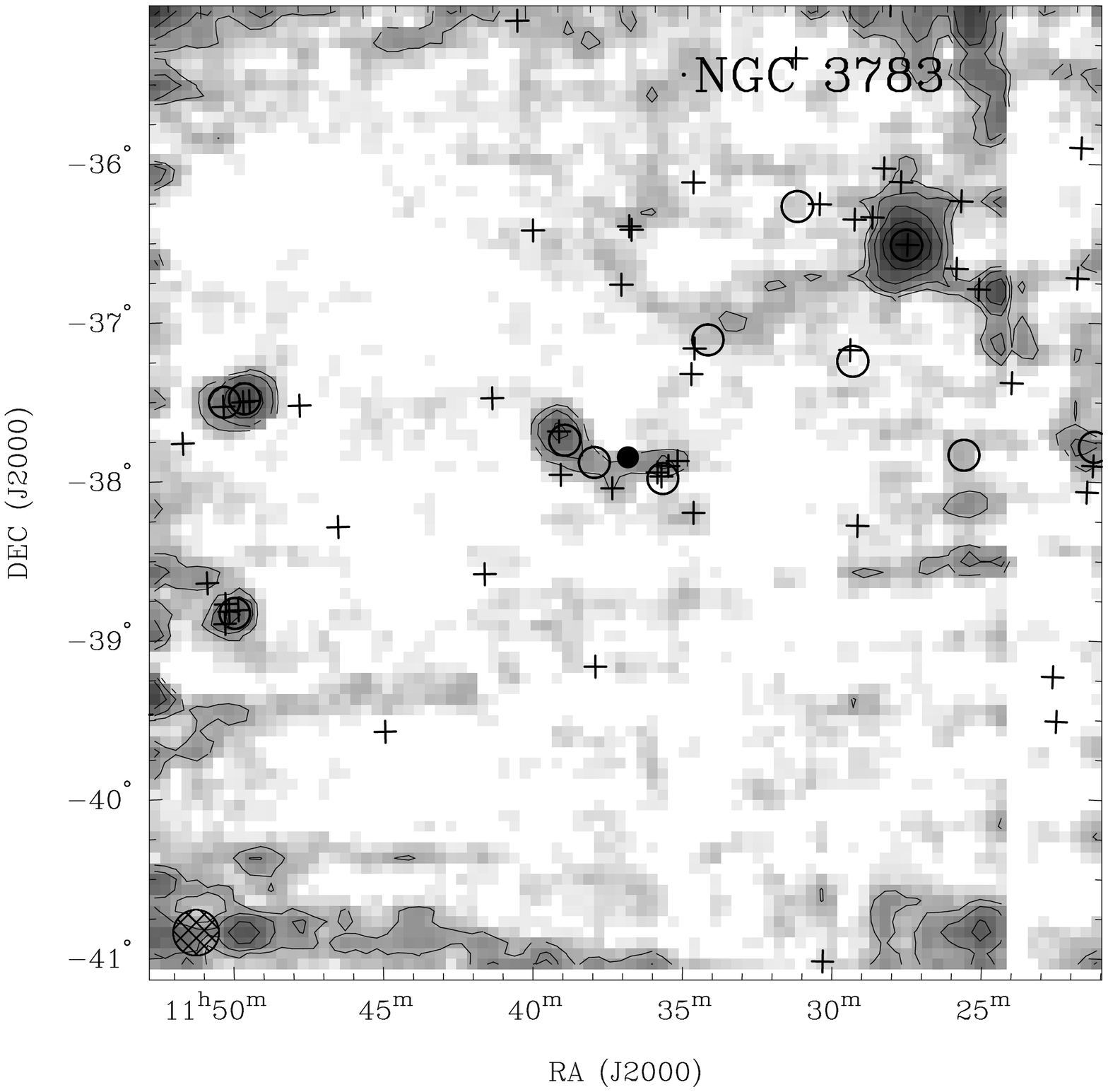,width=7.2cm, angle=-90}}\\
\end{tabular}
\caption{Groups with diffuse X-ray emission. The contour levels and
symbols are the same as in Figure~\ref{fig:xray1}.}
\label{fig:xray2}
\end{figure*}

\begin{figure*} 
\begin{tabular}{cc}
\mbox{\psfig{file=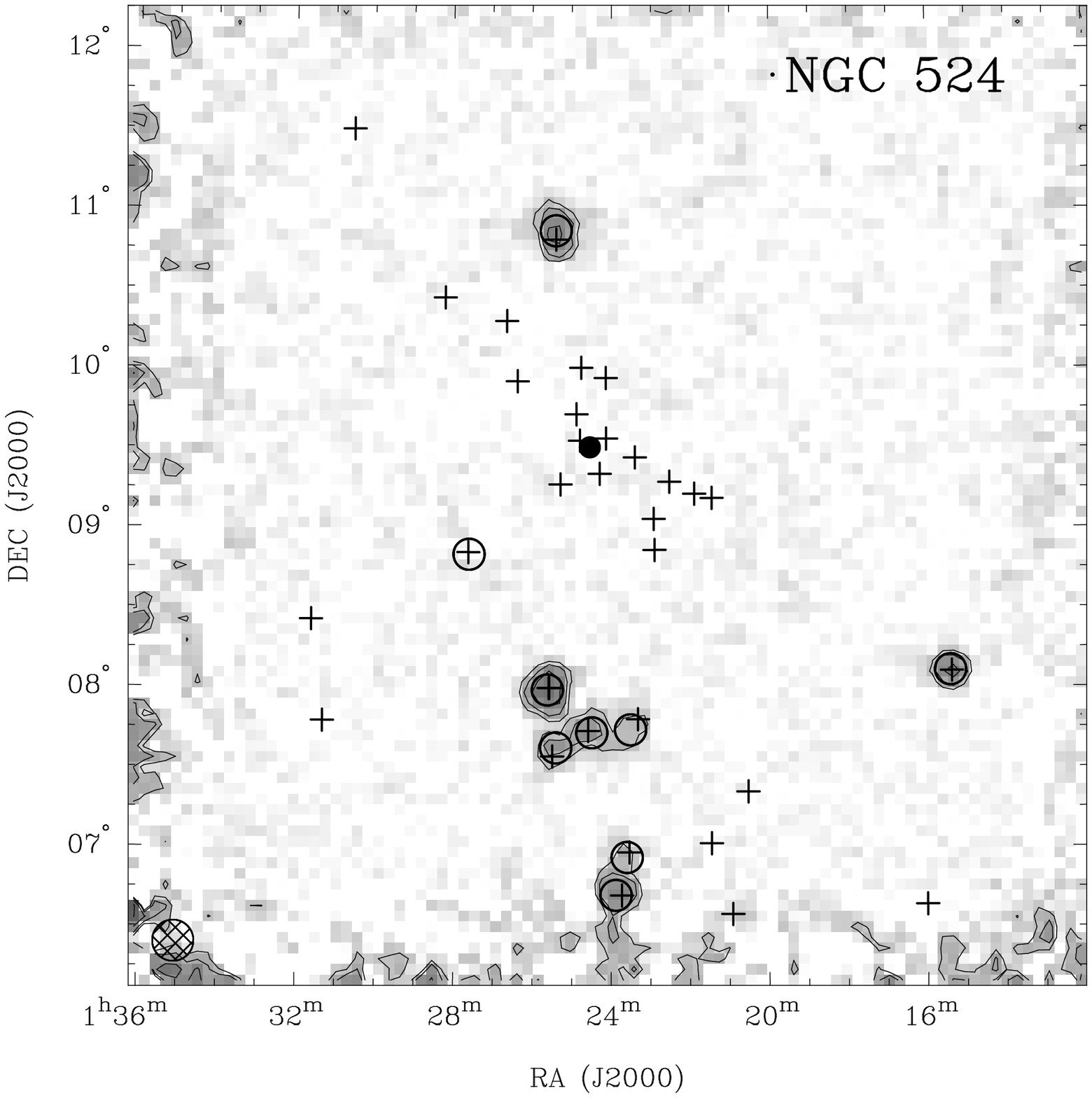,width=7.2cm, angle=-90}}&
\mbox{\psfig{file=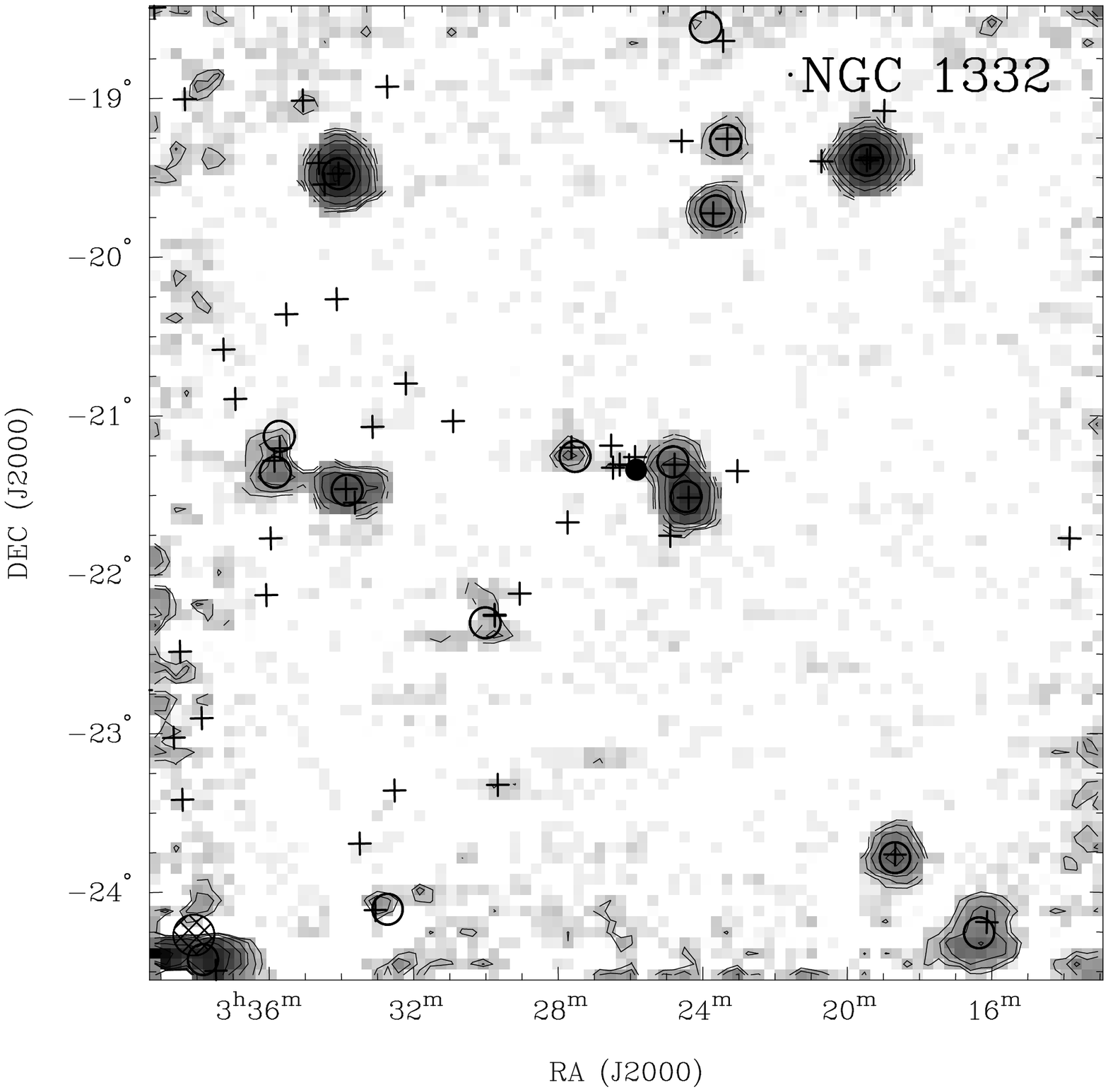,width=7.2cm, angle=-90}}\\
\mbox{\psfig{file=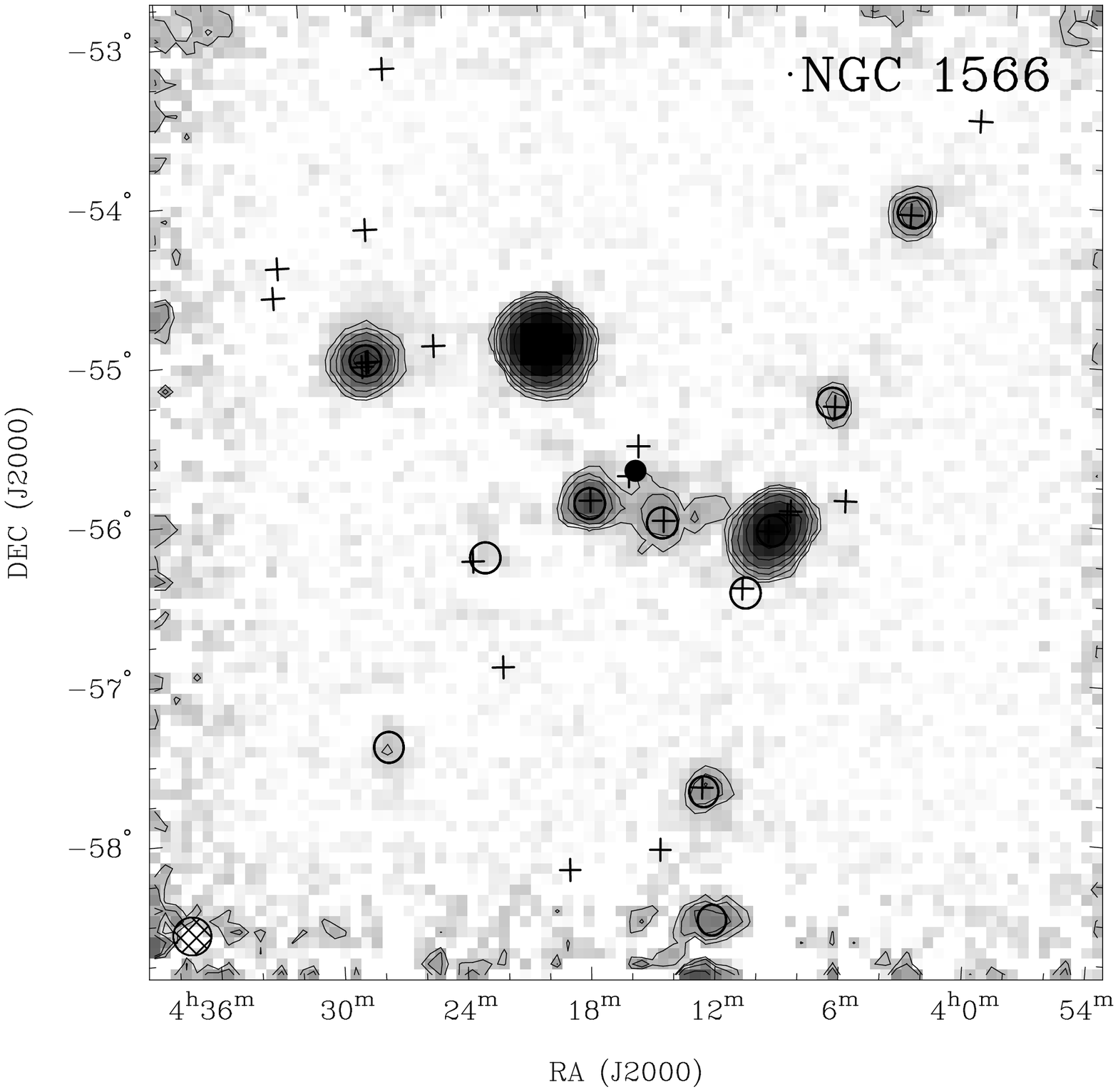,width=7.2cm, angle=-90}}&
\mbox{\psfig{file=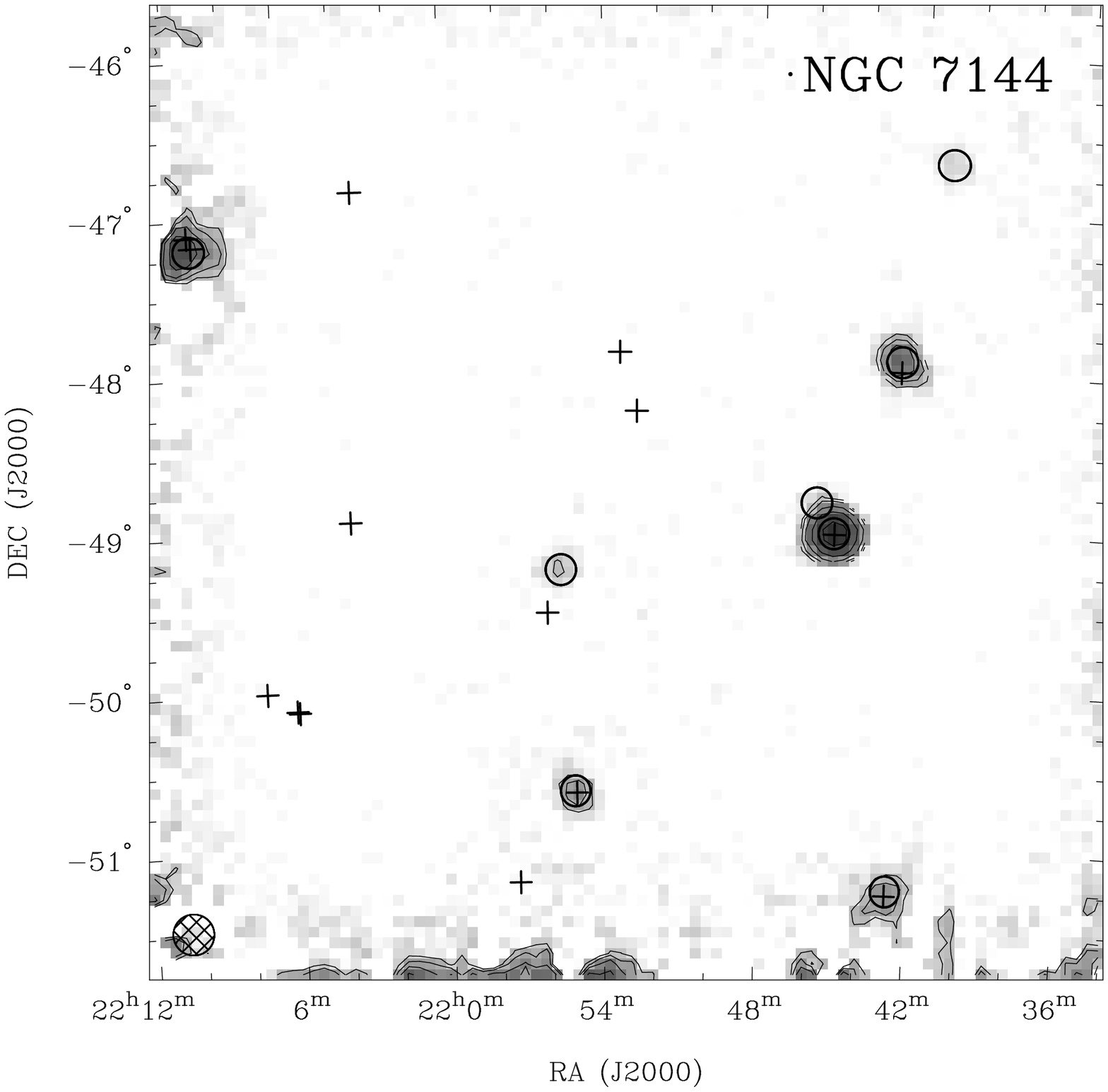,width=7.2cm, angle=-90}}\\
\mbox{\psfig{file=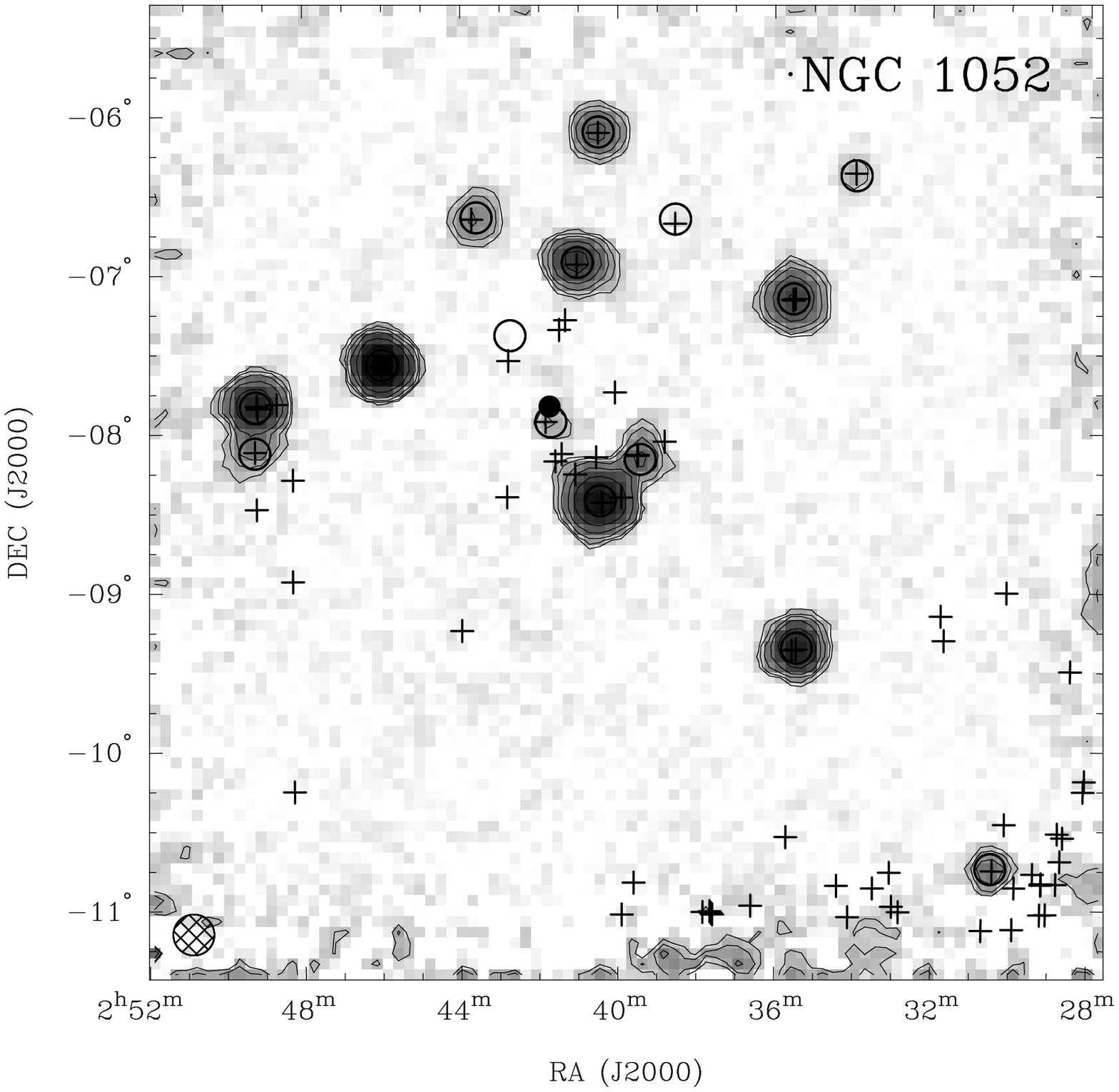,width=7.2cm, angle=-90}}&
\mbox{\psfig{file=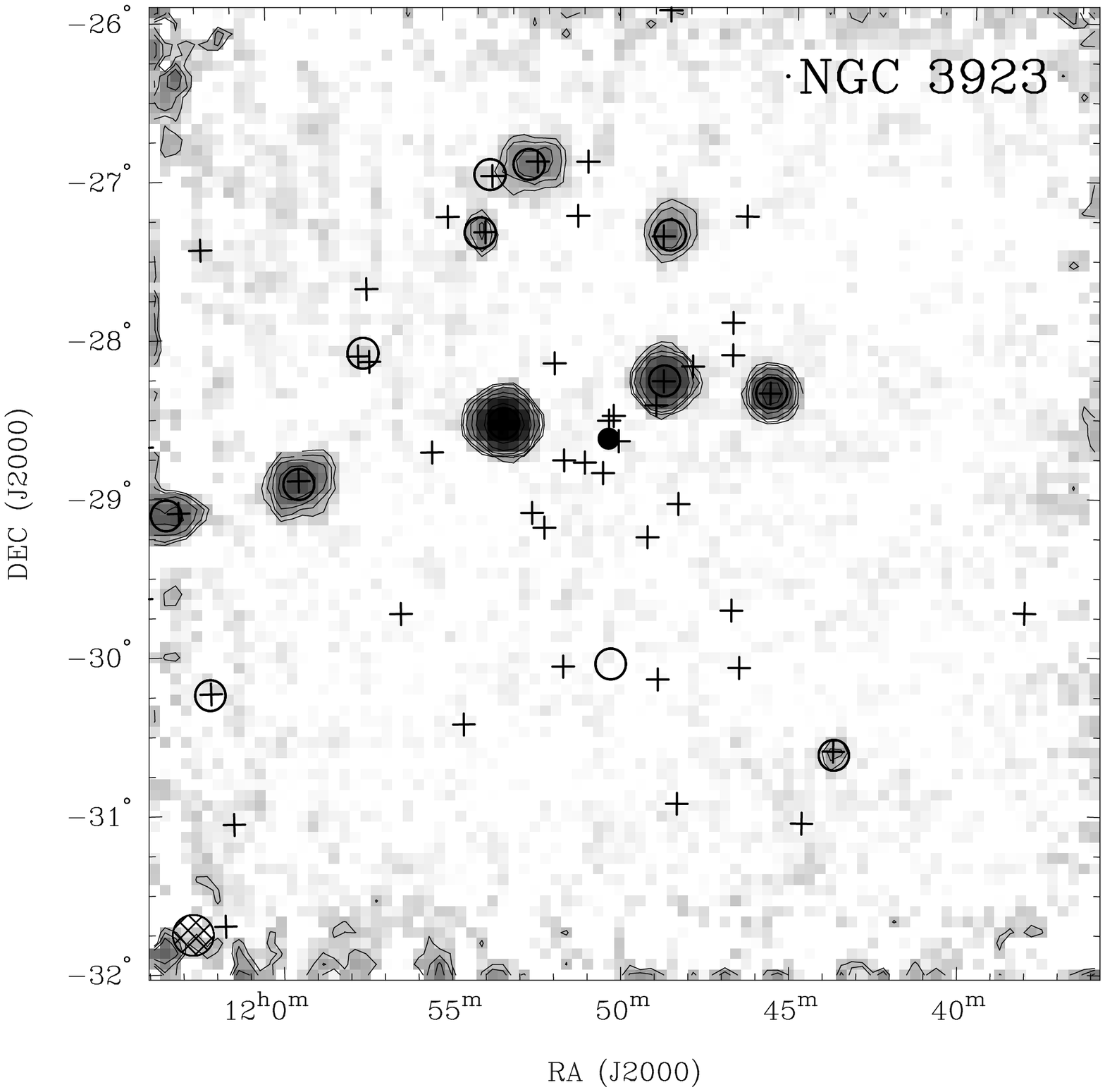,width=7.2cm, angle=-90}}\\
\end{tabular}
\caption{Groups with central galaxy only X-ray emission.  The contour levels and
symbols are the same as in Figure~\ref{fig:xray1}.}
\label{fig:centralxray}
\end{figure*}

\begin{figure*}
\begin{tabular}{cc}
\mbox{\psfig{file=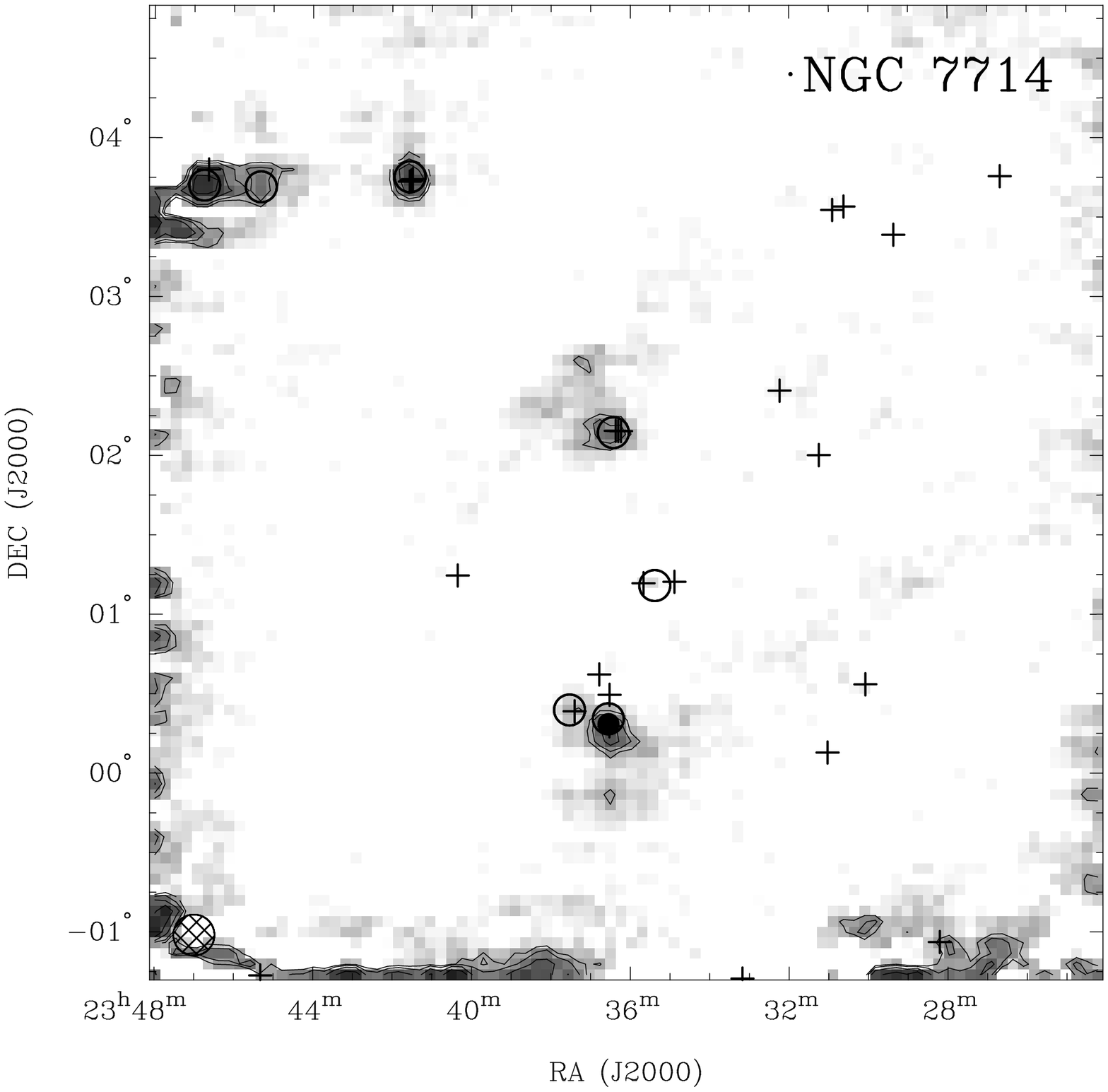,width=7.2cm, angle=-90}}&
\mbox{\psfig{file=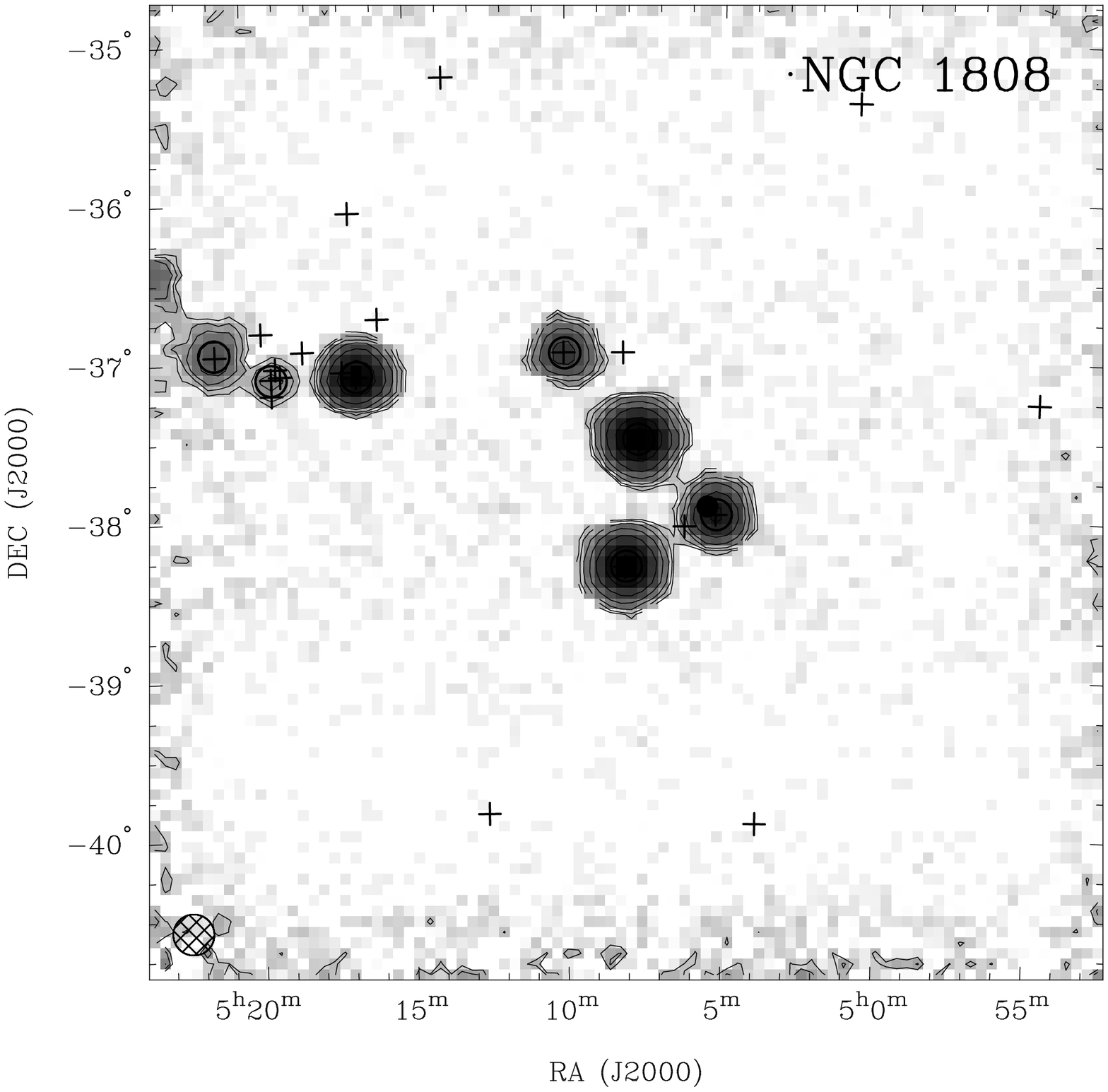,width=7.2cm, angle=-90}}\\
\end{tabular}
\caption{Groups undetected in X-rays. The contour levels are 1, 2, 4,
8, 16, 32 Jy beam$^{-1}$, apart from NGC 7714, where the contour levels
start at 2 Jy beam$^{-1}$. The symbols are the same as in Figure~\ref{fig:xray1}.}
\label{fig:noxray}
\end{figure*}

\section{\HI\ Distribution of the Groups}

Figures~\ref{fig:xray1}, ~\ref{fig:xray2}, ~\ref{fig:centralxray} and
~\ref{fig:noxray} show the integrated \HI\ distribution as measured
with the Parkes 64-m telescope for each of the 16 groups, divided by
group X-ray properties. The \HI\ distribution of the groups is varied,
from \HI\ being detected throughout the group, to \HI\ only being
detected in the outer regions of the group. As we have no distance
information for galaxies in the data, these \HI\ maps include the full
velocity range of the \HI\ data. In some cases it is likely that the
foreground or background \HI\ emission will appear to be part of the
group.

\begin{figure*}
\begin{tabular}{cc}
\mbox{\psfig{file=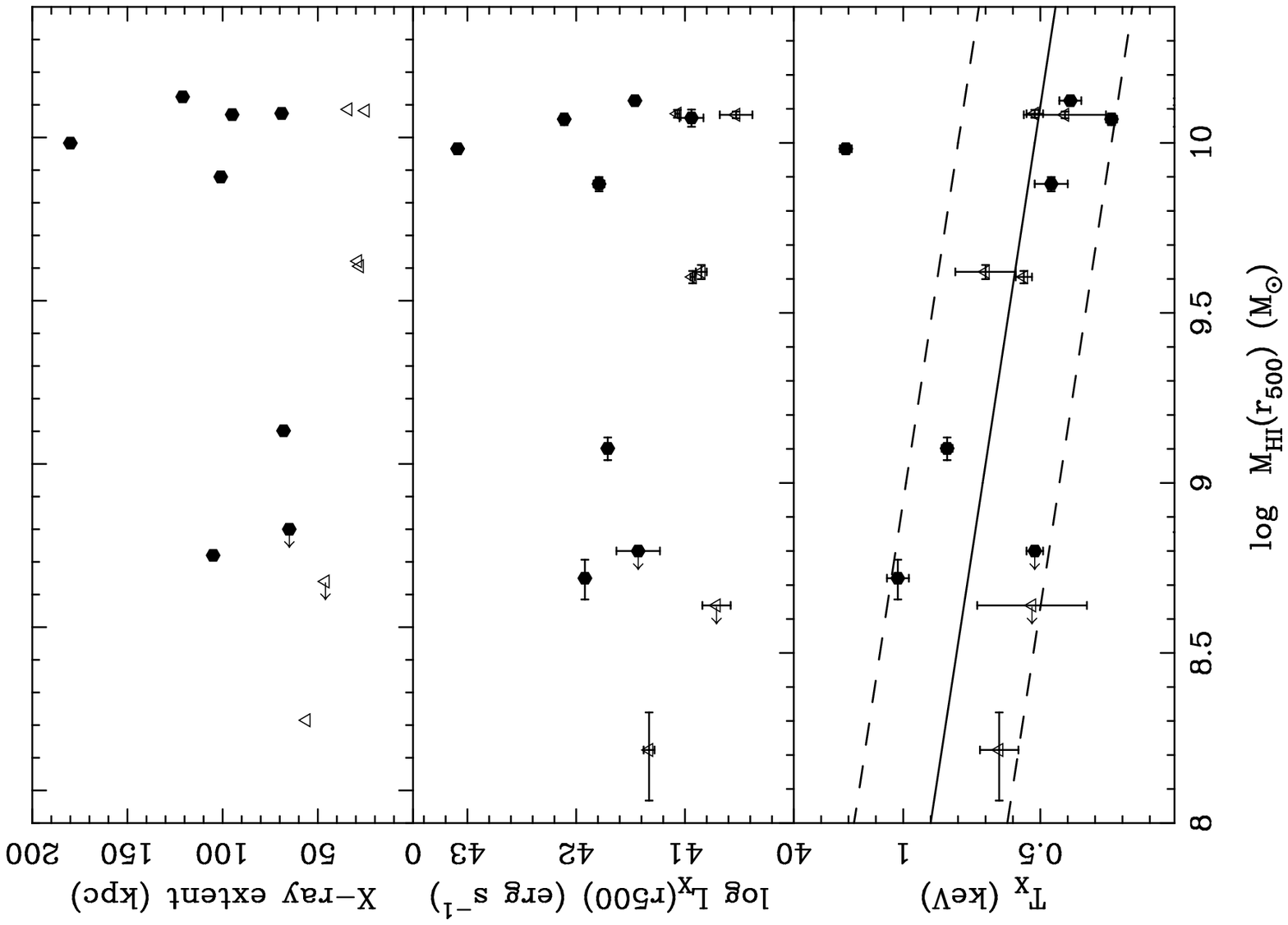,width=10.0cm, angle=-90}}&
\mbox{\psfig{file=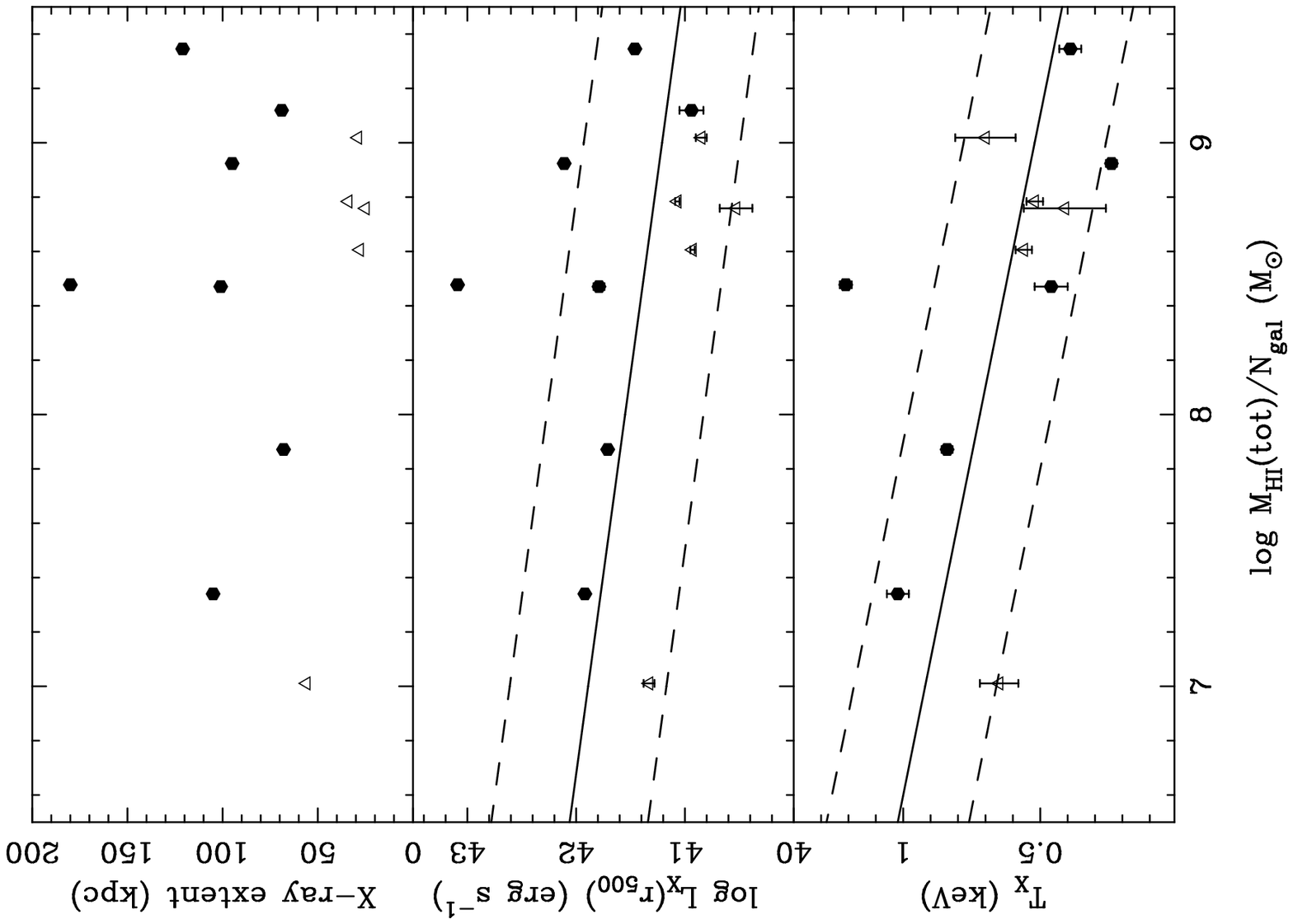,width=10.0cm, angle=-90}}\\
\end{tabular}
\caption{Total group \HI\ mass compared with X-ray properties. Left: The top panel shows the total \HI\ mass of the group (within the r$_{500}$ radius) against the extent of the X-ray emission in the group. The middle (bottom) panel plots the total \HI\ mass of the group against the X-ray luminosity (temperature). Right: The top panel shows the normalised \HI\ mass of the group against the extent of the X-ray emission in the group. The middle (bottom) panel plots the normalised \HI\ mass of the group against the X-ray  luminosity (temperature). The solid lines show a Buckley-James regression fit to the data, and the dashed lines show the 1-$\sigma$ standard deviation on the fit. The filled circles are groups with a detected extended X-ray emission, and the triangles are groups with no extended X-rays emission. }
\label{fig:HI_Xray}
\end{figure*}

\subsection{Comparison with X-ray emission}

Looking at the \HI\ content and distribution of groups with varying
X-ray properties can tell us about the evolutionary stage of the
group. We might expect those groups with extended X-ray halos to
contain the least \HI\ in them, as is often found in the more dense
cluster environment (e.g. \citealt{solanes2001};
\citealt{cayatte1990}). In some of the groups with extended X-ray
emission (see Figures~\ref{fig:xray1} and ~\ref{fig:xray2}) we do see
that there is a lack of \HI\ in the group (e.g. NGC 1407, NGC 4636),
however this trend is not so obvious in other groups with extended
X-ray emission (e.g. NGC 3783, IC 1459). Looking at the groups where
X-rays were only detected in the central galaxy
(Figure~\ref{fig:centralxray}), once again the \HI\ distribution is
varied. None of the galaxies in the central region of the NGC 524
group were detected in \HI\, but in other groups such as NGC 1052 and
NGC 1566, the majority of galaxies were detected. The two groups
without detected X-rays shown in Figure~\ref{fig:noxray} do not
present a consistent picture in \HI\ emission, with few galaxies
detected in the NGC 7714 group, and most galaxies detected in \HI\ in
the NGC 1808 group.

To obtain a quantative comparison of the \HI\ content of the groups
compared to X-ray properties of the groups we look at the \HI\ mass
detected in the groups. We determine the total \HI\ mass, and the
\HI\
 contained within r$_{500}$ for the groups. r$_{500}$ is the
radius corresponding to an overdensity of 500 times the critical
density, and is used in \cite{o4} and \cite{brough06b} for analysis of
GEMS groups.  Here and in subsequent analysis of the groups, we refer
specifically to the members of the groups defined by the
friends-of-friends algorithm in Paper I \citep{brough06b}. We use the
Kendall's rank correlation probabilities for each group parameter
pair, using the IRAF/STSDAS/STATISTICS package routine, as was used in
\citet{brough06b}. As explained in \citet{brough06b}, Kendall's rank
correlation is more reliable for samples where $N < 30$ than the
Spearman rank correlation, and it is non-parametric. Where available
we use upper limits for X-ray luminosities, and \HI\ masses, and the
survival analysis tasks available in IRAF take these into account in
the correlation. We also use the IRAF Buckley-James algorithm to fit
straight lines to our data.

Figure~\ref{fig:HI_Xray} shows the group \HI\ mass (within r$_{500}$),
and the normalised \HI\ mass ($M_{HI,tot}/N_{gals}$) plotted against
the X-ray properties (X-ray extent, $L_X$ and $T_X$) of the
groups. No significant correlation is seen between the extent of the
group X-ray emission and the group \HI\ mass, or between the X-ray
luminosity and total group \HI\ mass. A trend is
seen however between the group \HI\ mass and the X-ray temperature,
with a probability of 87.8\%, calculated using the Kendall's Tau
method described above. Fitting a straight line to these points using
the Buckely-James algorithm gives the functional fit of

\begin{equation}
T_X = -0.18^{\pm 0.14}M_{HI,r_{500}} + 2.4
\end{equation}

\noindent
with a standard deviation on the regression of 0.28. 

Using the group \HI\ mass within $r_{500}$ to compare with X-ray
properties is prone to bias due to the low number statistics in
individual groups. For example, just one gas-rich galaxy can make a
group appear to be gas-rich, when in fact the majority of the galaxies
in the group are gas-poor. To better understand the global \HI\
content of a group, we have determined the {\it normalised group \HI\
mass}, which is the {\it total} \HI\ mass in the group divided by the
total number of group members. 

When using this quantity, we again find no correlation between the group
X-ray extent and the normalised group \HI\ mass, however a weak trend
is seen between the X-ray luminosity and the normalised group \HI\
mass with a probability of 80\%, and a stronger trend between the
X-ray temperature and normalised group \HI\ mass with a probability of
93\% (Figure~\ref{fig:HI_Xray}). The Buckley-James fit to the X-ray
luminosity and temperature is

\begin{equation}
L_X = -0.54^{\pm 0.3}M_{HI,tot}/N_{gal} + 44.3
\end{equation}

\noindent
with a standard deviation on the regression of 0.7, and

\begin{equation}
T_X = -0.2^{\pm 0.12}M_{HI,tot}/N_{gal} + 2.3
\end{equation}

\noindent
with a standard deviation on the regression of 0.26. Using the
normalised group \HI\ mass in comparison to the X-ray properties of a
group have shown that the presence of an extended intra-group medium
and the depth of the potential well as indicated by the X-ray
temperature of the group is more important to the \HI\ content of
galaxies in groups, than the physical extent of the hot intra-group
medium.

\subsection{Dependence on total mass and morphology}

\begin{figure}
\begin{tabular}{c}
\mbox{\psfig{file=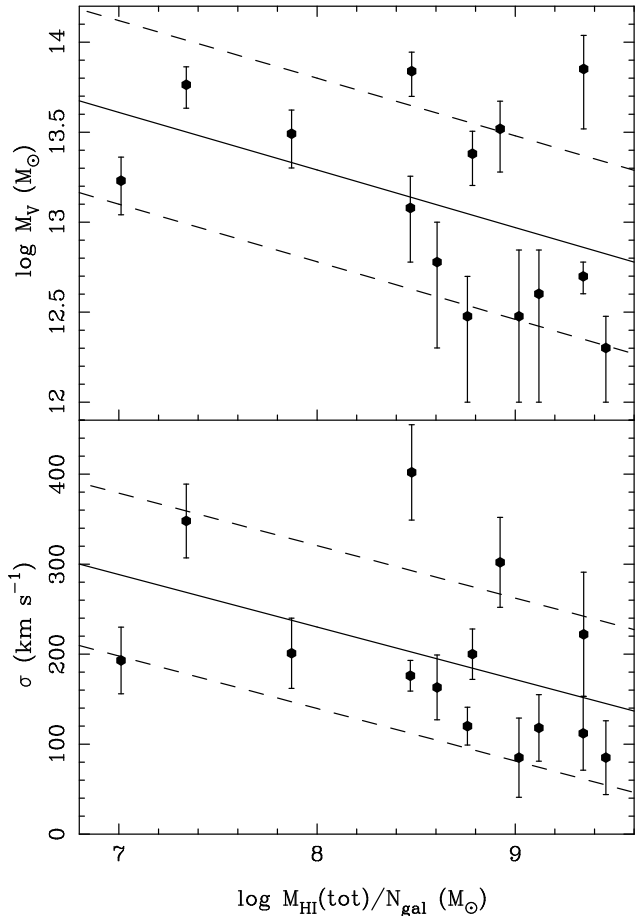,width=12cm, angle=-90}}
\end{tabular}
\caption{Normalised group \HI\ mass versus the group virial mass (top) and the velocity dispersion of group (bottom). The solid lines show a Buckley-James regression fit to the data, and the dashed lines show the standard deviation on the fit.}
\label{fig:HImasses}
\end{figure}

We examined the relationships between the total \HI\  mass of the
groups, and their respective virial mass and velocity dispersion. We
found no correlation in these parameters. However, when we compared
the normalised \HI\ group mass with these parameters, we did find
trends within the data. Figure~\ref{fig:HImasses} shows the
corresponding trends. Using the Kendall's Tau statistic, we
found that the probability there was a correlation between the group
virial mass and the normalised \HI\ mass was 81.2\%, with a
Buckley-James regression of

\begin{equation}
M_V = -0.32^{\pm 0.19}M_{HI,tot}/N_{gal} + 15.8,
\end{equation}

\noindent
with a standard deviation on the regression of 0.5. There was a
stronger correlation between the normalised \HI\ mass of the groups
and the velocity dispersion of 95.2\%, with a Buckley-James regression
of

\begin{equation}
\sigma_v = -58.3^{\pm 33.9}M_{HI,tot}/N_{gal} + 696.5,
\end{equation}

\noindent 
with a standard deviation on the regression of 90.5. These
results compare well with the finding of a trend between the
normalised \HI\ mass and the X-ray temperature and luminosity, as
these quantities all relate to how evolved a group is, and the depth
of the potential well.

We also looked for a correlation between the total K-band luminosity
and the total \HI\ group mass and normalised \HI\ group mass, however we
found none.

\begin{figure} 
\begin{tabular}{c}
\mbox{\psfig{file=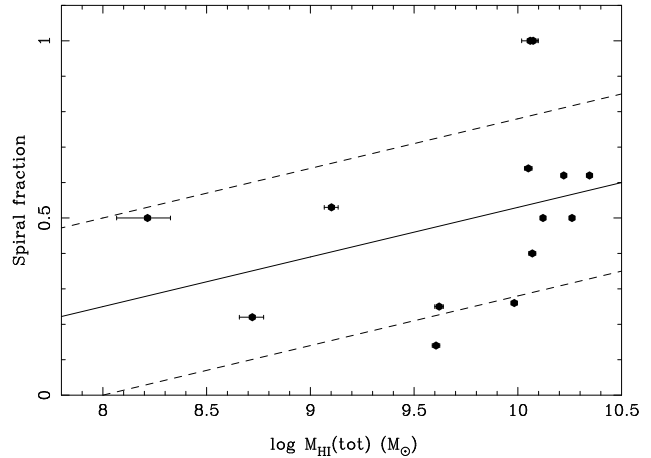,width=6cm,angle=-90}}
\end{tabular}
\caption{Total group \HI\ versus group spiral fraction. The solid
lines show a Buckley-James regression fit to the data, and the dashed
lines show the standard deviation on the fit. A trend is seen for
groups with a higher spiral fraction to be more \HI\ rich. }
\label{fig:hisp}
\end{figure}

It is well known that the \HI\ content of galaxies depends on if they
are early or late-type, with later types tending to be
the more \HI-rich. Figure~\ref{fig:hisp} shows the total group \HI\
mass plotted against the group spiral fraction. There is a 92.3\%
probability that the two parameter sets are correlated, with a Buckley-James regression of 

\begin{equation}
F_{sp} = 0.14^{\pm .11}M_{HI,tot} -0.87,
\end{equation}

\noindent
with a standard deviation on the regression of 0.25. There was a
weaker trend (81\% probability) between the normalised \HI\
group mass and the spiral fraction of the groups. This result is as
expected, showing that the groups with a higher spiral fraction also
contain the most neutral hydrogen, although the relationship does show
some considerable scatter.

Using the normalised group \HI\ mass, we have shown that the \HI\
content of groups is dependent on the X-ray temperature, and to a
lesser extent the X-ray luminosity. However, physical size of the
X-ray emission region does not have any effect on the the \HI\ content
of the groups. There was a strong trend such that the normalised \HI\
content of groups was lower for groups with a higher velocity
dispersion. Groups with a higher spiral fraction contained more
\HI. These results show a consistent picture, that the more evolved
groups, that have lower spiral fractions, higher velocity dispersions
and higher X-ray temperatures also contain the least neutral hydrogen
per group member.

\section{\HI\ content of galaxies in groups}

As we have seen in Figure~\ref{fig:hisp}, the \HI\ content of groups
is dependent on the make-up of their morphological types. To further
investigate the processes that are occuring in the group environment,
we now consider the \HI\ content of the individual galaxies, and compare
them to similar galaxies in the non-group (i.e. `field') environment.

We investigate whether the galaxies in groups are undergoing
transformations, by looking at their relative \HI\ content. We
determine the `expected' \HI\ from the optical properties of a galaxy,
and compare it with the detected \HI\ emission from our survey. To
compare the two values, we determine an `\HI\ deficiency parameter',
{\it Def}  \citep{haynes1984}, which is the difference between the
expected \HI\ mass and the observed \HI\ mass:

\begin{equation}
{\it Def} = {\rm log}\, M_{HI}(expected) - {\rm log}\, M_{HI}(obs).
\end{equation}

We calculate the expected \HI\ for a galaxy using its optical
properties, ie optical diameter, and morphological type. We obtained
morphological types and optical diameters for our group galaxies from
the LEDA\footnote{http://leda.univ-lyon1.fr/} database. The
relationships we used to calculate the expected \HI\ content are from Kilborn
et al. (2005), which also uses the LEDA database for optical
parameters. There is a large scatter in the relationship for early type
galaxies, so we limit our investigation to spiral galaxies of type Sa
and later. Due to the large beamsize of the Parkes telescope ($\sim 15
\arcmin$), there are a number of \HI\ sources with uncertain
optical counterparts. For the \HI\ deficiency analysis, we only
include galaxies that have one single optically matched counterpart at
the same redshift as the source. It is possible that we will still
include multiple galaxies in our analysis, with galaxies of unknown
redshift in the Parkes beam. This will have the effect of making a
particular galaxy appear more \HI\ rich than it actually is.

\begin{figure} 
\begin{tabular}{c}
\mbox{\psfig{file=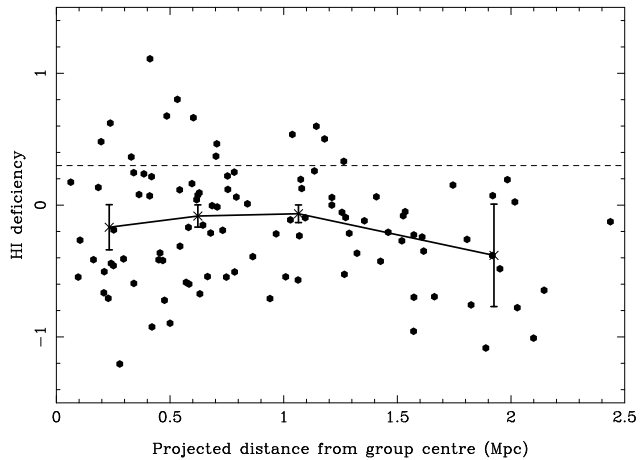,width=6cm,angle=-90}}
\end{tabular}
\caption{\HI\ deficiency versus distance from group centre. Galaxies
that are \HI\ deficient have a positive value for the \HI\ deficiency
parameter. The crosses indicate the mean \HI\ deficiency, with equal
numbers of galaxies in each bin, whilst the error bars show the
standard error on this mean value. Points above the dashed line are
defined as \HI\ deficient, containing less than half of the \HI\ mass
expected.}
\label{fig:hidef_dist}
\end{figure}

Figure~\ref{fig:hidef_dist} shows the \HI\ deficiency of non interacting
galaxies in our sample with available optical data as described above,
versus their projected distance from the group centre. We consider
galaxies with {\it Def} $>$ 0.3 to be \HI\ deficient (i.e. they contain
less than half \HI\ as expected). Overall, there is no average trend for
galaxies to be more \HI\ deficient in the centres of groups. However,
those galaxies that are \HI\ deficient tend to lie closer to the
centre of the groups rather than on the edges.

\begin{figure} 
\begin{tabular}{c}
\mbox{\psfig{file=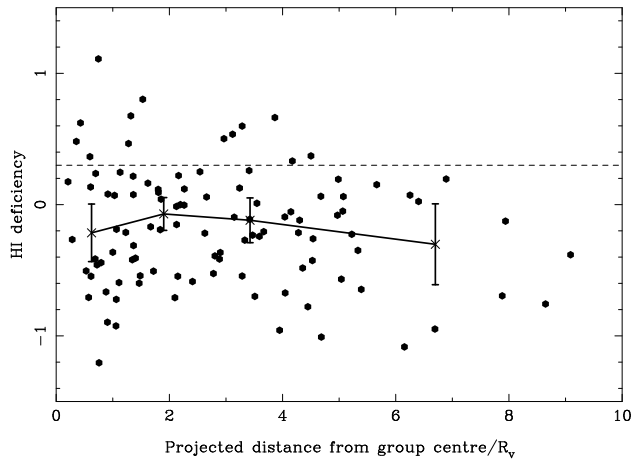,width=6cm,angle=-90}}
\end{tabular}
\caption{\HI\ deficiency versus distance from group centre, in units
of virial radii. Galaxies that are \HI\ deficient have a positive
value for the \HI\ deficiency parameter. The crosses indicate the mean
\HI\ deficiency, with equal numbers of galaxies in each bin, whilst
the error bars show the standard error on this mean value. Points
above the dashed line are defined as \HI\ deficient, containing less
than half of the \HI\ mass expected.}
\label{fig:hidef_Rv}
\end{figure}

Figure~\ref{fig:hidef_Rv} shows the \HI\ deficiency of galaxies versus their
projected distance from the group centre, normalised by the virial
radius of the group. This shows that almost half
of the \HI\ deficient galaxies lie 3 - 4 virial radii from the group
centre, whilst the others lie within 2 virial radii of the group
centre.

\begin{figure} 
\begin{tabular}{c}
\mbox{\psfig{file=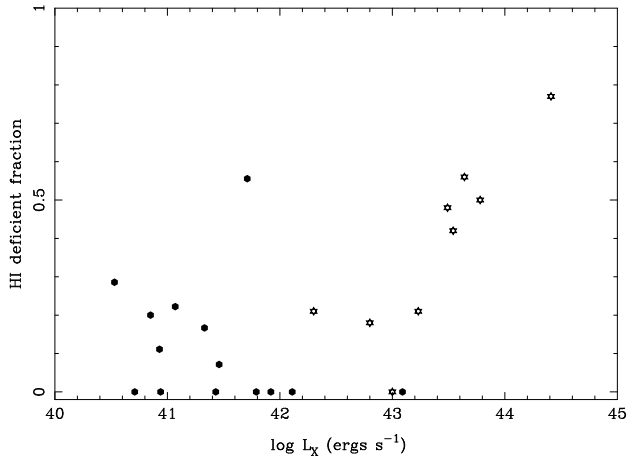,width=6cm,angle=-90}}
\end{tabular}
\caption{\HI\ deficient fraction versus X-ray luminosity. The filled
circles represent the fraction of \HI\ deficient galaxies in the GEMS
groups, and the stars show this for clusters \citep{giovanelli85}.}
\label{fig:hidef_Lx}
\end{figure}

Figure~\ref{fig:hidef_Lx} shows the proportion of \HI\ deficient
galaxies compared to the number of group members (with \HI\ deficiency
measurement available), versus the X-ray luminosity of the group. For
comparison, these quantities are also plotted for clusters, taken from
\citet{giovanelli85}. We find no correlation between X-ray luminosity
and the fraction of \HI\ deficient spirals. Contrary to our results,
\citet{giovanelli85} find an apparent trend between the \HI\ deficient
fraction of clusters and the X-ray luminosity, however with a larger
sample \citet{solanes2001} found no clear trend.

\begin{figure} 
\begin{tabular}{c}
\mbox{\psfig{file=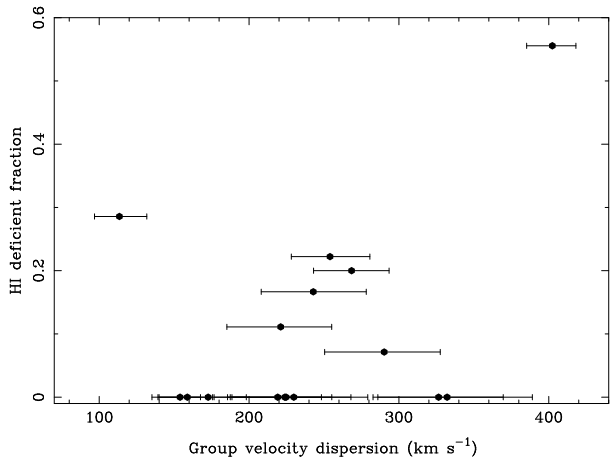,width=6cm,angle=-90}}
\end{tabular}
\caption{\HI\ deficient fraction versus group velocity dispersion. }
\label{fig:hidef_sigma}
\end{figure}

Figure~\ref{fig:hidef_sigma} shows the \HI\ deficient fraction of galaxies
compared to the group velocity dispersion. No correlation is seen
between these two parameters, as might have been expected if the \HI\
deficiencies seen in groups are due to ram pressure stripping.

\subsection{Removal of gas from galaxies in groups}

We have found that some galaxies in the group environment display an
\HI\ deficiency. What is the mechanism for the removal of the gas from
these galaxies, and what effect does gas removal have on the galaxy
itself?  Previous studies have found that galaxy interactions, and
tidal stripping are the dominant mechanism in loose groups, as the
\HI\ deficiency of galaxies appears to increase with galaxy density,
and is inversly correlated with radial velocity \citep{omar2005}, and
\HI\ deficient galaxies are found in groups without an intra-group
medium \citep{kilborn05}. In compact groups, the stripping mechanism
is not so clear, as the most \HI-deficient groups tend to show
intragroup X-ray emission \citep{verdes01}.

The group we found that had the highest number of \HI\ deficient
galaxies is NGC 4636 - a dynamically evolved group, which lies on the
outskirts of the Virgo cluster. However, most of the other \HI\
deficient galaxies in our groups lie in groups that do not have a hot
intra-group medium, and thus ram pressure stripping cannot be
attributed to the gas-loss from the galaxies in these groups.

While a hot IGM does not seem to make a difference to the gas-loss of
galaxies in groups, the position in the group does - we find that two thirds 
of the \HI\ deficient galaxies lie within 1 Mpc projected distance of the group centre
(see Figure~\ref{fig:hidef_dist}), although not every \HI\ deficient
galaxy is classified as a member of the group. 

Our results are consistent with the scenario that the dominant
mechanism of gas removal in the group environment is via tidal
stripping through galaxy-galaxy or galaxy-group interactions, rather
than ram pressure stripping, although we note that the group with the
most \HI\ deficient members also has the highest velocity dispersion.

\section{\HI\ Mass function in groups} 
 
 \begin{figure}
\begin{tabular}{c}
 \mbox{\psfig{file=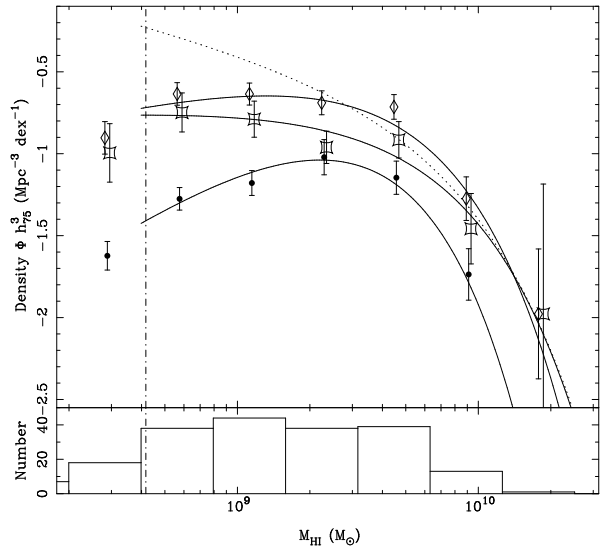,width=7.2cm,
angle=-90}}
 \end{tabular}
 \caption{\HI\ mass function for the
composite group. The open diamonds
 show the \HI\ mass function for
all galaxies we detected in the \HI\
 cubes, the filled circles are
for galaxies determined to be members of
 groups, and the open squares
are the galaxies that are not members of
 the groups. The solid lines
show a least-squares fit of a Schecter
 function to the binned data,
and the dashed line is the \HI\ mass
 function from HIPASS for
comparison \citep{zwaan2005}. The histogram
 shows the \HI\
distribution of all detected galaxies. The Schecter fit
 parameters
are listed in Table~\ref{tab:himf}. The errors on each point are
statistical errors.}
 \label{fig:himf}
 
\end{figure}

 To investigate the distribution of \HI\ mass in groups, we construct
an \HI\ mass function, to compare the \HI\ content of our groups to
that of the field.

Due to the low number statistics for the individual groups, we combine
the data from each group together to determine a composite \HI\ mass
function for the groups. We normalised the number of galaxies in each
\HI\ cube by the available volume in the \HI\ cube according to the
redshift range covered by the cube. 

Figure~\ref{fig:himf} shows the resulting \HI\ mass function for the
composite group.  We show  the \HI\ mass  function for the
full sample, one for those galaxies found in the groups
according to \cite{brough06b}, and the \HI\ mass function for galaxies found
in the \HI\ cubes, but not in the groups. 

To parameterise our data, we fit a Schechter function
\citep{schechter1976} to each data set:

\begin{equation}
\Psi({M_{HI} \over M^*})d({M_{HI} \over M^*}) = \theta^*({M_{HI} \over M^*})^{- \alpha} \times {\rm exp} ({- M_{HI} \over M^*})d({M_{HI} \over M^*})
\end{equation}

\noindent 
where $M^*$ is the characteristic \HI\ mass, $\alpha$ is the
low-mass slope, and $\theta^*$ is the normalisation of the mass
function. The solid lines show least-square fits of a Schechter
function to the binned data. We do not fit for data below the
completeness limit of the least sensitive group observations, at an
\HI\ mass of $5 \times 10^8$\Msun.

We show the \cite{zwaan2005} \HI\ mass function from HIPASS for
comparison (normalised to our data), with log $M^* = 9.8$\Msun, and $\alpha
= 1.37$. Schechter parameters found for the group and
non-group galaxies are summarised in Table~\ref{tab:himf}.

We find that the overall \HI\ mass function in the general group
environment (i.e. the groups, and galaxies surrounding the groups) is
very flat, with a low-mass slope of $\alpha = 0.65\pm0.17$. When the
sample is divided into galaxies that are members of the groups, and
those that are not, we find a noticeable difference in the shape of
the \HI\ mass function. For those galaxies that are confirmed members
of a group, the \HI\ mass function shows a very shallow low-mass slope
($\alpha =0.0\pm0.18$). We find a steeper slope for those galaxies
surrounding the groups ($\alpha =0.92\pm0.23$), which is closer to,
but still significantly shallower than, the global result from
\citet{zwaan2005} of $\alpha = 1.37$, perhaps indicating that group
processes may apply to galaxies nearby groups as well.

Galaxies of low \HI\ mass could be dwarf galaxies, early-type galaxies
(which typically contain less \HI\ than spiral galaxies), or
gas-deficient spiral galaxies. The group \HI\ mass function also
shows, that the highest \HI\ mass galaxies are not group members, with
the highest \HI\ mass being in the log $M_{HI} = 9.75 $ \Msun\ bin. So
not only are the least massive galaxies found in fewer numbers in the
group environment, but we also do not find the most massive galaxies
either. This may be an effect of stripping of the most gas-rich
galaxies in the group environment, similar as is seen in the Coma
cluster \citep{gavazzi06}.

Our results on the \HI\ mass function for groups show similarity to
that of the Virgo cluster, which is found to peak at an \HI\ mass of
$\sim 4 \times 10^8$\Msun, and decline thereafter
\citep{davies2004,gavazzi05}. \citet{gavazzi05} used a Monte Carlo
simulation to model the effect of \HI\ deficiency on the shape of the
\HI\ mass function for the Virgo cluster. They found a good agreement
between their simulations and the \HI\ mass function, showing that it
is possible to transform the field \HI\ mass function into that seen
in a cluster by galaxies losing their neutral hydrogen content. Our
result of finding \HI\ deficient galaxies in groups, combined with the
observation of a flat \HI\ mass function in groups suggests a similar
process could be occuring in groups.

\begin{table}
\caption{\HI\ mass function Schechter parameters. Values for the low-mass slope of the \HI\ mass function,
$\alpha$, and the characteristic \HI\ mass, $M^{\ast}$ for group,
non-group and all detected galaxies, and HIPASS for comparison \citep{zwaan2005}.}
\label{tab:himf} 
\begin{tabular}{lll}
\hline
Sample    & $\alpha$ & log $M^{\ast}$\Msun \\
\hline
All sample & $0.65 \pm 0.17$ & $9.6 \pm 0.1$\\
Group galaxies & $ 0.0 \pm 0.18$ & $9.7 \pm 0.2$\\
Non-group galaxies & $0.92 \pm 0.23$ & $9.4 \pm 0.06$ \\
HIPASS & $1.37 \pm 0.03 $ & $ 9.8 \pm 0.03 $ \\
 \hline
\end{tabular}
\end{table}

\section{Discussion}

The group environment appears to play an important part in
transforming galaxies in the local universe, with many of the
processes seen in the much denser cluster environment also apparent in
the less extreme group environment. This includes the removal of gas
from spiral galaxies, the discovery of X-ray halos in galaxy groups
indicating a deep potential well and the virialisation of galaxy
groups.

When comparing the \HI\ content of galaxy groups to the X-ray
properties of the groups, we found that the groups with the highest
X-ray temperatures and luminosities typically contained less neutral
hydrogen. These results were strengthened when we examined the
normalised \HI\ content of the groups, where the \HI\ content was
normalised by the number of group members. We also found that the
groups with the highest virial masses and velocity dispersions also
contained less \HI. These results are consistent with the picture that
groups lose their \HI\ as they become more evolved, grow their
potential wells and increase in X-ray brightness, and also increase in
virial mass and velocity dispersion. However, these results do not
prove that the group environment has caused the galaxies to contain
less \HI\ - to investigate this question,we looked at the \HI\
deficiency of galaxies within the group environment, compared to a
much larger field sample.

We found that a number of the galaxies in our sample were \HI\
deficient. About two-thirds of the \HI\ deficient galaxies lie within
a projected distance of 1 Mpc from the group centre. This is
consistent with previous results for the cluster environment
\citep{solanes2001}, although the percentage of \HI\ deficient
galaxies found in the loose groups is much lower than that seen in
clusters. While we observe \HI\ deficient galaxies within the centres
of X-ray bright groups, they are also present in groups that have no
detected intra-group medium. Thus the mechanism for this \HI\
deficiency is not clear - in the case where no IGM is present, ram
pressure stripping cannot be the cause of the \HI\ deficiency, and the
loss of gas is more easily explained by tidal stripping. Our high
resolution images of interacting galaxies in groups (see
\citealt{kern07}) show that \HI\ can be removed from galaxies in
interactions and mergers, although the ultimate fate of the gas
(whether it is lost to the IGM, or falls back onto one or both of the
galaxies) is not certain. \citet{gomez2003} find there is a `break' in
the star-formation-density relationship for clusters and groups at a
density of $~\sim 1\, h^{-2}_{75}\, {\rm Mpc}^{-2}$, which corresponds
to clustercentric radius of 3--4 virial radii. This corresponds to the
maximum radius we find \HI\ deficient galaxies in our groups, perhaps
indicating a correlation between the loss of gas of galaxies in dense
environments, and the decrease in their star-formation rate. This
suggests a pre-processing in the group environment.

Comparing the \HI\ mass function from our combined groups sample with
previous studies, we find that the low-mass slope is flatter than
found in HIPASS, with a low-mass slope of $ 0.0 \pm 0.18$,
compared with the field HIPASS low-mass slope of $\alpha$ = 1.37
\citep{zwaan2005}.  Our result goes against the trend found by
\citet{zwaan2005} for a steeper low-mass slope in denser regions, but
is more consistent with the result of \citet{springob2005} and
\citet{rosenberg2002} who find tentative evidence for flattening of
the low-mass slope in denser envrionments, and \cite{freeland09} who
find a flat low-mass slope in groups.

Our \HI\ results show that the evolutionary status of the group, and
 the morphological make-up are important in determining the \HI\
 content of a group. This result is consistent with
 \citet{giovanelli85}, who find in a study of the \HI\ content of
 clusters, those clusters that are not detected in X-rays, and that contain a higher proportion of spiral galaxies, are not \HI\ deficient.

 \section{Conclusions}We have investigated the evolution of
 galaxies in loose groups through an \HI\ and X-ray survey of 16
 groups. We find that the X-ray emission and \HI\ content of the
 groups is not simply correlated, and that while some groups that
 show extended X-ray emission have little \HI\ content, others are
 \HI\ rich. We find the total \HI\ content is anti-correlated with
 the X-ray temperature, and to a lesser extent the X-ray luminosity,
 and similarly the \HI\ content was correlated with the velocity
 dispersion or virial mass of the group. Not surprisingly, groups
 with a higher fraction of spirals also contained the most \HI. These
 results indicate that those groups that are more evolved contain
 less \HI\ than younger groups. We determined the \HI\ deficiency of
 galaxies within the groups, and found that around two thirds of \HI\
 deficient galaxies were preferentially located at a projected
 distance of less than 1 Mpc from the group centre. There was no
 correlation found between the fraction of \HI\ deficient galaxies in
 a group and the X-ray luminosity. We calculated a combined \HI\ mass
 function for the groups, and found that it is very shallow compared
 to the global HIPASS \HI\ mass function.

\section*{Acknowledgments}
We thank Ruth Musgrave for help with searching the Parkes \HI\ cubes,
and Nuria McKay for helping with the Parkes observations. We thank the
staff at the Parkes and Narrabri observatories for their help in the
project. We acknowledge the expertise of Mark Calabretta, and thank
him for his assistance with AIPS++.

\newpage
\appendixpage
\appendix
\section{\HI\ parameters for all GEMS \HI\ detections.}
\begin{table*}
\label{tab:app1}
\centering
\caption{\HI\ detections of the galaxies in groups. The columns are as
follows: (1)GEMS \HI\ catalogue designation; (2) $\alpha$(J2000) [h m s];(3)
$\delta$ (J2000)[$^{\rm o}$ \arcmin \arcsec]; (4) Systemic velocity,
$V_{sys}$, of source in \kms; (5) error on $V_{sys}$ , in \kms; (6)
\HI\ line width of source at 20\% of the peak, $V_{20}$, in \kms; (7)
error on $V_{20}$, in \kms; (8) \HI\ line width of source at 50\% of
the peak, $V_{50}$, in \kms; (9) error on $V_{50}$, in \kms; (10) Peak
flux, $S_p$ in Jy; (11) error on $S_p$ in Jy; (12) total \HI\ flux,
$S$, in Jy \kms; (13) error on $S$, in Jy \kms; (14) \HI\ mass, \MHI,
assuming group distance as given in Table~\ref{tab:groups}, in units
of $10^8$\Msun; (15) error on \MHI, in units of $10^8$\Msun.  }
\begin{tabular}{lccrrrrrrrrrrrrr}
(1)&(2)&(3)&(4)&(5)&(6)&(7)&(8)&(9)&(10)&(11)&(12)&(13)&(14)&(15)\\
\hline
GEMS\_N524\_1  &1:25:36.989 &7:58:41.185 &2898 &3 &138 &9 &56 &6 &0.113 &0.007 &8.8 &0.68 &10.31 &0.79 &\\ 
GEMS\_N524\_2  &1:25:24.242 &7:36:53.923 &2778 &3 &169 &9 &145 &6 &0.054 &0.006 &5.9 &0.63 &6.91 &0.74 &\\ 
GEMS\_N524\_3  &1:24:29.564 &7:42:35.913 &2740 &2 &231 &6 &211 &4 &0.064 &0.005 &8.2 &0.61 &9.61 &0.72 &\\ 
GEMS\_N524\_4  &1:23:35.983 &6:55:45.128 &2728 &2 &61 &6 &44 &4 &0.051 &0.005 &2.2 &0.35 &2.58 &0.41 &\\   
GEMS\_N524\_5  &1:25:24.181 &10:51:08.303 &2549 &3 &136 &9 &96 &6 &0.082 &0.007 &6.9 &0.63 &8.08 &0.74 &\\ 
GEMS\_N524\_6  &1:15:25.003 &8:06:18.193 &2348 &2 &155 &6 &144 &4 &0.056 &0.005 &5.8 &0.52 &6.80 &0.61 &\\ 
GEMS\_N524\_7  &1:23:52.770 &6:41:30.747 &2335 &1 &43 &3 &34 &2 &0.121 &0.008 &4.2 &0.49 &4.92 &0.58 &\\   
GEMS\_N524\_8  &1:23:30.768 &7:43:42.491 &2689 &3 &126 &9 &104 &6 &0.039 &0.005 &2.5 &0.40 &2.93 &0.47 &\\ 
GEMS\_N524\_9  &1:27:36.255 &8:49:36.256 &2430 &4 &74 &12 &61 &8 &0.029 &0.006 &1.4 &0.41 &1.64 &0.48 &\\

\hline
GEMS\_N720\_1 &1:41:40.196 &-16:08:58.995 &1635 &2 &120 &6 &111 &4 &0.056 &0.006 &4.2 &0.50 &6.54 &0.78 &\\    
GEMS\_N720\_2 &1:42:58.502 &-16:07:44.993 &1154 &9 &174 &27 &147 &18 &0.019 &0.006 &1.4 &0.53 &2.18 &0.82 &\\  
GEMS\_N720\_3 &1:46:53.731 &-14:26:35.705 &1742 &5 &145 &15 &71 &10 &0.047 &0.005 &3.3 &0.46 &5.14 &0.72 &\\   
GEMS\_N720\_4 &1:48:22.883 &-12:22:30.845 &1613 &2 &108 &6 &82 &4 &0.182 &0.010 &15.0 &0.96 &23.34 &1.49 &\\   
GEMS\_N720\_5 &1:56:57.619 &-11:46:50.562 &1865 &2 &193 &6 &164 &4 &0.078 &0.006 &10.6 &0.75 &16.49 &1.17 &\\  
GEMS\_N720\_6 &1:48:58.649 &-12:48:18.681 &1770 &7 &220 &21 &165 &14 &0.033 &0.005 &3.7 &0.59 &5.76 &0.92 &\\  
GEMS\_N720\_7 &1:49:35.781 &-13:29:37.625 &1450 &8 &252 &24 &165 &16 &0.032 &0.005 &3.7 &0.54 &5.76 &0.84 &\\  
GEMS\_N720\_8 &1:44:59.449 &-16:26:18.797 &1585 &7 &152 &21 &117 &14 &0.029 &0.006 &2.2 &0.57 &3.42 &0.89 &\\

\hline
GEMS\_N1052\_1  &2:39:25.731 &-8:09:40.445 &1266 &3 &299 &9 &244 &6 &0.070 &0.006 &13.1 &0.80 &10.00 &0.61 &\\ 
GEMS\_N1052\_2 	&2:41:01.566 &-6:55:18.063 &1297 &1 &194 &3 &183 &2 &0.174 &0.010 &21.8 &1.10 &16.64 &0.84 &\\ 
GEMS\_N1052\_3 	&2:45:59.555 &-7:34:03.815 &1405 &2 &332 &6 &293 &4 &0.251 &0.013 &59.6 &2.09 &45.50 &1.60 &\\ 
GEMS\_N1052\_4 	&2:49:12.555 &-7:49:54.822 &1329 &1 &190 &3 &172 &2 &0.192 &0.011 &29.2 &1.38 &22.29 &1.05 &\\ 
GEMS\_N1052\_5 	&2:40:29.765 &-6:05:59.947 &1325 &1 &97 &3 &84 &2 &0.135 &0.008 &10.0 &0.68 &7.63 &0.52 &\\    
GEMS\_N1052\_6 	&2:41:42.039 &-7:55:29.289 &1373 &4 &136 &12 &81 &8 &0.047 &0.005 &3.8 &0.45 &2.90 &0.35 &\\   
GEMS\_N1052\_7 	&2:40:26.810 &-8:25:17.977 &1371 &1 &118 &3 &100 &2 &0.463 &0.024 &44.2 &2.37 &33.74 &1.81 &\\ 
GEMS\_N1052\_8 	&2:35:26.590 &-9:20:53.242 &1504 &1 &278 &3 &256 &2 &0.183 &0.010 &32.6 &1.40 &24.89 &1.07 &\\ 
GEMS\_N1052\_9 	&2:33:56.139 &-6:22:24.736 &1411 &2 &57 &6 &40 &4 &0.063 &0.005 &2.6 &0.33 &1.98 &0.25 &\\     
GEMS\_N1052\_10	 &2:49:13.865 &-8:07:17.779 &1419 &4 &155 &12 &80 &8 &0.080 &0.007 &7.6 &0.68 &5.80 &0.52 &\\  
GEMS\_N1052\_11	 &2:35:31.894 &-7:08:58.737 &1532 &1 &91 &3 &74 &2 &0.243 &0.013 &17.6 &1.10 &13.43 &0.84 &\\  
GEMS\_N1052\_12	 &2:30:28.038 &-10:44:12.632 &2100 &2 &139 &6 &114 &4 &0.082 &0.006 &7.3 &0.60 &5.57 &0.46 &\\ 
GEMS\_N1052\_13	 &2:42:44.220 &-7:22:56.063 &1420 &3 &55 &9 &48 &6 &0.025 &0.005 &0.9 &0.30 &0.69 &0.23 &\\    
GEMS\_N1052\_14	 &2:43:35.611 &-6:38:23.618 &1408 &1 &56 &3 &41 &2 &0.129 &0.008 &5.0 &0.49 &3.82 &0.38 &\\    
GEMS\_N1052\_15	 &2:38:32.127 &-6:38:55.517 &1470 &6 &168 &18 &141 &12 &0.023 &0.005 &2.7 &0.57 &2.06 &0.44 &\\

\hline
GEMS\_N1332\_1 &3:16:22.347 &-24:16:07.601 &2087 &3 &113 &9 &76 &6 &0.159 &0.015 &9.7 &1.19 &9.98 &1.22 &\\	   
GEMS\_N1332\_2$^1$ &3:33:47.980 &-19:29:33.799 &1964 &2 &233 &6 &200 &4 &0.216 &0.012 &38.6 &1.67 &39.72 &1.72 &\\   
GEMS\_N1332\_3$^1$ &3:33:40.004 &-21:29:10.175 &1860 &2 &212 &6 &196 &4 &0.100 &0.007 &13.0 &0.84 &13.38 &0.86 &\\   
GEMS\_N1332\_4 &3:23:42.628 &-19:44:14.250 &1846 &2 &126 &6 &113 &4 &0.087 &0.008 &8.9 &0.80 &9.16 &0.82 &\\	   
GEMS\_N1332\_5$^1$ &3:35:36.163 &-21:22:01.989 &1808 &3 &170 &9 &132 &6 &0.057 &0.005 &6.1 &0.52 &6.28 &0.54 &\\	   
GEMS\_N1332\_6 &3:29:57.124 &-22:19:45.366 &1767 &4 &101 &12 &60 &8 &0.074 &0.008 &3.9 &0.60 &4.01 &0.61 &\\	   
GEMS\_N1332\_7 &3:24:29.555 &-21:32:14.609 &1588 &1 &346 &3 &330 &2 &0.142 &0.009 &26.6 &1.31 &27.37 &1.35 &\\   
GEMS\_N1332\_8 &3:27:29.867 &-21:17:03.795 &1696 &10 &364 &30 &188 &20 &0.036 &0.005 &6.0 &0.69 &6.17 &0.71 &\\  
GEMS\_N1332\_9 &3:19:40.437 &-19:24:43.686 &1577 &1 &285 &3 &268 &2 &0.225 &0.013 &36.9 &1.65 &37.97 &1.70 &\\   
GEMS\_N1332\_10 &3:18:43.358 &-23:48:08.206 &1538 &1 &98 &3 &82 &2 &0.140 &0.009 &9.8 &0.74 &10.09 &0.76 &\\	   
GEMS\_N1332\_11 &3:24:00.101 &-18:34:48.091 &1453 &9 &293 &27 &181 &18 &0.044 &0.007 &4.6 &0.73 &4.73 &0.75 &\\  
GEMS\_N1332\_12 &3:24:51.240 &-21:19:05.317 &1333 &2 &68 &6 &45 &4 &0.132 &0.010 &6.6 &0.70 &6.79 &0.72 &\\	   
GEMS\_N1332\_13 &3:37:49.595 &-24:25:43.446 &1512 &4 &209 &12 &157 &8 &0.163 &0.016 &20.9 &1.86 &21.51 &1.92 &\\ 
GEMS\_N1332\_14 &3:32:42.148 &-24:07:42.582 &1913 &3 &194 &9 &164 &6 &0.061 &0.007 &4.2 &0.57 &4.32 &0.58 &\\	   
GEMS\_N1332\_15 &3:35:28.893 &-21:08:27.717 &1514 &9 &138 &27 &80 &18 &0.028 &0.006 &1.6 &0.48 &1.65 &0.49 &\\   
GEMS\_N1332\_16 &3:23:27.444 &-19:17:31.585 &1553 &3 &52 &9 &33 &6 &0.092 &0.010 &3.1 &0.59 &3.19 &0.60 &\\      
\hline
\end{tabular}
\end{table*}

\newpage
\begin{table*}
\centering
\caption{\HI\ detections of the galaxies in groups. The columns are
the same as Table~\ref{tab:app1}1.}
\begin{tabular}{lccrrrrrrrrrrrrr}
(1)&(2)&(3)&(4)&(5)&(6)&(7)&(8)&(9)&(10)&(11)&(12)&(13)&(14)&(15)\\
\hline
GEMS\_N1407\_1$^1$ &3:33:45.965 &-19:29:22.958 &1969 &2 &227 &6 &199 &4 &0.224 &0.013 &37.4 &1.69 &38.49 &1.74 &\\
GEMS\_N1407\_2 &3:32:11.129 &-17:41:02.037 &1943 &3 &200 &9 &152 &6 &0.084 &0.006 &11.5 &0.70 &11.83 &0.72 &\\
GEMS\_N1407\_3$^1$ &3:33:39.753 &-21:30:30.801 &1869 &2 &229 &6 &212 &4 &0.089 &0.008 &15.4 &1.06 &15.85 &1.10 &\\
GEMS\_N1407\_4$^1$ &3:35:22.788 &-21:18:38.901 &1816 &3 &182 &9 &148 &6 &0.071 &0.007 &7.9 &0.74 &8.13 &0.76 &\\  
GEMS\_N1407\_5 &3:42:36.925 &-17:26:04.314 &1712 &5 &223 &15 &152 &10 &0.058 &0.007 &6.9 &0.78 &7.10 &0.80 &\\
GEMS\_N1407\_6 &3:34:42.734 &-19:04:25.312 &1604 &2 &370 &6 &363 &4 &0.034 &0.005 &4.9 &0.57 &5.04 &0.59 &\\  
GEMS\_N1407\_7 &3:29:32.760 &-17:47:04.322 &1530 &1 &146 &3 &130 &2 &0.134 &0.008 &14.6 &0.85 &15.03 &0.87 &\\
GEMS\_N1407\_8 &3:38:11.118 &-18:54:57.297 &1192 &7 &245 &21 &161 &14 &0.043 &0.006 &5.1 &0.68 &5.25 &0.70 &\\

\hline
GEMS\_N1566\_1  &4:12:10.61 &-58:34:17.40 &1466 &3 &102 &9 &55 &6 &0.106 &0.009 &6.1 &0.73 &6.34 &0.76 &\\	
GEMS\_N1566\_2 	&4:10:51.16 &-56:30:31.51 &1310 &6 &129 &18 &104 &12 &0.028 &0.006 &1.9 &0.53 &1.97 &0.55 &\\	
GEMS\_N1566\_3 	&4:12:38.84 &-57:45:49.62 &1176 &2 &214 &6 &189 &4 &0.076 &0.006 &8.7 &0.63 &9.04 &0.66 &\\	
GEMS\_N1566\_4 	&4:27:25.22 &-57:27:21.05 &1215 &5 &137 &15 &75 &10 &0.048 &0.006 &3.5 &0.49 &3.64 &0.51 &\\	
GEMS\_N1566\_5 	&4:09:41.61 &-56:06:55.80 &796 &3 &324 &9 &248 &6 &0.382 &0.022 &73.5 &3.18 &76.37 &3.31 &\\	
GEMS\_N1566\_6 	&4:22:41.62 &-56:16:58.66 &1350 &3 &76 &9 &52 &6 &0.049 &0.005 &2.5 &0.40 &2.60 &0.42 &\\		
GEMS\_N1566\_7 	&4:14:40.37 &-56:04:25.17 &1298 &6 &344 &18 &189 &12 &0.140 &0.013 &23.2 &1.69 &24.10 &1.76 &\\	
GEMS\_N1566\_8 	&4:17:56.28 &-55:57:06.31 &1370 &2 &218 &6 &193 &4 &0.124 &0.008 &17.0 &0.91 &17.66 &0.95 &\\	
GEMS\_N1566\_9 	&4:07:10.06 &-55:18:01.11 &1068 &2 &119 &6 &99 &4 &0.073 &0.006 &5.9 &0.54 &6.13 &0.56 &\\	
GEMS\_N1566\_10	 &4:19:53.33 &-54:56:45.66 &1502 &1 &223 &3 &200 &2 &1.040 &0.053 &149.4 &6.50 &155.23 &6.75 &\\	
GEMS\_N1566\_11	 &4:27:45.38 &-55:01:07.73 &1572 &3 &174 &9 &94 &6 &0.283 &0.017 &32.0 &1.81 &33.25 &1.88 &\\	
GEMS\_N1566\_12	 &4:03:56.23 &-54:05:04.20 &1180 &2 &381 &6 &357 &4 &0.101 &0.007 &15.9 &0.86 &16.52 &0.90 &\\	
GEMS\_N1566\_13	 &4:05:41.94 &-52:41:00.59 &902 &2 &104 &6 &88 &4 &0.079 &0.008 &4.9 &0.66 &5.09 &0.69 &\\       

\hline
GEMS\_N1808\_1 &5:16:36.579 &-37:05:56.048 &1344 &1 &293 &3 &275 &2 &0.297 &0.016 &55.6 &2.22 &37.86 &1.51 &\\
GEMS\_N1808\_2 &5:19:18.508 &-37:06:35.334 &1340 &2 &91 &6 &76 &4 &0.090 &0.007 &5.0 &0.51 &3.40 &0.35 &\\	
GEMS\_N1808\_3 &5:21:05.551 &-36:56:41.001 &1292 &2 &160 &6 &133 &4 &0.118 &0.008 &13.3 &0.87 &9.06 &0.59 &\\	
GEMS\_N1808\_4 &5:05:13.040 &-37:58:49.413 &1206 &2 &324 &6 &298 &4 &0.152 &0.009 &37.8 &1.46 &25.74 &0.99 &\\
GEMS\_N1808\_5 &5:07:42.077 &-37:30:23.888 &1002 &2 &311 &6 &254 &4 &0.307 &0.016 &70.4 &2.48 &47.93 &1.69 &\\
GEMS\_N1808\_6 &5:10:01.901 &-36:57:41.750 &1041 &2 &189 &6 &164 &4 &0.156 &0.009 &21.4 &1.06 &14.57 &0.72 &\\
GEMS\_N1808\_7 &5:08:05.366 &-38:18:19.704 &1020 &1 &132 &3 &115 &2 &0.620 &0.031 &64.6 &3.28 &43.99 &2.23 &\\

\hline
GEMS\_N3557\_1  &11:10:52.547 &-37:26:30.563 &2450 &2 &321 &6 &280 &4 &0.119 &0.007 &27.6 &1.09 &117.45 &4.63 &\\	 
GEMS\_N3557\_2 	&11:11:14.459 &-36:57:19.026 &2470 &6 &286 &18 &130 &12 &0.068 &0.006 &9.9 &0.70 &42.13 &2.97 &\\	 
GEMS\_N3557\_3 	&11:17:01.328 &-34:56:42.410 &2607 &1 &247 &3 &226 &2 &0.136 &0.008 &23.7 &1.08 &100.86 &4.61 &\\	 
GEMS\_N3557\_4 	&11:10:49.023 &-35:59:25.033 &2768 &3 &296 &9 &264 &6 &0.071 &0.006 &11.8 &0.81 &50.22 &3.46 &\\	 
GEMS\_N3557\_5 	&11:03:47.797 &-38:45:41.909 &2690 &2 &146 &6 &126 &4 &0.048 &0.005 &3.8 &0.43 &16.17 &1.82 &\\	 
GEMS\_N3557\_6 	&11:21:56.051 &-37:52:27.196 &2754 &2 &142 &6 &120 &4 &0.062 &0.005 &6.6 &0.56 &28.09 &2.39 &\\	 
GEMS\_N3557\_7 	&11:12:14.333 &-38:04:24.742 &2730 &3 &149 &9 &128 &6 &0.033 &0.004 &2.7 &0.40 &11.49 &1.71 &\\	 
GEMS\_N3557\_8 	&11:12:23.782 &-36:28:02.396 &2933 &3 &229 &9 &199 &6 &0.049 &0.005 &7.8 &0.71 &33.19 &3.02 &\\	 
GEMS\_N3557\_9 	&10:59:40.443 &-37:23:15.073 &2922 &6 &144 &18 &112 &12 &0.025 &0.004 &2.3 &0.44 &9.79 &1.88 &\\	 
GEMS\_N3557\_10	 &11:07:05.243 &-37:10:29.839 &3119 &4 &472 &12 &395 &8 &0.059 &0.005 &11.9 &0.76 &50.64 &3.24 &\\ 
GEMS\_N3557\_11	 &11:18:07.861 &-40:35:34.768 &3106 &3 &307 &9 &253 &6 &0.115 &0.009 &20.3 &1.19 &86.39 &5.09 &\\	 
GEMS\_N3557\_12	 &11:01:02.095 &-38:00:12.367 &3591 &2 &159 &6 &129 &4 &0.110 &0.007 &11.1 &0.77 &47.24 &3.26 &\\  
GEMS\_N3557\_13 &11:11:11.135 &-36:45:56.109 &2443 &3 &293 &9 &273 &6 &0.055 &0.008 &7.360 &0.89 &31.320 &3.800 &\\

\hline
GEMS\_N3783\_1   &11:34:18.389 &-37:09:39.085 &3202 &7 &340 &21 &250 &14 &0.058 &0.008 &11.1 &1.18 &33.89 &3.61 &\\  
GEMS\_N3783\_2 	 &11:31:32.407 &-36:18:53.054 &2733 &11 &142 &33 &36 &22 &0.041 &0.008 &1.9 &0.57 &5.80 &1.74 &\\    
GEMS\_N3783\_3 	 &11:28:05.772 &-36:32:44.950 &3018 &2 &284 &6 &253 &4 &0.229 &0.013 &38.0 &1.77 &116.03 &5.40 &\\   
GEMS\_N3783\_4 	 &11:49:39.599 &-37:30:49.655 &3055 &2 &53 &6 &36 &4 &0.201 &0.014 &7.4 &0.89 &22.59 &2.71 &\\	     
GEMS\_N3783\_5 	 &11:49:00.993 &-37:29:47.332 &2933 &3 &157 &9 &119 &6 &0.150 &0.013 &16.6 &1.37 &50.69 &4.18 &\\    
GEMS\_N3783\_6 	 &11:49:33.388 &-38:50:39.854 &2989 &3 &178 &9 &155 &6 &0.095 &0.009 &12.5 &1.08 &38.17 &3.29 &\\    
GEMS\_N3783\_7 	 &11:38:50.962 &-37:47:37.741 &2923 &2 &171 &6 &141 &4 &0.121 &0.010 &12.4 &1.01 &37.86 &3.09 &\\    
GEMS\_N3783\_8 	 &11:37:54.347 &-37:56:04.265 &2947 &6 &215 &18 &158 &12 &0.064 &0.009 &6.9 &0.95 &21.07 &2.90 &\\   
GEMS\_N3783\_9 	 &11:35:42.955 &-38:02:07.807 &2705 &6 &521 &18 &464 &12 &0.069 &0.010 &19.5 &1.80 &59.54 &5.48 &\\  
GEMS\_N3783\_10	  &11:26:06.212 &-37:51:26.206 &2810 &5 &99 &15 &53 &10 &0.056 &0.008 &2.5 &0.56 &7.63 &1.70 &\\     
GEMS\_N3783\_11	  &11:21:57.181 &-37:46:44.550 &2741 &10 &141 &30 &82 &20 &0.062 &0.015 &4.0 &1.26 &12.21 &3.86 &\\  
GEMS\_N3783\_12	  &11:29:42.717 &-37:16:59.336 &3049 &7 &235 &21 &212 &14 &0.026 &0.007 &3.7 &0.80 &11.30 &2.45 &\\  
\hline
\end{tabular}
$^1$ Note three sources from the NGC 1407 group were also detected in
the NGC 1332 group: GEMS\_N1407\_1 = GEMS\_N1332\_2, GEMS\_N1407\_3 =
GEMS\_N1332\_3, GEMS\_N1407\_4 = GEMS\_N1332\_5. These sources are
listed in both groups so the reader can compare and contrast the derived values.

\end{table*}

\begin{table*}
\centering
\caption{\HI\ detections of the galaxies in groups. The columns are
the same as Table~\ref{tab:app1}1.}
\begin{tabular}{lccrrrrrrrrrrrrr}
(1)&(2)&(3)&(4)&(5)&(6)&(7)&(8)&(9)&(10)&(11)&(12)&(13)&(14)&(15)\\
\hline

GEMS\_N3923\_1   &11:53:21.485 &-28:33:23.599 &1706 &1 &287 &3 &272 &2 &0.313 &0.017 &63.7 &2.45 &68.09 &2.62 &\\	 
GEMS\_N3923\_2 	 &11:53:59.368 &-27:21:12.797 &1634 &2 &132 &6 &113 &4 &0.081 &0.007 &6.7 &0.62 &7.16 &0.66 &\\	 
GEMS\_N3923\_3 	 &11:45:40.039 &-28:21:51.750 &1842 &1 &196 &3 &175 &2 &0.160 &0.009 &23.1 &1.16 &24.69 &1.25 &\\	 
GEMS\_N3923\_4 	 &11:48:35.358 &-27:22:15.794 &1875 &3 &192 &9 &110 &6 &0.109 &0.007 &12.6 &0.77 &13.47 &0.83 &\\	 
GEMS\_N3923\_5 	 &11:53:41.920 &-26:59:10.496 &1908 &5 &148 &15 &105 &10 &0.050 &0.007 &5.0 &0.70 &5.34 &0.75 &\\	 
GEMS\_N3923\_6 	 &11:52:35.295 &-26:55:26.144 &2027 &5 &373 &15 &303 &10 &0.143 &0.016 &32.4 &2.40 &34.63 &2.57 &\\ 
GEMS\_N3923\_7 	 &11:59:16.997 &-28:55:30.793 &2022 &1 &192 &3 &176 &2 &0.186 &0.011 &20.6 &1.21 &22.02 &1.30 &\\	 
GEMS\_N3923\_8 	 &11:43:44.599 &-30:38:36.454 &1987 &2 &149 &6 &135 &4 &0.069 &0.006 &6.7 &0.60 &7.16 &0.64 &\\	 
GEMS\_N3923\_9 	 &12:01:59.521 &-30:14:36.611 &2059 &4 &98 &12 &75 &8 &0.042 &0.007 &2.1 &0.52 &2.24 &0.55 &\\	 
GEMS\_N3923\_10	  &12:03:08.055 &-29:06:10.678 &2202 &3 &121 &9 &89 &6 &0.139 &0.012 &10.7 &1.05 &11.44 &1.13 &\\	 
GEMS\_N3923\_11	  &11:50:17.353 &-30:04:33.781 &1600 &3 &70 &9 &61 &6 &0.026 &0.005 &1.3 &0.36 &1.39 &0.39 &\\	 
GEMS\_N3923\_12	  &11:48:44.460 &-28:17:24.017 &1938 &1 &118 &3 &104 &2 &0.252 &0.013 &25.5 &1.37 &27.26 &1.47 &\\	 
GEMS\_N3923\_13	  &11:57:22.427 &-28:06:15.341 &2131 &1 &37 &3 &28 &2 &0.067 &0.006 &1.7 &0.29 &1.82 &0.31 &\\

\hline
GEMS\_N4636\_1    &12:47:48.791 &4:20:46.733 &986 &1 &65 &3 &44 &2 &0.637 &0.032 &29.0 &2.22 &12.64 &0.97 &\\	 
GEMS\_N4636\_2 	  &12:32:46.834 &0:06:04.089 &1127 &1 &317 &3 &301 &2 &0.402 &0.021 &87.5 &3.11 &38.13 &1.36 &\\	 
GEMS\_N4636\_3 	  &12:49:46.875 &2:51:58.984 &1158 &1 &223 &3 &214 &2 &0.053 &0.004 &7.5 &0.55 &3.27 &0.24 &\\	 
GEMS\_N4636\_4 	  &12:44:29.212 &0:29:24.114 &1180 &1 &120 &3 &107 &2 &0.139 &0.008 &12.7 &0.79 &5.53 &0.34 &\\	 
GEMS\_N4636\_5 	  &12:33:51.088 &3:36:41.541 &1134 &2 &41 &6 &29 &4 &0.052 &0.005 &1.5 &0.26 &0.65 &0.11 &\\	 
GEMS\_N4636\_6 	  &12:39:49.367 &1:39:51.479 &1222 &4 &169 &12 &101 &8 &0.049 &0.005 &4.8 &0.47 &2.09 &0.20 &\\	 
GEMS\_N4636\_7 	  &12:53:31.313 &2:04:43.774 &1217 &4 &110 &12 &38 &8 &0.064 &0.006 &3.4 &0.43 &1.48 &0.19 &\\	 
GEMS\_N4636\_8 	  &12:33:02.136 &4:35:35.262 &1223 &4 &101 &12 &62 &8 &0.038 &0.004 &2.1 &0.34 &0.92 &0.15 &\\	 
GEMS\_N4636\_9 	  &12:36:40.018 &3:07:30.296 &1441 &1 &128 &3 &117 &2 &0.067 &0.005 &6.6 &0.48 &2.88 &0.21 &\\	 
GEMS\_N4636\_10	   &12:32:28.618 &0:24:20.325 &1529 &1 &167 &3 &154 &2 &0.297 &0.015 &34.2 &1.70 &14.90 &0.74 &\\	 
GEMS\_N4636\_11	   &12:34:08.975 &2:38:59.839 &1737 &1 &376 &3 &356 &2 &0.413 &0.021 &96.2 &3.29 &41.92 &1.43 &\\	 
GEMS\_N4636\_12	   &12:42:34.107 &-0:04:51.599 &1722 &1 &246 &3 &223 &2 &0.239 &0.013 &46.3 &1.84 &20.18 &0.80 &\\	 
GEMS\_N4636\_13	   &12:41:13.093 &1:25:08.767 &1701 &3 &168 &9 &104 &6 &0.079 &0.006 &8.6 &0.63 &3.75 &0.27 &\\	 
GEMS\_N4636\_14	   &12:34:27.211 &2:13:13.437 &1806 &1 &342 &3 &324 &2 &0.329 &0.017 &73.3 &2.61 &31.94 &1.14 &\\	 
GEMS\_N4636\_15	   &12:31:38.089 &3:56:40.982 &1733 &1 &175 &3 &154 &2 &0.334 &0.017 &43.5 &2.03 &18.96 &0.89 &\\	 
GEMS\_N4636\_16	   &12:32:43.880 &2:41:14.371 &1731 &2 &164 &6 &147 &4 &0.099 &0.007 &12.1 &0.75 &5.27 &0.33 &\\	 
GEMS\_N4636\_17	   &12:31:17.108 &0:59:13.442 &2296 &4 &67 &12 &30 &8 &0.052 &0.007 &1.7 &0.38 &0.74 &0.17 &\\	 
GEMS\_N4636\_18	   &12:34:13.033 &1:45:40.391 &2472 &4 &105 &12 &94 &8 &0.024 &0.005 &1.4 &0.37 &0.61 &0.16 &\\	 
GEMS\_N4636\_19	   &12:37:32.697 &4:44:20.324 &1632 &7 &60 &21 &27 &14 &0.022 &0.005 &0.7 &0.27 &0.31 &0.12 &\\	 
GEMS\_N4636\_20	   &12:39:22.679 &-0:36:01.896 &1069 &1 &214 &3 &198 &2 &1.131 &0.060 &172.0 &7.62 &74.95 &3.32 &\\	 
GEMS\_N4636\_21	   &12:45:18.952 &-0:29:00.355 &1518 &2 &397 &6 &324 &4 &0.380 &0.020 &84.7 &3.09 &36.91 &1.35 &\\	 
GEMS\_N4636\_22    &12:55:05.564 &0:07:09.054 &1320 &1 &186 &3 &175 &2 &0.182 &0.014 &21.2 &1.51 &9.24 &0.66 &\\

\hline
GEMS\_N5044\_1    &13:14:09.927 &-16:41:41.635 &3075 &5 &204 &15 &148 &10 &0.039 &0.005 &4.4 &0.53 &8.72 &1.06 &\\    
GEMS\_N5044\_2 	  &13:12:36.812 &-15:51:07.902 &2928 &6 &183 &18 &32 &12 &0.087 &0.007 &6.6 &0.66 &13.08 &1.30 &\\    
GEMS\_N5044\_3 	  &13:18:29.023 &-14:36:24.290 &2897 &3 &236 &9 &196 &6 &0.086 &0.007 &11.2 &0.87 &22.19 &1.72 &\\    
GEMS\_N5044\_4 	  &13:03:09.633 &-17:23:16.021 &2967 &1 &56 &3 &33 &2 &0.208 &0.012 &8.1 &0.76 &16.05 &1.51 &\\	     
GEMS\_N5044\_5 	  &13:05:37.270 &-15:46:58.451 &2911 &3 &118 &9 &93 &6 &0.049 &0.005 &3.9 &0.49 &7.73 &0.98 &\\	     
GEMS\_N5044\_6 	  &13:19:19.577 &-14:49:02.816 &2749 &2 &401 &6 &380 &4 &0.069 &0.006 &16.5 &0.88 &32.69 &1.74 &\\    
GEMS\_N5044\_7 	  &13:10:59.343 &-15:20:44.288 &2828 &8 &386 &24 &315 &16 &0.026 &0.005 &4.3 &0.63 &8.52 &1.25 &\\    
GEMS\_N5044\_8 	  &13:12:38.338 &-17:32:18.269 &2753 &4 &396 &12 &346 &8 &0.051 &0.005 &11.2 &0.75 &22.19 &1.48 &\\   
GEMS\_N5044\_9 	  &13:15:08.786 &-17:59:08.810 &2773 &2 &63 &6 &40 &4 &0.090 &0.007 &3.9 &0.49 &7.73 &0.96 &\\	     
GEMS\_N5044\_10	   &13:20:14.216 &-14:25:41.013 &2748 &2 &64 &6 &42 &4 &0.108 &0.008 &4.9 &0.56 &9.71 &1.11 &\\	     
GEMS\_N5044\_11	   &13:19:58.070 &-17:17:43.872 &2685 &3 &122 &9 &97 &6 &0.063 &0.006 &4.4 &0.51 &8.72 &1.01 &\\	     
GEMS\_N5044\_12	   &13:16:58.645 &-16:17:44.175 &2624 &3 &310 &9 &277 &6 &0.047 &0.005 &9.2 &0.70 &18.23 &1.40 &\\    
GEMS\_N5044\_13	   &13:13:21.838 &-16:05:40.043 &2694 &4 &152 &12 &26 &8 &0.133 &0.009 &10.3 &0.79 &20.41 &1.57 &\\   
GEMS\_N5044\_14	   &13:08:05.016 &-14:44:53.255 &2599 &1 &112 &3 &101 &2 &0.072 &0.006 &5.6 &0.54 &11.10 &1.06 &\\    
GEMS\_N5044\_15	   &13:03:18.888 &-14:45:52.821 &2572 &4 &137 &12 &73 &8 &0.075 &0.006 &5.2 &0.56 &10.30 &1.10 &\\    
GEMS\_N5044\_16	   &13:13:36.158 &-15:25:54.235 &2502 &1 &91 &3 &73 &2 &0.207 &0.012 &15.4 &1.05 &30.51 &2.07 &\\     
GEMS\_N5044\_17	   &13:18:58.428 &-17:37:14.239 &2494 &3 &108 &9 &68 &6 &0.067 &0.006 &4.7 &0.52 &9.31 &1.03 &\\	     
GEMS\_N5044\_18	   &13:11:36.133 &-14:40:40.125 &2488 &4 &141 &12 &100 &8 &0.040 &0.005 &3.2 &0.46 &6.34 &0.91 &\\    
GEMS\_N5044\_19	   &13:17:18.638 &-17:17:59.994 &2520 &2 &230 &6 &219 &4 &0.053 &0.006 &7.3 &0.67 &14.46 &1.33 &\\    
GEMS\_N5044\_20	   &13:16:57.879 &-16:36:21.407 &1836 &2 &141 &6 &107 &4 &0.085 &0.006 &6.8 &0.58 &13.47 &1.15 &\\    
GEMS\_N5044\_21	   &13:24:52.255 &-19:41:24.120 &1947 &5 &243 &15 &172 &10 &0.244 &0.025 &29.3 &2.85 &58.05 &5.65 &\\ 
GEMS\_N5044\_22	   &13:05:20.342 &-16:55:08.371 &2834 &3 &182 &9 &156 &6 &0.056 &0.006 &4.1 &0.50 &8.12 &1.00 &\\     
GEMS\_N5044\_23    &13:09:33.913 &-16:36:24.423 &2576 &2 &354 &6 &333 &4 &0.114 &0.007 &28.8 &1.12 &57.06 &2.22 &\\   
\hline\\

\end{tabular}
\end{table*}

\begin{table*}
\centering
\caption{\HI\ detections of the galaxies in groups. The columns are
the same as Table~\ref{tab:app1}1.}
\begin{tabular}{lccrrrrrrrrrrrrr}
(1)&(2)&(3)&(4)&(5)&(6)&(7)&(8)&(9)&(10)&(11)&(12)&(13)&(14)&(15)\\
\hline
GEMS\_N7144\_1 &21:45:08.395 &-49:00:43.150 &1598 &2 &114 &6 &71 &4 &0.315 &0.017 &21.40 &1.46 &26.21 &1.79 &\\
GEMS\_N7144\_2 &22:09:20.739 &-47:11:18.717 &1722 &7 &460 &21 &210 &14 &0.109 &0.009 &21.89 &1.29 &26.81 &1.58 &\\
GEMS\_N7144\_3 &21:45:49.101 &-48:49:13.536 &1617 &3 &150 &9 &116 &6 &0.096 &0.008 &8.73 &0.83 &10.69 &1.02 &\\
GEMS\_N7144\_4 &21:55:07.416 &-50:38:42.157 &1871 &3 &218 &9 &196 &6 &0.077 &0.008 &10.25 &0.95 &12.55 &1.16 &\\
GEMS\_N7144\_5 &21:55:38.108 &-49:15:06.927 &1860 &2 &97 &6 &83 &4 &0.066 &0.008 &4.83 &0.68 &5.92 &0.83 &\\
GEMS\_N7144\_6 &21:42:42.971 &-47:55:24.345 &1956 &3 &248 &9 &196 &6 &0.141 &0.010 &20.26 &1.22 &24.81 &1.50 &\\
GEMS\_N7144\_7 &21:41:02.448 &-46:40:18.955 &1958 &4 &78 &12 &27 &8 &0.073 &0.008 &2.44 &0.47 &2.99 &0.57 &\\
GEMS\_N7144\_8 &21:42:44.625 &-51:15:25.309 &2075 &3 &152 &9 &129 &6 &0.076 &0.008 &7.00 &0.79 &8.57 &0.96 &\\

\hline
GEMS\_N7714\_1 &23:41:33.273 &3:45:17.985 &2891 &4 &261 &12 &152 &8 &0.169 &0.011 &25.75 &1.39 &92.27 &4.99 &\\
GEMS\_N7714\_2 &23:36:25.686 &2:08:44.429 &2721 &5 &325 &15 &205 &10 &0.112 &0.009 &19.53 &1.22 &69.99 &4.36 &\\
GEMS\_N7714\_3 &23:37:31.755 &0:23:59.861 &2674 &3 &164 &9 &141 &6 &0.084 &0.008 &9.32 &0.88 &33.40 &3.17 &\\
GEMS\_N7714\_4 &23:46:44.349 &3:42:03.825 &2901 &3 &300 &9 &235 &6 &0.162 &0.011 &25.56 &1.38 &91.59 &4.95 &\\
GEMS\_N7714\_5 &23:36:33.185 &0:20:41.739 &2588 &3 &286 &9 &238 &6 &0.123 &0.009 &23.57 &1.32 &84.46 &4.75 &\\
GEMS\_N7714\_6 &23:35:22.987 &1:10:59.750 &2599 &2 &103 &6 &84 &4 &0.082 &0.008 &5.63 &0.69 &20.17 &2.48 &\\
GEMS\_N7714\_7 &23:45:18.649 &3:41:28.458 &2755 &7 &489 &21 &212 &14 &0.131 &0.010 &24.01 &1.33 &86.04 &4.78 &\\

\hline
GEMS\_HCG90\_1 &22:01:19.670 &-32:36:12.937 &2271 &2 &181 &6 &165 &4 &0.081 &0.008 &8.10 &0.83 &24.73 &2.54 &\\
GEMS\_HCG90\_2 &22:01:28.086 &-31:33:10.938 &2353 &6 &370 &18 &282 &12 &0.060 &0.008 &12.05 &1.11 &36.79 &3.39 &\\
GEMS\_HCG90\_3 &22:03:49.062 &-32:18:59.783 &2550 &3 &288 &9 &266 &6 &0.070 &0.008 &13.34 &1.11 &40.73 &3.39 &\\
GEMS\_HCG90\_4 &22:07:02.414 &-31:03:03.066 &2583 &5 &294 &15 &200 &10 &0.084 &0.008 &11.82 &0.99 &36.09 &3.04 &\\
GEMS\_HCG90\_5 &21:55:35.591 &-34:44:08.965 &2614 &2 &169 &6 &133 &4 &0.156 &0.010 &17.96 &1.16 &54.84 &3.53 &\\
GEMS\_HCG90\_6 &21:56:48.419 &-34:34:30.401 &2656 &6 &374 &18 &230 &12 &0.087 &0.008 &15.52 &1.13 &47.39 &3.45 &\\
GEMS\_HCG90\_7 &22:01:27.658 &-35:14:35.759 &2515 &2 &200 &6 &185 &4 &0.088 &0.008 &6.97 &0.76 &21.28 &2.31 &\\
GEMS\_HCG90\_8 &22:08:37.651 &-34:17:58.435 &2670 &5 &207 &15 &80 &10 &0.115 &0.009 &11.12 &0.91 &33.95 &2.79 &\\
GEMS\_HCG90\_9 &22:03:08.521 &-33:53:03.327 &2720 &6 &453 &18 &367 &12 &0.060 &0.008 &11.94 &1.10 &36.46 &3.37 &\\
GEMS\_HCG90\_10 &22:08:43.330 &-30:56:29.547 &2560 &3 &58 &9 &46 &6 &0.044 &0.007 &1.57 &0.45 &4.79 &1.37 &\\
GEMS\_HCG90\_11 &22:06:19.336 &-31:04:41.189 &2596 &4 &183 &12 &44 &8 &0.144 &0.010 &7.54 &0.75 &23.02 &2.28 &\\
GEMS\_HCG90\_12 &22:01:48.903 &-31:54:43.334 &2316 &12 &104 &36 &34 &24 &0.028 &0.007 &2.97 &0.76 &9.07 &2.31 &\\

\hline
GEMS\_IC1459\_1 &23:02:12.289 &-39:34:04.659 &1198 &1 &249 &3 &234 &2 &0.31 &0.02 &52.06 &2.26 &90.74 &3.94 &\\
GEMS\_IC1459\_2 &22:59:30.939 &-37:41:37.029 &1299 &2 &226 &6 &196 &4 &0.14 &0.01 &21.83 &1.27 &38.05 &2.22 &\\
GEMS\_IC1459\_3 &23:00:18.564 &-37:11:21.982 &1239 &3 &74 &9 &42 &6 &0.087 &0.008 &3.62 &0.55 &6.31 &0.95 &\\
GEMS\_IC1459\_4 &22:55:43.115 &-34:34:17.800 &1282 &8 &195 &24 &83 &16 &0.052 &0.007 &4.04 &0.68 &7.04 &1.18 &\\
GEMS\_IC1459\_5 &23:02:15.438 &-37:04:55.632 &1372 &2 &137 &6 &121 &4 &0.104 &0.009 &11.04 &0.92 &19.24 &1.61 &\\
GEMS\_IC1459\_6 &22:56:28.173 &-37:01:55.086 &1446 &2 &233 &6 &207 &4 &0.180 &0.011 &28.00 &1.46 &48.81 &2.55 &\\
GEMS\_IC1459\_7 &22:56:42.596 &-36:14:16.044 &1667 &2 &242 &6 &228 &4 &0.123 &0.009 &17.25 &1.13 &30.07 &1.98 &\\
GEMS\_IC1459\_8 &22:56:47.404 &-37:19:42.368 &1789 &3 &141 &9 &129 &6 &0.052 &0.007 &4.30 &0.70 &7.50 &1.22 &\\
GEMS\_IC1459\_9 &22:57:21.286 &-36:35:09.619 &1790 &6 &192 &18 &96 &12 &0.072 &0.008 &5.59 &0.71 &9.74 &1.24 &\\
GEMS\_IC1459\_10 &23:00:49.061 &-35:22:07.512 &1773 &4 &214 &12 &172 &8 &0.062 &0.008 &8.41 &0.92 &14.66 &1.60 &\\
GEMS\_IC1459\_11 &22:58:05.708 &-35:48:43.725 &1950 &3 &199 &9 &101 &6 &0.390 &0.021 &38.19 &2.11 &66.57 &3.67 &\\
GEMS\_IC1459\_12 &22:56:41.430 &-36:44:14.421 &2101 &2 &216 &6 &180 &4 &0.164 &0.011 &23.49 &1.33 &40.94 &2.31 &\\
GEMS\_IC1459\_13 &22:45:42.249 &-39:21:09.628 &2279 &5 &233 &15 &137 &10 &0.087 &0.008 &8.87 &0.85 &15.46 &1.49 &\\
GEMS\_IC1459\_14 &22:43:37.768 &-39:54:04.091 &2147 &2 &169 &6 &145 &4 &0.219 &0.013 &24.26 &1.41 &42.29 &2.45 &\\
GEMS\_IC1459\_15 &22:58:03.586 &-33:45:54.083 &1718 &6 &382 &18 &346 &12 &0.043 &0.007 &9.00 &1.09 &15.69 &1.90 &\\
GEMS\_IC1459\_16 &23:09:02.363 &-39:48:58.259 &1787 &4 &96 &12 &80 &8 &0.037 &0.007 &2.32 &0.59 &4.04 &1.03 &\\
GEMS\_IC1459\_17 &22:53:19.129 &-38:47:56.731 &879 &5 &72 &15 &40 &10 &0.041 &0.007 &1.52 &0.46 &2.65 &0.79 &\\
GEMS\_IC1459\_18 &23:01:38.433 &-36:30:33.653 &1668 &12 &149 &36 &83 &24 &0.027 &0.007 &1.63 &0.57 &2.84 &0.99 &\\
\hline
\end{tabular}
\end{table*}

\section{Optical counterparts to the \HI\ detections.  }

\begin{table*}
\label{tab:app2}
\centering
\caption{Optical counterparts to the \HI\ detections. The columns are
as follows: (1) GEMS \HI\ catalogue designation; (2) Systemic velocity,
$V_{sys}$, derived from the \HI\ spectrum, in \kms; (3) distance of
\HI\ source position from optically catalogued galaxy position, in
arcmin; (4) name of optically catalogued galaxy; (5) 6dFGS name; (6)
\& (7) Position of optically catalogued source, $\alpha$,$\delta$
(J2000) [h m s][$^{\rm o}$ \arcmin \arcsec]; (8) Systemic velocity of
optically catalogued source in \kms.}
\begin{tabular}{lcclrrrr}
(1)&(2)&(3)&(4)&(5)&(6)&(7)&(8)\\
\hline
GEMS\_N524\_1 & 2898 &   0.71 & UGC 00993 NED01                &  $\cdots$           & 01:25:34.20 & 07:59:24.0 &  2926 \\ 
GEMS\_N524\_1 & 2898 &   0.76 & UGC 00993 NED02                &  $\cdots$           & 01:25:35.12 & 07:59:27.0 &  2928 \\ 
GEMS\_N524\_2 & 2778 &   3.45 & UGC 00989                      &  $\cdots$           & 01:25:29.60 & 07:33:41.0 &  2788 \\ 
GEMS\_N524\_3 & 2740 &   7.95 & VV 730                         &  $\cdots$           & 01:23:19.50 & 07:47:42.0 &  2700 \\ 
GEMS\_N524\_3 & 2740 &   0.82 & UGC 00964                      &  $\cdots$           & 01:24:35.06 & 07:43:16.2 &  2737 \\ 
GEMS\_N524\_4 & 2728 &   1.91 & UGC 00941                      &  $\cdots$           & 01:23:32.42 & 06:57:38.0 &  2728 \\ 
GEMS\_N524\_5 & 2549 &   3.32 & UGC 00990                      &  $\cdots$           & 01:25:24.00 & 10:47:49.0 &  2536 \\ 
GEMS\_N524\_6 & 2348 &   0.56 & UGC 00803                      &  $\cdots$           & 01:15:22.91 & 08:05:57.5 &  2342 \\ 
GEMS\_N524\_7 & 2335 &   2.24 & UGC 00942                      &  $\cdots$           & 01:23:43.70 & 06:41:29.0 &  2335 \\ 
GEMS\_N524\_8 & 2687 &   4.87 & VV 730                         &  $\cdots$           & 01:23:19.50 & 07:47:42.0 &  2700 \\ 
GEMS\_N524\_9 & 2430 &   0.84 & LEDA 093841                    &  $\cdots$           & 01:27:37.30 & 08:50:24.0 &  2435 \\ 

\hline
GEMS\_N720\_1 & 1635 &   1.04 & MCG -03-05-014                 & $\cdots$            & 01:41:35.92 & -16:08:49.6 &  1637 \\ 
GEMS\_N720\_2 & 1154 &   0.00 & MCG -03-05-017                 & $\cdots$            & 01:42:58.50 & -16:07:45.0 &  1154 \\ 
GEMS\_N720\_3 & 1742 &   0.68 & KUG 0144-147                   & J0146553-142710     & 01:46:55.27 & -14:27:09.5 &  1711 \\ 
GEMS\_N720\_4 & 1613 &   0.79 & ARP 004                        & $\cdots$            & 01:48:25.67 & -12:22:55.3 &  1614 \\ 
GEMS\_N720\_5 & 1865 &   0.38 & UGCA 022                       & $\cdots$            & 01:56:59.14 & -11:46:47.8 &  1853 \\ 
GEMS\_N720\_6 & 1770 &   $\cdots$ & $\cdots$                   & $\cdots$            & $\cdots$ & $\cdots$ &  $\cdots$ \\ 
GEMS\_N720\_7 & 1450 &   4.66 & KUG 0147-138                   & $\cdots$            & 01:49:37.92 & -13:34:15.4 &  1449 \\ 
GEMS\_N720\_8 & 1585 &   1.08 & APMUKS(BJ) B014238.60-164056.8 & J0145037-162556     & 01:45:03.67 & -16:25:55.7 &  1588 \\ 

\hline
GEMS\_N1052\_1 & 1266 &   1.89 & NGC 1035                       & $\cdots$            & 02:39:29.09 & -08:07:58.6 &  1241 \\ 
GEMS\_N1052\_1 & 1266 &   1.68 & 6dF J0239299-080821            & J0239299-080821     & 02:39:29.92 & -08:08:21.1 &  1393 \\ 
GEMS\_N1052\_2 & 1297 &   0.88 & NGC 0961                       & $\cdots$            & 02:41:02.46 & -06:56:09.1 &  1295 \\ 
GEMS\_N1052\_3 & 1405 &   0.66 & NGC 1084                       & $\cdots$            & 02:45:59.93 & -07:34:43.1 &  1407 \\ 
GEMS\_N1052\_4$^1$ & 1329 &   0.81 & NGC 1110              & $\cdots$            & 02:49:09.57 & -07:50:15.2 &  1333 \\ 
GEMS\_N1052\_5 & 1325 &   0.40 & UGCA 038                       & $\cdots$            & 02:40:30.19 & -06:06:23.0 &  1327 \\ 
GEMS\_N1052\_6 & 1351 &   1.96 & SDSS J024149.95-075530.1       & $\cdots$            & 02:41:49.96 & -07:55:30.0 &  1372 \\ 
GEMS\_N1052\_7 & 1371 &   1.00 & NGC 1042                       & J0240240-082601 & 02:40:23.97 & -08:26:01.0 &  1411 \\ 
GEMS\_N1052\_8 & 1504 &   0.55 & NGC 0988                       & J0235277-092122 & 02:35:27.73 & -09:21:21.6 &  1551 \\ 
GEMS\_N1052\_9 & 1411 &   0.84 & [RC3] 0231.5-0635              & $\cdots$            & 02:33:57.04 & -06:21:36.2 &  1410 \\ 
GEMS\_N1052\_10$^1$ & 1418 &   0.42 & [MMB2004] J0249-0806  & $\cdots$            & 02:49:13.42 & -08:06:53.2 &  $\cdots$ \\ 
GEMS\_N1052\_11 & 1532 &   0.77 & USGC S092 NED09                & $\cdots$            & 02:35:28.79 & -07:08:59.0 &  1532 \\ 
GEMS\_N1052\_11 & 1532 &   0.63 & 6dF J0235320-070936            & J0235320-070936 & 02:35:32.00 & -07:09:36.3 &  1554 \\ 
GEMS\_N1052\_11 & 1532 &   0.35 & NGC 0991                       & $\cdots$            & 02:35:32.69 & -07:09:16.0 &  1532 \\ 
GEMS\_N1052\_12 & 2100 &   0.96 & DDO 023                        & $\cdots$            & 02:30:25.46 & -10:44:55.6 &  2110 \\ 
GEMS\_N1052\_13$^{**}$ & 1420 &   9.59 & SDSS J024246.84-073230.3& $\cdots$            & 02:42:46.84 & -07:32:30.4 &  1344 \\ 
GEMS\_N1052\_14 & 1408 &   1.91 & MCG -01-08-001                 & $\cdots$            & 02:43:42.80 & -06:39:05.0 &  1410 \\ 
GEMS\_N1052\_15 & 1466 &   1.73 & NGC 1022                       & $\cdots$            & 02:38:32.70 & -06:40:38.7 &  1453 \\ 
\hline

\end{tabular}
\end{table*}

\begin{table*}
\centering
\caption{Optical counterparts to the \HI\ detections. The columns are the same as in Table~\ref{tab:app2}1.}
\begin{tabular}{lcclrrrr}
(1)&(2)&(3)&(4)&(5)&(6)&(7)&(8)\\
\hline
GEMS\_N1332\_1 & 2087 &   5.01 & UGCA 063                       & $\cdots$            & 03:16:09.66 & -24:12:02.2 &  2087 \\ 
GEMS\_N1332\_2 & 1964 &   0.08 & NGC 1359                       & $\cdots$            & 03:33:47.71 & -19:29:31.4 &  1973 \\ 
GEMS\_N1332\_2 & 1964 &   6.61 & ESO 548- G 043                 & $\cdots$            & 03:34:10.53 & -19:33:30.1 &  1931 \\ 
GEMS\_N1332\_3$^4$ & 1860 &   0.63 & IC 1953               & $\cdots$            & 03:33:41.87 & -21:28:43.1 &  1867 \\ 
GEMS\_N1332\_4 & 1846 &   1.43 & UGCA 068                       & $\cdots$            & 03:23:47.25 & -19:45:10.2 &  1838 \\ 
GEMS\_N1332\_5 & 1808 &   4.40 & IC 1962                        & $\cdots$            & 03:35:37.49 & -21:17:38.6 &  1806 \\ 
GEMS\_N1332\_6 & 1767 &   4.48 & NGC 1347                       & $\cdots$            & 03:29:41.50 & -22:17:06.0 &  1759 \\ 
GEMS\_N1332\_7 & 1588 &   1.01 & NGC 1325                       & J0324256-213238 & 03:24:25.57 & -21:32:38.3 &  1588 \\ 
GEMS\_N1332\_8 & 1696 &   3.63 & 2MASX J03273556-2113417        & J0327356-211341 & 03:27:35.57 & -21:13:41.4 &  1743 \\ 
GEMS\_N1332\_9 & 1577 &   1.44 & 2MASXi J0319359-192344         & J0319360-192345 & 03:19:35.99 & -19:23:44.8 &  1696 \\ 
GEMS\_N1332\_9 & 1577 &   0.16 & NGC 1300                       & $\cdots$            & 03:19:41.08 & -19:24:40.9 &  1577 \\ 
GEMS\_N1332\_10 & 1538 &   1.19 & UGCA 065                       & $\cdots$            & 03:18:43.14 & -23:46:57.1 &  1538 \\ 
GEMS\_N1332\_11$^{**}$ & 1453 &   8.40 & ESO 548- G 004          & $\cdots$            & 03:23:32.22 & -18:39:59.7 &  1528 \\ 
GEMS\_N1332\_12 & 1333 &   1.25 & NGC 1325A                      & $\cdots$            & 03:24:48.50 & -21:20:10.0 &  1333 \\ 
GEMS\_N1332\_13$^4$ & 1513 &   6.51 & NGC 1385              & J0337283-243005 & 03:37:28.32 & -24:30:04.7 &  1511 \\ 
GEMS\_N1332\_14 & 1913 &   4.59 & ESO 482- G 005                 & $\cdots$            & 03:33:02.25 & -24:07:58.3 &  1915 \\ 
GEMS\_N1332\_15 & 1514 &   4.56 & ESO 548- G 049                 & $\cdots$            & 03:35:28.07 & -21:13:01.3 &  1510 \\ 
GEMS\_N1332\_16 & 1553 &   0.76 & SGC 0321.2-1929                & $\cdots$            & 03:23:25.10 & -19:17:00.0 &  1545 \\ 

\hline
GEMS\_N1407\_1 & 1969 &   0.43 & NGC 1359                       & $\cdots$            & 03:33:47.71 & -19:29:31.4 &  1973 \\ 
GEMS\_N1407\_1 & 1969 &   7.11 & ESO 548- G 043                 & $\cdots$            & 03:34:10.53 & -19:33:30.1 &  1931 \\ 
GEMS\_N1407\_2 & 1943 &   2.80 & UGCA 077                       & $\cdots$            & 03:32:19.17 & -17:43:04.6 &  1961 \\ 
GEMS\_N1407\_3$^4$ & 1869 &   1.86 & IC 1953               & $\cdots$            & 03:33:41.87 & -21:28:43.1 &  1867 \\ 
GEMS\_N1407\_4 & 1816 &   3.57 & IC 1962                        & $\cdots$            & 03:35:37.49 & -21:17:38.6 &  1806 \\ 
GEMS\_N1407\_5 & 1712 &   4.62 & ESO 548- G 082                 & $\cdots$            & 03:42:43.27 & -17:30:26.0 &  1716 \\ 
GEMS\_N1407\_6 & 1551 &   2.69 & ESO 548- G 047                 & $\cdots$            & 03:34:43.48 & -19:01:44.1 &  1606 \\ 
GEMS\_N1407\_7 & 1530 &   0.45 & NGC 1345                       & J0329317-174642 & 03:29:31.69 & -17:46:42.2 &  1556 \\ 
GEMS\_N1407\_8 & 1192 &   7.13 & NGC 1390                       & $\cdots$            & 03:37:52.17 & -19:00:30.1 &  1207 \\ 

\hline
GEMS\_N1566\_1 & 1466 &   1.20 & IC2049                         &$\cdots$             & 04:12:04.30  & -58:33:25.0 & 1469\\ 
GEMS\_N1566\_2 & 1310 &   2.08 & NGC 1536                       & J0410599-562850 & 04:10:59.86 & -56:28:49.6 &  1274 \\ 
GEMS\_N1566\_3 & 1176 &   1.66 & NGC 1543                       & J0412432-574416 & 04:12:43.20 & -57:44:16.4 &  1149 \\ 
GEMS\_N1566\_4$^2$ & 1215 &   0.00 & LSBGF157-081          & $\cdots$            & 04:27:25.22 & -57:27:21.0 &  $\cdots$ \\ 
GEMS\_N1566\_5 & 796 &   1.44 & NGC 1533                        & J0409518-560706 & 04:09:51.84 & -56:07:06.4 &  764 \\ 
GEMS\_N1566\_6$^2$ & 1350 &   0.00 & APMBGC157+016+068     & $\cdots$            & 04:22:41.62 & -56:16:58.7 &  $\cdots$ \\ 
GEMS\_N1566\_7 & 1298 &   0.94 & NGC 1546                       & J0414364-560340 & 04:14:36.38 & -56:03:39.5 &  1238 \\ 
GEMS\_N1566\_8 & 1370 &   1.16 & IC 2058                        & $\cdots$            & 04:17:54.35 & -55:55:58.4 &  1379 \\ 
GEMS\_N1566\_9 & 1068 &   1.73 & IC 2032                        & $\cdots$            & 04:07:03.04 & -55:19:25.8 &  859 \\ 
GEMS\_N1566\_10 & 1502 &   1.13 & NGC 1566                       & $\cdots$            & 04:20:00.42 & -54:56:16.1 &  1504 \\ 
GEMS\_N1566\_11 & 1572 &   1.17 & NGC 1596                       & $\cdots$            & 04:27:38.11 & -55:01:40.1 &  1510 \\ 
GEMS\_N1566\_11 & 1572 &   0.84 & AM 0426-550                    & $\cdots$            & 04:27:44.00 & -55:01:57.0 &  1572 \\ 
GEMS\_N1566\_11 & 1572 &   2.71 & NGC 1602                       & $\cdots$            & 04:27:54.97 & -55:03:27.8 &  1568 \\ 
GEMS\_N1566\_12 & 1180 &   1.33 & NGC 1515                       & $\cdots$            & 04:04:02.72 & -54:06:00.2 &  1094 \\ 
GEMS\_N1566\_13 & 902 &   4.04 & NGC 1522                       & $\cdots$            & 04:06:07.92 & -52:40:06.3 &  905 \\ 

\hline
GEMS\_N1808\_1$^4$ & 1344 &   0.49 & ESO 362- G 011        & $\cdots$            & 05:16:38.80 & -37:06:09.1 &  1367 \\ 
GEMS\_N1808\_2 & 1339 &   0.30 & ESO 362- G 016                 & $\cdots$            & 05:19:18.48 & -37:06:17.4 &  1344 \\ 
GEMS\_N1808\_3 & 1292 &   0.78 & ESO 362- G 019                 & $\cdots$            & 05:21:04.18 & -36:57:24.7 &  1299 \\ 
GEMS\_N1808\_4 & 1206 &   0.27 & NGC 1792                       & J0505144-375851 & 05:05:14.41 & -37:58:50.5 &  1175 \\ 
GEMS\_N1808\_5 & 1002 &   0.37 & NGC 1808:[AB70] C              & J0507423-373046 & 05:07:42.34 & -37:30:46.1 &  969 \\ 
GEMS\_N1808\_6 & 1041 &   0.46 & NGC 1827                       & J0510046-365737 & 05:10:04.11 & -36:57:34.9 &  1037 \\ 
GEMS\_N1808\_7 & 1020 &   0.50 & ESO 305- G 009                 & $\cdots$            & 05:08:07.62 & -38:18:33.5 &  1021 \\ 
\hline\\

\end{tabular}
\end{table*}

\begin{table*}
\centering
\caption{Optical counterparts to the \HI\ detections. The columns are the same as in Table~\ref{tab:app2}1.}
\begin{tabular}{lcclrrrr}

(1)&(2)&(3)&(4)&(5)&(6)&(7)&(8)\\

\hline
GEMS\_N3557\_1 & 2453 &   7.89 & NGC 3557:[ZM2000] 0017         & $\cdots$            & 11:10:13.60 & -37:24:55.0 &  2447 \\ 
GEMS\_N3557\_1 & 2453 &   0.87 & NGC 3568                       & J1110486-372652 & 11:10:48.57 & -37:26:52.3 &  2440 \\ 
GEMS\_N3557\_2 & 2470 &   4.86 & NGC 3573                       & J1111186-365232 & 11:11:18.57 & -36:52:31.9 &  2410 \\ 
GEMS\_N3557\_3 & 2607 &   0.89 & ESO 377- G 034                 & J1117046-345719 & 11:17:04.50 & -34:57:19.0 &  2586 \\ 
GEMS\_N3557\_4 & 2768 &   1.57 & ESO 377- G 021                 & J1110565-355900 & 11:10:56.54 & -35:59:00.9 &  2750 \\ 
GEMS\_N3557\_5 & 2690 &   0.00 & ESO 318- G 033                 & $\cdots$            & 11:03:47.80 & -38:45:41.9 &  $\cdots$ \\ 
GEMS\_N3557\_6 & 2754 &   1.55 & ESO 319- G 015                 & $\cdots$            & 11:21:57.93 & -37:53:57.8 &  2737 \\ 
GEMS\_N3557\_7 & 2730 &   0.00 & ESO 319- G 007                 & $\cdots$            & 11:12:14.33 & -38:04:24.7 &  $\cdots$ \\ 
GEMS\_N3557\_8 & 2933 &   3.13 & ESO 377- G 024                 & J1112340-362541 & 11:12:33.99 & -36:25:40.6 &  2836 \\ 
GEMS\_N3557\_9 & 2921 &   0.00 & NVSS J105929-372102            & $\cdots$            & 10:59:40.44 & -37:23:15.1 &  $\cdots$ \\ 
GEMS\_N3557\_10 & 3119 &   0.48 & NGC 3533                       & J1107076-371022 & 11:07:07.55 & -37:10:21.5 &  3165 \\ 
GEMS\_N3557\_11 & 3106 &   2.75 & ESO 319- G 011                 & J1117534-403537 & 11:17:53.38 & -40:35:37.1 &  3115 \\ 
GEMS\_N3557\_12 & 3591 &   0.29 & ESO 318- G 030                 & J1101022-375955 & 11:01:02.19 & -37:59:54.9 &  3636 \\ 
GEMS\_N3557\_13 & 2443 &   6.76 & NGC 3573                       & J1111186-365232 & 11:11:18.57 & -36:52:31.9 &  2410 \\

\hline
GEMS\_N3783\_1$^3$ & 3202 &   6.01 & ESO 378- G 011        & J1134436-371257 & 11:34:43.57 & -37:12:57.4 &  3245 \\ 
GEMS\_N3783\_2$^3$ & 2733 & $\cdots$ & $\cdots$            & $\cdots$            & $\cdots$ & $\cdots$ &  $\cdots$ \\ 
GEMS\_N3783\_3$^3$ & 3018 &   0.40 & ESO 378- G 003        & $\cdots$            & 11:28:04.01 & -36:32:33.8 &  3022 \\ 
GEMS\_N3783\_4 & 3055 &   7.15 & NGC 3903                       & J1149035-373102 & 11:49:03.54 & -37:31:01.6 &  2983 \\ 
GEMS\_N3783\_4 & 3055 &   1.74 & AM 1147-371                    & $\cdots$            & 11:49:41.70 & -37:32:31.0 &  2964 \\ 
GEMS\_N3783\_5 & 2933 &   2.03 & ESO 378- G 023                 & J1148518-373042 & 11:48:51.83 & -37:30:42.0 &  2932 \\ 
GEMS\_N3783\_5 & 2933 &   1.34 & NGC 3903                       & J1149035-373102 & 11:49:03.54 & -37:31:01.6 &  2983 \\ 
GEMS\_N3783\_6 & 2991 &   1.81 & ESO 320- G 024                 & J1149261-384933 & 11:49:26.03 & -38:49:33.4 &  3037 \\  
GEMS\_N3783\_6 & 2991 &   4.88 & ESO 320- G 026                 & J1149503-384704 & 11:49:50.30 & -38:47:03.9 &  2717 \\ 
GEMS\_N3783\_6 & 2991 &   5.42 & 6dF J1149529-385431            & J1149529-385431 & 11:49:52.93 & -38:54:31.1 &  2707\\ 
GEMS\_N3783\_7 & 2923 &   3.94 & NGC 3783                       & J1139017-374419 & 11:39:01.72 & -37:44:18.9 &  2817 \\ 
GEMS\_N3783\_8$^3$ & 2947 & $\cdots$ & $\cdots$            & $\cdots$            & $\cdots$ & $\cdots$ &  $\cdots$ \\ 
GEMS\_N3783\_9 & 2705 &   5.17 & NGC 3742                       & J1135325-375723 & 11:35:32.51 & -37:57:23.0 &  2715 \\ 
GEMS\_N3783\_9 & 2705 &   0.96 & AM 1133-374                    & $\cdots$            & 11:35:45.70 & -38:01:20.0 &  2870 \\ 
GEMS\_N3783\_9 & 2705 &   3.05 & NGC 3749                       & J1135532-375951 & 11:35:53.21 & -37:59:50.5 &  2742 \\ 
GEMS\_N3783\_10 & 2810 &   0.04 & ESO 319- G 020                 & $\cdots$           & 11:26:06.00 & -37:51:26.0 &  $\cdots$ \\ 
GEMS\_N3783\_11 & 2740 &   7.22 & ESO 319- G 015                 & $\cdots$            & 11:21:57.93 & -37:53:57.8 &  2737 \\ 
GEMS\_N3783\_12 & 3034 &   4.30 & ESO 378- G 007                 & J1129480-371249 & 11:29:48.01 & -37:12:49.2 &  3041 \\ 

\hline
GEMS\_N3923\_1 & 1706 &   0.60 & UGCA 250                       & J1153241-283311 & 11:53:24.06 & -28:33:11.4 &  1699 \\ 
GEMS\_N3923\_1 & 1706 &   1.30 & 2MASX J11532725-2833064        & J1153273-283306 & 11:53:27.25 & -28:33:06.0 &  1664 \\ 
GEMS\_N3923\_2 & 1634 &   1.95 & ESO 504- G 025                 & $\cdots$            & 11:53:50.64 & -27:21:00.0 &  1637 \\ 
GEMS\_N3923\_3 & 1842 &   0.43 & ESO 440- G 004                 & $\cdots$            & 11:45:41.88 & -28:21:59.5 &  1842 \\ 
GEMS\_N3923\_4 & 1875 &   2.48 & ESO 504- G 017                 & J1148464-272245 & 11:48:46.31 & -27:22:45.0 &  1874 \\ 
GEMS\_N3923\_5 & 1907 &   1.07 & ESO 504- G 024                 & $\cdots$            & 11:53:37.89 & -26:59:44.9 &  1894 \\ 
GEMS\_N3923\_6 & 2027 &   3.45 & NGC 3936                       & J1152206-265421 & 11:52:20.59 & -26:54:21.2 &  2011 \\ 
GEMS\_N3923\_7 & 2022 &   1.22 & ESO 440- G 037                 & J1159171-285418 & 11:59:17.09 & -28:54:17.6 &  2009 \\ 
GEMS\_N3923\_8$^4$ & 1986 &   1.41 & ESO 439- G 025        & J1143458-303713 & 11:43:45.76 & -30:37:13.4 &  1984 \\ 
GEMS\_N3923\_9 & 2059 &   0.54 & ESO 440- G 039                 & $\cdots$            & 12:01:57.87 & -30:14:12.5 &  2045 \\ 
GEMS\_N3923\_10 & 2202 &   4.77 & ESO 440- G 044                & $\cdots$            & 12:02:46.40 & -29:05:33.0 &  2198 \\ 
GEMS\_N3923\_11$^4$ & 1600 &   $\cdots$ & $\cdots$      & $\cdots$  & $\cdots$ & $\cdots$ & $\cdots$ \\ 
GEMS\_N3923\_12 & 1938 &   0.31 & UGCA 247                      & J1148456-281734 & 11:48:45.62 & -28:17:34.9 &  1978 \\ 
GEMS\_N3923\_13$^4$ & 2131 &   2.20 & [KK2000] 47          & $\cdots$            & 11:57:30.77 & -28:07:27.2 &  2125$^*$ \\ 
\hline\\

\end{tabular}
\end{table*}

\begin{table*}
\centering
\caption{Optical counterparts to the \HI\ detections. The columns are the same as in Table~\ref{tab:app2}1.}
\begin{tabular}{lcclrrrr}

(1)&(2)&(3)&(4)&(5)&(6)&(7)&(8)\\

\hline
GEMS\_N4636\_1 & 986 &   0.85 & NGC 4688                       & $\cdots$            & 12:47:46.46 & 04:20:09.9 &    986 \\ 
GEMS\_N4636\_1 & 986 &   0.91 & SDSS J124747.00+041959.3       & $\cdots$            & 12:47:47.00 & 04:19:59.3 &    968 \\ 
GEMS\_N4636\_1 & 986 &   5.92 & CGCG 043-029 NED01             & $\cdots$            & 12:47:59.70 & 04:26:02.0 &   1041 \\ 
GEMS\_N4636\_2 & 1127 &   6.92 & UZC-CG 173                     & $\cdots$            & 12:32:42.00 & 00:12:53.0 &  1394 \\ 
GEMS\_N4636\_2 & 1127 &   0.89 & NGC 4517                       & $\cdots$            & 12:32:45.59 & 00:06:54.1 &  1131 \\ 
GEMS\_N4636\_3 & 1158 &   1.16 & UGC 07982                      & $\cdots$            & 12:49:50.19 & 02:51:10.4 &  1155 \\ 
GEMS\_N4636\_4 & 1180 &   1.32 & UGC 07911                      & $\cdots$            & 12:44:28.77 & 00:28:05.0 &  1183 \\ 
GEMS\_N4636\_4 & 1180 &   5.59 & VLA J124445.6-002536           & $\cdots$            & 12:44:45.60 & 00:25:36.0 &  1375 \\ 
GEMS\_N4636\_5 & 1134 &   4.09 & UGC 07715                      & $\cdots$            & 12:33:55.69 & 03:32:46.2 &  1138 \\ 
GEMS\_N4636\_6 & 1222 &   0.55 & UGC 07824                      & $\cdots$            & 12:39:50.31 & 01:40:21.5 &  1227 \\ 
GEMS\_N4636\_7 & 1217 &   5.71 & SDSS J125328.31+021023.4       & $\cdots$            & 12:53:28.31 & 02:10:23.5 &  977 \\ 
GEMS\_N4636\_7 & 1217 &   5.40 & NGC 4772                       & $\cdots$            & 12:53:29.16 & 02:10:06.2 &  1040 \\ 
GEMS\_N4636\_8 & 1232 &   1.56 & VCC 1468                       & $\cdots$            & 12:32:56.86 & 04:34:44.5 &  1233 \\ 
GEMS\_N4636\_8 & 1232 &   1.18 & SDSS J123258.82+043445.0       & $\cdots$            & 12:32:58.82 & 04:34:45.0 &  1207 \\ 
GEMS\_N4636\_9 & 1441 &   1.13 & UGC 07780                      & $\cdots$            & 12:36:42.08 & 03:06:30.2 &  1441 \\ 
GEMS\_N4636\_10 & 1529 &   0.97 & NGC 4517A                      & $\cdots$            & 12:32:28.15 & 00:23:22.8 &  1509 \\ 
GEMS\_N4636\_11 & 1737 &   0.26 & NGC 4527                       & $\cdots$            & 12:34:08.50 & 02:39:13.7 &  1736 \\ 
GEMS\_N4636\_12 & 1722 &   2.53 & SDSS J124227.74+000253.5       & $\cdots$            & 12:42:27.75 & 00:02:53.6 &  1517 \\ 
GEMS\_N4636\_12 & 1722 &   0.34 & NGC 4632                       & $\cdots$            & 12:42:32.80 & 00:04:47.0 &  1723 \\ 
GEMS\_N4636\_13 & 1701 &   0.65 & UGC 07841                      & $\cdots$            & 12:41:11.60 & 01:24:37.0 &  1737 \\ 
GEMS\_N4636\_14 & 1806 &   6.43 & NGC 4533                       & $\cdots$            & 12:34:22.02 & 02:19:31.3 &  1753 \\ 
GEMS\_N4636\_14 & 1806 &   1.95 & NGC 4536                       & $\cdots$            & 12:34:27.13 & 02:11:16.4 &  1808 \\ 
GEMS\_N4636\_15 & 1733 &   0.42 & NGC 4496A                      & $\cdots$            & 12:31:39.21 & 03:56:22.1 &  1730 \\ 
GEMS\_N4636\_15 & 1733 &   0.50 & SDSS J123139.98+035631.5       & $\cdots$            & 12:31:39.98 & 03:56:31.5 &  1747 \\ 
GEMS\_N4636\_15 & 1733 &   0.86 & NGC 4496                       & $\cdots$            & 12:31:40.10 & 03:55:59.0 &  1738 \\ 
GEMS\_N4636\_16 & 1731 &   2.40 & IC 3474                        & $\cdots$            & 12:32:36.51 & 02:39:41.5 &  1727 \\ 
GEMS\_N4636\_17 & 2296 &   0.44 & [ISI96] 1228+0116              & $\cdots$            & 12:31:16.69 & 00:59:39.3 &  2289 \\ 
GEMS\_N4636\_18 & 2472 &   2.75 & SDSS J123402.06+014555.7       & $\cdots$            & 12:34:02.06 & 01:45:55.7 &  2502 \\ 
GEMS\_N4636\_19 & 1632 &   1.17 & VCC 1713                       & $\cdots$            & 12:37:29.10 & 04:45:05.1 &  1655 \\ 
GEMS\_N4636\_20 & 1070 &   4.23 & NGC 4592                       & $\cdots$            & 12:39:18.73 & 00:31:55.2 &  1069 \\ 
GEMS\_N4636\_21 & 1518 &   3.69 & 2MASXi J1245041-002851         & $\cdots$            & 12:45:04.20 & 00:28:51.5 &  1679 \\ 
GEMS\_N4636\_21 & 1518 &   3.22 & [FNC2004] J124507.23-002740.3  & $\cdots$            & 12:45:07.23 & 00:27:40.3 &  1559 \\ 
GEMS\_N4636\_21 & 1518 &   2.87 & NGC 4666                       & $\cdots$            & 12:45:08.68 & 00:27:42.9 &  1533 \\ 
GEMS\_N4636\_21 & 1518 &   4.52 & NGC 4668                       & $\cdots$            & 12:45:31.99 & 00:32:08.6 &  1623 \\ 
GEMS\_N4636\_21 & 1518 &   5.14 & 2QZ J124532.4-003253           & $\cdots$            & 12:45:32.46 & 00:32:52.9 &  1499 \\ 
GEMS\_N4636\_21 & 1518 &   7.86 & SDSS J124547.90-002556.1       & $\cdots$            & 12:45:47.90 & 00:25:56.1 &  1657 \\ 
GEMS\_N4636\_22 & 1331 &   1.78 & UGC 08041                      & $\cdots$            & 12:55:12.65 & 00:07:00.0 &  1359 \\ 
\hline\\

\end{tabular}
\end{table*}

\begin{table*}
\centering
\caption{Optical counterparts to the \HI\ detections. The columns are the same as in Table~\ref{tab:app2}1.}
\begin{tabular}{lcclrrrr}
(1)&(2)&(3)&(4)&(5)&(6)&(7)&(8)\\

\hline
GEMS\_N5044\_1$^4$ & 3074 &   1.60 & LEDA 083818           & $\cdots$            & 13:14:09.93 & -16:41:41.6 &  $\cdots$ \\ 
GEMS\_N5044\_2 & 2928 &   4.12 & NGC 5010                       & $\cdots$            & 13:12:26.35 & -15:47:52.3 &  2975 \\ 
GEMS\_N5044\_3 & 2897 &   0.35 & IC 4221                        & $\cdots$            & 13:18:30.35 & -14:36:32.0 &  2895 \\ 
GEMS\_N5044\_4 & 2967 &   0.84 & MCG -03-33-029                 & $\cdots$            & 13:03:11.51 & -17:22:33.4 &  2964 \\ 
GEMS\_N5044\_5$^4$ & 2912 &   2.10 & [RC3] 1303.0-1530     & $\cdots$            & 13:05:37.27 & -15:46:58.5 & $\cdots$ \\ 
GEMS\_N5044\_6 & 2749 &   1.65 & NGC 5073                       & $\cdots$            & 13:19:20.65 & -14:50:40.3 &  2744 \\ 
GEMS\_N5044\_7 & 2828 &   $\cdots$ & $\cdots$                   & $\cdots$            & $\cdots$ & $\cdots$ & $\cdots$ \\ 
GEMS\_N5044\_8 & 2753 &   0.73 & MCG -03-34-014                 & $\cdots$            & 13:12:35.43 & -17:32:32.7 &  2760 \\ 
GEMS\_N5044\_9 & 2773 &   1.49 & ESO 576- G 017                 & J1315128-175800 & 13:15:12.80 & -17:58:00.4 &  3006 \\ 
GEMS\_N5044\_10$^{1,4}$ & 2748 &   1.87 & [MMB2004] J1320-1427 & $\cdots$            & 13:20:13.00 & -14:27:32.0 &  $\cdots$ \\ 
GEMS\_N5044\_11 & 2688 &   1.43 & SGC 1317.2-1702                & $\cdots$            & 13:19:54.83 & -17:18:55.8 &  2689 \\ 
GEMS\_N5044\_12 & 2624 &   3.77 & 2MASX J13164875-1620397        & J1316488-162040 & 13:16:48.75 & -16:20:39.7 &  2619 \\ 
GEMS\_N5044\_12 & 2624 &   3.16 & MCG -03-34-041                 & $\cdots$            & 13:17:06.13 & -16:15:07.9 &  2628 \\ 
GEMS\_N5044\_13 & 2694 &   3.12 & MCG -03-34-020                 & $\cdots$            & 13:13:12.48 & -16:07:50.1 &  2663 \\ 
GEMS\_N5044\_14$^4$ & 2599 & 1.80 & [RC3] 1305.5-1430     & $\cdots$            & 13:08:05.02 & -14:44:53.3 &  $\cdots$ \\ 
GEMS\_N5044\_15 & 2572 &   1.9  & PGC 045101                     & $\cdots$            & 13:03:26.57 & -14:46:08.0  & 2555 \\ 
GEMS\_N5044\_16 & 2502 &   0.44 & UGCA 338                       & $\cdots$            & 13:13:34.34 & -15:25:55.2 &  2503 \\ 
GEMS\_N5044\_17 & 2494 &   0.98 & SGC 1316.2-1722                & $\cdots$            & 13:18:56.50 & -17:38:06.0 &  2495 \\ 
GEMS\_N5044\_18$^4$ & 2472 & $\cdots$ & $\cdots$            & $\cdots$            & $\cdots$ & $\cdots$ &  $\cdots$ \\ 
GEMS\_N5044\_19 & 2520 &   3.11 & IC 0863                        & J1317124-171516 & 13:17:12.40 & -17:15:16.1 &  2514 \\ 
GEMS\_N5044\_20 & 1834 &   0.87 & MCG -03-34-040                 & J1316562-163535 & 13:16:56.23 & -16:35:34.7 &  2112 \\ 
GEMS\_N5044\_20 & 1834 &   1.74 & NGC 5054                       & $\cdots$            & 13:16:58.49 & -16:38:05.5 &  1741 \\ 
GEMS\_N5044\_21 & 1924 &    2.4 &  UGCA 353                      & J1324418-194145 & 13:24:42.10 & -19:41:49.8 &  1984 \\  
GEMS\_N5044\_22 & 2834 &   2.15 & MCG -03-33-031                 & $\cdots$            & 13:05:15.60 & -16:53:19.0 &  2842 \\ 
GEMS\_N5044\_23 & 2576 &   2.45 & MCG -03-34-004                 & $\cdots$            & 13:09:44.07 & -16:36:07.6 &  2619 \\ 
GEMS\_N5044\_23 & 2576 &   6.94 & NGC 4997                       & $\cdots$            & 13:09:51.70 & -16:30:55.7 &  2376 \\ 
GEMS\_N5044\_23 & 2576 &   7.13 & 6dF J1309535-163102            & J1309535-163102 & 13:09:53.47 & -16:31:01.6 &  2376 \\ 
\hline
\end{tabular}
\end{table*}
\begin{table*}
\centering
\caption{Optical counterparts to the \HI\ detections. The columns are the same as in Table~\ref{tab:app2}1.}
\begin{tabular}{lcclrrrr}

(1)&(2)&(3)&(4)&(5)&(6)&(7)&(8)\\

\hline
GEMS\_N7144\_1$^4$ & 1598 &   1.0  & ESO 236- G 039        & $\cdots$            & 21:45:14.50 & -49:00:32.0 &  $\cdots$\\
GEMS\_N7144\_1$^4$ & 1598 &   1.2  & KTS 65                & $\cdots$            & 21:45:01.10 & -49:00:36.0 &  $\cdots$\\
GEMS\_N7144\_2 & 1722 &   1.52 & NGC 7213                       & $\cdots$            & 22:09:16.25 & -47:10:00.0 &  1784 \\ 
GEMS\_N7144\_2 & 1722 &   5.15 & 2MASX J22092496-4706129        & J2209250-470613 & 22:09:24.98 & -47:06:12.8 &  1787 \\ 
GEMS\_N7144\_3 & 1617 &   0.00 & ESO 236- G 041                 & $\cdots$            & 21:45:49.10 & -48:49:13.5 &  $\cdots$ \\ 
GEMS\_N7144\_4 & 1871 &   0.89 & NGC 7151                       & $\cdots$            & 21:55:03.72 & -50:39:22.5 &  1859 \\ 
GEMS\_N7144\_5$^4$ & 1860 &   3.10 & APMUKS(BJ) B215242.56-492853.8 & $\cdots$   & 21:55:56.92 & -49:14:38.3 &  $\cdots$ \\ 
GEMS\_N7144\_6$^4$ & 1956 &   3.97 & ESO 236- G 035        & $\cdots$            & 21:42:43.71 & -47:59:22.3 &  2086 \\ 
GEMS\_N7144\_6$^4$ & 1956 &   4.30 & ESO 236- G 036        & $\cdots$            & 21:42:53.10 & -47:51:28.0 &  $\cdots$ \\
GEMS\_N7144\_7$^4$ & 1958 &   0.00 & APMUKS(BJ) B213754.44-465227.8 & $\cdots$   & 21:41:09.10 & -46:38:48.0 &  $\cdots$ \\ 
GEMS\_N7144\_8 & 2075 &   1.81 & ESO 236- G 034                 & $\cdots$            & 21:42:46.36 & -51:17:12.4 &  2257 \\ 

\hline
GEMS\_N7714\_1 & 2891 &   1.38 & NGC 7731                       & $\cdots$            & 23:41:29.07 & 03:44:24.1 &  2886 \\ 
GEMS\_N7714\_1 & 2891 &   1.43 & KPG 590                        & $\cdots$            & 23:41:31.50 & 03:43:56.5 &  2801 \\ 
GEMS\_N7714\_1 & 2891 &   1.81 & NGC 7732                       & $\cdots$            & 23:41:33.87 & 03:43:29.8 &  2907 \\ 
GEMS\_N7714\_2 & 2721 &   2.95 & NGC 7714                       & $\cdots$            & 23:36:14.10 & 02:09:18.6 &  2798 \\ 
GEMS\_N7714\_2 & 2721 &   1.99 & ARP 284                        & $\cdots$            & 23:36:18.11 & 02:09:21.3 &  2795 \\ 
GEMS\_N7714\_2 & 2721 &   1.12 & NGC 7715                       & $\cdots$            & 23:36:22.08 & 02:09:24.1 &  2770 \\ 
GEMS\_N7714\_3 & 2674 &   2.00 & UGC 12709                      & $\cdots$            & 23:37:24.02 & 00:23:30.1 &  2682 \\ 
GEMS\_N7714\_4 & 2901 &   6.14 & NGC 7750                       & $\cdots$            & 23:46:37.84 & 03:47:59.3 &  2938 \\ 
GEMS\_N7714\_5 & 2588 &   2.89 & NGC 7716                       & $\cdots$            & 23:36:31.46 & 00:17:50.3 &  2571 \\ 
GEMS\_N7714\_6 & 2599 &   7.55 & APMUKS(BJ) B233219.69+005549.2 & $\cdots$            & 23:34:53.31 & 01:12:24.8 &  2602 \\ 
GEMS\_N7714\_7 & 2755 &   $\cdots$ & $\cdots$                   & $\cdots$            & $\cdots$    & $\cdots$   &  $\cdots$ \\

\hline
GEMS\_HCG90\_1 & 2271 &   2.49 & ESO 404- G 018                 & $\cdots$            & 22:01:10.16 & -32:34:43.7 &  2268 \\ 
GEMS\_HCG90\_2 & 2353 &   2.15 & ESO 466- G 036                 & $\cdots$            & 22:01:20.46 & -31:31:46.9 &  2559 \\ 
GEMS\_HCG90\_2 & 2353 &   1.92 & 2dFGRS S408Z202                & $\cdots$            & 22:01:21.44 & -31:31:52.9 &  2457 \\ 
GEMS\_HCG90\_3 & 2550 &   1.91 & ESO 404- G 027                 & J2203478-321706 & 22:03:47.84 & -32:17:06.0 &  2611 \\ 
GEMS\_HCG90\_4 & 2583 &   1.76 & NGC 7204                       & $\cdots$            & 22:06:54.20 & -31:03:05.0 &  2590 \\ 
GEMS\_HCG90\_4 & 2583 &   1.54 & NGC 7204B                      & $\cdots$            & 22:06:55.26 & -31:03:11.1 &  2590 \\ 
GEMS\_HCG90\_5 & 2614 &   5.57 & NGC 7154                       & $\cdots$            & 21:55:21.04 & -34:48:50.9 &  2616 \\ 
GEMS\_HCG90\_6 & 2656 &   3.89 & ESO 404- G 012                 & J2157072-343456 & 21:57:07.20 & -34:34:55.7 &  2646 \\ 
GEMS\_HCG90\_7 & 2515 &   7.00 & ESO 404- G 017                 & J2200559-351713 & 22:00:55.87 & -35:17:13.2 &  2478 \\ 
GEMS\_HCG90\_8 & 2670 &   1.64 & ESO 404- G?033                 & $\cdots$            & 22:08:31.00 & -34:17:04.2 &  2596 \\ 
GEMS\_HCG90\_8 & 2670 &   1.50 & 2MASX J22083186-3417070        & J2208317-341706 & 22:08:31.74 & -34:17:05.9 &  2709 \\ 
GEMS\_HCG90\_9 & 2720 &   3.05 & IC 5156                        & J2203149-335018 & 22:03:14.87 & -33:50:18.4 &  2709 \\ 
GEMS\_HCG90\_10 & 2560 &   3.53 & DUKST 467-039                 & $\cdots$            & 22:08:38.24 & -30:53:08.3 &  2613 \\ 
GEMS\_HCG90\_11$^4$ & 2596 &   2.17 & ESO 467- G 002       & J2206096-310518 & 22:06:09.60 & -31:05:17.7 &  2521 \\ 
GEMS\_HCG90\_11$^4$ & 2596 &   7.64 & NGC 7204             & $\cdots$            & 22:06:54.20 & -31:03:05.0 &  2590 \\ 
GEMS\_HCG90\_12 & 2316 &   3.74 & NGC 7172                       & J2202019-315211 & 22:02:01.87 & -31:52:11.1 &  2557 \\ 
GEMS\_HCG90\_12 & 2316 &   4.78 & NGC 7173                       & $\cdots$            & 22:02:03.19 & -31:58:25.3 &  2497 \\ 
GEMS\_HCG90\_12 & 2316 &   4.20 & 2dFGRS S407Z097                & $\cdots$            & 22:02:04.81 & -31:52:13.5 &  2384 \\ 
GEMS\_HCG90\_12 & 2316 &   6.24 & NGC 7176                       & J2202085-315923 & 22:02:08.45 & -31:59:23.3 &  2503   \\ 
GEMS\_HCG90\_12 & 2316 &   6.27 & 2dFGRS S407Z090                & $\cdots$            & 22:02:16.07 & -31:57:11.8 &  2541 \\ 
\hline\\
\end{tabular}
\end{table*}

\begin{table*}
\centering
\caption{Optical counterparts to the \HI\ detections. The columns are the same as in Table~\ref{tab:app2}1.}
\begin{tabular}{lcclrrrr}
(1)&(2)&(3)&(4)&(5)&(6)&(7)&(8)\\
\hline
GEMS\_IC1459\_1 & 1198 &   0.38 & NGC 7456                       & J2302104-393410 & 23:02:10.37 & -39:34:09.8 &  1187 \\ 
GEMS\_IC1459\_2 & 1299 &   1.01 & IC 5273                        & J2259267-374210 & 22:59:26.70 & -37:42:10.4 &  1286 \\ 
GEMS\_IC1459\_3 & 1239 &   1.01 & ESO 406- G 040                 & $\cdots$            & 23:00:22.18 & -37:12:04.8 &  1248 \\ 
GEMS\_IC1459\_4 & 1282 &   2.19 & ESO 406- G 022                 & J2255526-343318 & 22:55:52.59 & -34:33:17.8 &  1285 \\ 
GEMS\_IC1459\_5 & 1372 &   0.26 & ESO 406- G 042                 & $\cdots$            & 23:02:14.21 & -37:05:01.4 &  1375 \\ 
GEMS\_IC1459\_6 & 1446 &   1.59 & NGC 7418                       & J2256361-370148 & 22:56:36.13 & -37:01:47.8 &  1417 \\ 
GEMS\_IC1459\_7 & 1667 &   1.39 & IC 5269B                       & J2256367-361459 & 22:56:36.72 & -36:14:59.1 &  1638 \\ 
GEMS\_IC1459\_8 & 1789 &   1.79 & NGC 7421                       & J2256543-372050 & 22:56:54.33 & -37:20:50.7 &  1801 \\ 
GEMS\_IC1459\_9 & 1790 &   5.98 & IC 5264                        & $\cdots$            & 22:56:53.04 & -36:33:15.0 &  1940 \\ 
GEMS\_IC1459\_9 & 1790 &   7.73 & IC 1459                        & J2257106-362744 & 22:57:10.61 & -36:27:44.2 &  1713 \\ 
GEMS\_IC1459\_9 & 1790 &   5.43 & 2MASX J22571092-3640103        & $\cdots$            & 22:57:10.92 & -36:40:10.4 &  1945 \\ 
GEMS\_IC1459\_10 & 1773 &   0.12 & IC 5269C                       & $\cdots$            & 23:00:48.55 & -35:22:10.5 &  1796 \\ 
GEMS\_IC1459\_11 & 1950 &   3.52 & IC 5270                        & J2257549-355129 & 22:57:54.86 & -35:51:28.5 &  1929 \\ 
GEMS\_IC1459\_12 & 2101 &   2.11 & NGC 7418A                      & $\cdots$            & 22:56:41.15 & -36:46:21.2 &  2102 \\ 
GEMS\_IC1459\_12 & 2101 &   7.17 & 2MASX J22571092-3640103        & $\cdots$            & 22:57:10.92 & -36:40:10.4 &  1945 \\ 
GEMS\_IC1459\_13 & 2279 &   2.14 & NGC 7368                       & J2245317-392031 & 22:45:31.68 & -39:20:30.7 &  2355 \\ 
GEMS\_IC1459\_14 & 2147 &   4.65 & ESO 345- G 046                 & $\cdots$            & 22:43:16.09 & -39:51:58.9 &  2149 \\ 
GEMS\_IC1459\_15 & 1718 &   1.42 & IC 5271                        & $\cdots$            & 22:58:01.82 & -33:44:32.0 &  1738 \\ 
GEMS\_IC1459\_16 & 1787 &   3.04 & ESO 346- G 033                 & $\cdots$            & 23:08:46.60 & -39:48:42.7 &  1759 \\ 
GEMS\_IC1459\_17 & 879 &   0.93 & ESO 346- G 007                 & $\cdots$            & 22:53:23.90 & -38:47:58.4 &  928 \\ 
GEMS\_IC1459\_18 & 1668 &   4.69 & DUKST 406-083                  & $\cdots$            & 23:02:00.49 & -36:29:01.9 &  1624 \\ 
\hline

\end{tabular}
\flushleft
$^{1-4}$ Sources have ATCA observations, and thus a confirmed optical counterpart (except in the case of GEMS\_N3783\_2 where no optical counterpart is seen).The references are: \\
$^1$ \citet{mckay04}\\
$^2$ \citet{kilborn05}\\
$^3$ \citet{kilborn06}\\
$^4$ \citet{kern07}\\
$^*$ Velocity is a previous \HI\ measurement associated with this galaxy.\\
$^{**}$ Previously catalogued optical counterpart is further than 8 arcminutes from the central \HI\ position. Confirmation of optical counterpart is required.

\end{table*}

\bsp

\label{lastpage}
\end{document}